\documentclass[sts]{imsart}

\RequirePackage{amsthm,amsmath,amsfonts,amssymb}
\RequirePackage[numbers]{natbib}
\usepackage{hyperref}
\hypersetup{citecolor=blue,urlcolor=blue}

\RequirePackage{graphicx,xcolor}

\startlocaldefs
\theoremstyle{plain}

\theoremstyle{remark}

\endlocaldefs

\begin{document}

\begin{frontmatter}
\title{Aitchison's Compositional Data Analysis 40 Years On: A Reappraisal}
\runtitle{Compositional data analysis: a reappraisal}

\begin{aug}
\author[A]{\fnms{Michael} \snm{Greenacre}\ead[label=e1]{michael.greenacre@upf.edu}\orcid{0000-0002-0054-3131}\ead[label=u1,url]{www.econ.upf.edu/~michael}},
\author[B]{\fnms{Eric} \snm{Grunsky}\ead[label=e2]{egrunsky@gmail.com}\orcid{0000-0003-4521-163X}},
\author[C]{\fnms{John} \snm{Bacon-Shone}\ead[label=e3]{johnbs@hku.hk}\orcid{0000-0002-9827-1815}},
\author[D]{\fnms{Ionas} \snm{Erb}\ead[label=e4]{ionas.erb@crg.eu}\orcid{0000-0002-2331-9714}},
\and
\author[E]{\fnms{Thomas} \snm{Quinn}\ead[label=e5]{contacttomquinn@gmail.com}\orcid{0000-0003-0286-6329}}.

\address[A]{Michael Greenacre is Professor, Department of Economics \& Business,
Universitat Pompeu Fabra, and Barcelona School of Management, Barcelona, Spain \printead{e1,u1}.}

\address[B]{\newline Eric Grunsky is Adjunct Professor,  Department of Earth \& Environmental Sciences, University of Waterloo, Canada \printead{e2}.}

\address[C]{\newline John Bacon-Shone is Honorary Professor, Faculty of Social Sciences,
Hong Kong University, Hong Kong \printead{e3}.}

\address[D]{\newline Ionas Erb is Researcher, Centre de Regulació Genòmica,
Barcelona, Spain \printead{e4}.}

\address[E]{\newline Thomas Quinn is Researcher, Applied Artificial Intelligence Institution (A2I2), Deakin University, Australia \printead{e5}.}
\end{aug}

\begin{abstract}
The development of John Aitchison's approach to compositional data analysis is followed since his paper read to the Royal Statistical Society in 1982.
Aitchison's logratio approach, which was proposed to solve the problematic aspects of working with data with a fixed sum constraint, is summarized and reappraised.
It is maintained that the properties on which this approach was originally built, the main one being subcompositional coherence, are not required to be satisfied exactly --- quasi-coherence is sufficient, that is near enough to being coherent for all practical purposes.
This opens up the field to using simpler data transformations, such as power transformations, that permit zero values in the data.
The  additional property of exact isometry, which was subsequently introduced and not in Aitchison's original conception, imposed the use of isometric logratio transformations, but these are complicated and problematic to interpret, involving ratios of geometric means.
If this property is regarded as important in certain analytical contexts, for example unsupervised learning, it can be relaxed by showing that regular pairwise logratios, as well as the alternative quasi-coherent transformations, can also be quasi-isometric, meaning they are close enough to exact isometry for all practical purposes.
It is concluded that the isometric and related logratio transformations such as pivot logratios are not a prerequisite for good practice, although many authors insist on their obligatory use.
This conclusion is fully supported here by case studies in geochemistry and in genomics, where the good performance is demonstrated of pairwise logratios, as originally proposed by Aitchison, or Box-Cox power transforms of the original compositions where no zero replacements are necessary.

\end{abstract}


\end{frontmatter}

\section{Introduction}

It is now 40 years since John Aitchison's read paper, with discussion, to the Royal Statistical Society (RSS), ``The Statistical Analysis of Compositional Data'' \cite{Aitchison:82}, read on 13 January, 1982, where he discussed the problems that arise when analysing data in the positive simplex, that is data subject to the unit-sum constraint.
Four years later Aitchison published his seminal book with the same title \cite{Aitchison:86}, which established a field of statistics based on the so-called \textit{logratio} transformation.
Since then, compositional data analysis (which we will refer to as CoDA) has experienced a development in several directions theoretically and in many different fields practically.
The present article aims to trace the development of Aitchison's approach and make some comments about the future of the subject, especially its practice.
Notice that we do not discuss several model-based approaches to CoDA that do not use logratios, which have experienced an independent development over the years, for example, approaches based on distributions other than the logistic normal \cite{ScealyWelsh:11, Smithson:22}.

In the original RSS paper, Aitchison introduced the concept of the logratio transformation and discussed applications to four compositional data sets, three in a geological context and one on household expenditure budgets, as a motivation for his approach. 
The fact that these data sets had modest numbers of components (10, 4, 3 and 7, respectively) is relevant to our present discussion, and it is a fact that, until recently, most of the development in this area has been concerned with data sets having less than 50 components.
However, the present-day use of CoDA is increasingly aimed at much higher-dimensional data sets, for example in ecology, microbiome research and omics studies in general, where the numbers of components are in the hundreds or even thousands, so that the properties of CoDA, often regarded as foundational principles, need to be reconsidered and evaluated in this new context.

Several books have been published that present developments in the theory and practice of CoDA based on logratios \cite{PawlowskyBuccianti:11, PawlowskyEtAl:15, Greenacre:18, Filzmoser:18}.
This article takes a more critical look at the development of this approach over the last decades, a development that has not been without controversy.
Section 2 starts by describing a typical medium-sized compositional data set, in the context of geochemistry, which is the field where most of the original work in CoDA has taken place.
In Section 3 the various concepts and properties of CoDA are introduced, always illustrated in the context of the present geochemical data set.
In Section 4 the data are analysed by various approaches, using logratio transformations as well as an alternative approach using the chi-square standardization of correspondence analysis, combined with the Box-Cox power transformation.  
Section 5 deals with the important topic of variable selection, since it is advantageous to reduce the number of variables, whether the objective is unsupervised or supervised. 
Section 6 deals with compositional data with very many parts, in the hundreds or thousands, a situation which is becoming increasingly common in biology. 
Section 7 poses the question whether the coherence and isometry properties of CODA are necessary to be satisfied exactly, and concludes that quasi-coherence and quasi-isometry suffice, that is are close enough for all practical purposes.
Section 8 wraps up the reappraisal of present-day CoDA, gives some indications of future practice and gives John Aitchison the last word.

\section{Compositional data: the Tellus geochemical data set}
The Tellus geochemical data set is a typical candidate for compositional data analysis, and will be used to introduce CoDA concepts as well as provide a series of pertinent results.
These data emanate from a survey in Northern Ireland \cite{Smyth:07} which was conducted in two time periods, 1994--1996 and 2004--2006.
The data of concern here are the geochemistry of the soil samples collected in the second period, analysed by X-ray fluorescence (XRF) spectrometry.
Specifically, 52 elements in 6799 samples constitute the data set, which has been transformed from the original values to cation counts.
In each of the 6799 samples the total number of cations is related to the amount of rock material being processed and is irrelevant --- hence, the data are expressed as relative values by dividing each sample's set of 52 cation counts by their total.
This relativization of the data is called \textit{closure}, or \textit{normalization}, and turns the rows of cation counts into \textit{compositions}, each composition having values summing to 1, called the \textit{unit-sum constraint}.
There are other contexts where the totals in the original data do have substantive relevance --- this will be commented on later but not specifically treated here. 

To fix some basic notation, we mostly use Aitchison's original definitions, with some variations.
The normalized compositional data are observed on $D$ components, called \textit{parts}, and the unit-sum constraint implies that the compositions ${\bf x} = (x_1,\ldots,x_D)$ lie in a simplex space $\mathbb{S}^d$, where $d = D-1$:
\begin{align}
  \mathbb{S}^d &= \{(x_1,\ldots,x_d,x_{d+1}): x_j \geq 0, j\in\{1,...,d+1\},  \nonumber \\
  &\hspace{3.7cm} x_1+\ldots+x_{d+1} = 1\} .
    \label{simplex}
\end{align}
where, again, $d+1 = D$.
The dimensionality of the simplex is $d$, one less than the number of parts, since the value of any one part can be known based on the values of the remaining parts.
For example, a three-part composition is in a two-dimensional triangular simplex, a four-part composition is in a three-dimensional tetrahedral simplex, and so on.
The term ``dimensionality" is the geometric equivalent of the mathematical term ``rank", referring to the minimum number of dimensions of the space containing the compositions.

In this application the number of samples $N=6799$ is much larger than the number of parts $D=52$, and the dimensionality is $d=51$.
In the wide matrices that are typical in other areas such as genomics and microbiome studies, where $N < D$, the dimensionality of the compositional vectors within the simplex will be $N-1$ (see Section 6).

The inclusion of zero values in the definition (\ref{simplex}) of a composition in a closed simplex is not universally accepted by CoDA researchers who use log-transformed data, but this definition is preferred for several reasons. 
First and foremost, most observed compositional data sets do have zero values, and in some cases very many zeros, so many observed compositions lie on the edges and faces of the multidimensional simplex.
In order to apply Aitchison's logratio approach these will have to be replaced by small values.
When the presence of zeros is very extensive, zero replacement can become more problematic --- see the genomics application in Section 6.
Some data sets may contain structural zeros, that is, parts that will never be observed in some samples no matter how deeply the simplex is sampled, so to replace a true zero with some artificial value seems erroneous.
However, zeros are usually due to measurements below detection limits, where replacement with small positive values seems less problematic.
Second, the definition of a composition should arguably not depend on the method that is subsequently developed for its analysis, and there are indeed methods closely related to the logratio approach which do admit zeros --- these will be introduced and applied in subsequent sections.

In the Tellus data set there are 3883 data zeros, representing a fairly low percentage, 1.1\%, of all data values, due to observations on 12 of the 52 cation elements falling below their detection limits. 
There are certain minerals/rock types where some elements are incompatible and structural zeros can occur, but in the case of the Tellus data, such events would be exceptionally rare. We can safely assume that there are no structural zeros.
For purposes of computing logratios, the zeros were replaced using the simplest strategy of substituting zeros with 2/3 the minimum positive value of the respective element, assuming a triangular distribution for the zero element below this minimum.
Using alternative imputation strategies \cite{Palarea:15} did not change the essential results to be presented here.

A \textit{subcomposition} refers to a subset of $C=c+1$ parts of a composition of $D=d+1$ parts, where $C < D$, which has been normalized with respect to its own subtotals, and thus resides in a $c$-dimensional simplex $\mathbb{S}^c$.
In practice, almost all observed compositions are \textit{de facto} subcompositions, since there are many unobserved (or unobservable) parts. 
For example, a biochemical laboratory can only identify a subset of the fatty acids that its equipment allows, and more sophisticated equipment could identify more, which would obviously affect the relative values in the normalized samples.
The Tellus data set, however, with its 11 major elements, 18 minor elements, 14 trace elements and 9 rare earth elements, consists of a geochemical composition as complete as one can find in practice.

Aitchison also defined \textit{amalgamations}, which are groups of parts that are merged by summation, and \textit{partitions}, which are all the parts amalgamated into subsets of parts.
A good example is in fatty acid research where all the fatty acids are routinely partitioned into three amalgamations: saturated, monounsaturated and polyunsaturated fatty acids.
The summing of parts is a natural transformation of the data, and again depends on the research question. 
As well as geo- and biochemical variables, species, budget items and diet categories, similarly have parts that form natural groups, for example, species of bacteria grouped into genera.
For the Tellus data there are groupings of elements such as Mafic (=\textit{\textsf{Mg+Fe+Mn}}) and Felsic (=\textit{\textsf{Na+Si+Al+K}}), which can be investigated, and creating such groupings can also alleviate the zeros problem. 

Aitchison also drew attention to the hierarchy of parts where the simplex of the amalgamated parts in a partition could be studied as well as the simplices of all the subcompositions formed by the partition's subsets.
In other words, it is the relationships between parts in the same subset and between different subset amalgamations that are relevant, not only the parts that are in different subsets of the partition.

The samples in the Tellus data set are classified into 10 different ``age brackets" and one objective of the analysis is to understand the geochemical differences between these classes.
``Uncategorized" samples may or may not contribute to the processes reflected in the data.
The choice of accepting/rejecting uncategorized data may be based on how the composition fits with the overall compositional variation of the categorized data.
In some cases, the compositions of unlabelled/uncategorized samples may be uncharacteristic relative to those samples that are categorized. 
These samples can be accepted or rejected on a case-by-case basis, depending on the objectives.

Before we proceed to analyse the Tellus data, the basic properties of Aitchison-style CoDA are reviewed in the context of this geochemical data set.

\section{Basic properties of C\NoCaseChange{o}DA}
Most of Aitchison's original paper dealt with distributional issues, for example, the quest to find a transformation of compositions into interval-scale multivariate vectors where the multivariate normal distribution, for example, could be validly used.
The various logistic transformations proposed to produce ``transformed-normal'' models all involved logarithmic transformations, which precluded zero data values, and hence there arose the issue of zero replacement, which is still a debated issue today even if glossed over in publications.

Through Aitchison's early work, several properties were laid down as fundamental to the good practice of CoDA, which are dealt with here in turn, as well as some additional ones introduced subsequently.

\subsection{Scale invariance}  
This is the formal property that enunciates the requirement that only relative values are of interest for compositional data.
The results should be the same for data matrices $\bf X$ and $\bf Y$ if ${\bf Y} = {\bf D}_q {\bf X}$ for any diagonal matrix  ${\bf D}_q$ of positive scalars $q_1, q_2,\ldots,q_N$.
That is, the data vectors in the rows of $\bf X$ and $\bf Y$ are compositionally equivalent: ${\rm C}({\bf x}_i) = {\rm C}({\bf y}_i)$ for all rows $i \in  \{1,\ldots,N\}$, where C is the operation of closure (normalization).
Hence, if one of the original Tellus cation samples were a positive multiple of another one (i.e., they are simply of different overall sizes), then they are compositionally equivalent.
This justifies restricting the definition of a composition to be a non-negative data vector divided by its total, in which case the original total is referred to as the ``size'' of the vector and the composition, summing to 1, as its ``shape'' \cite{Greenacre:17}.
As mentioned before, in many cases the size of the vector will be of no interest, while in a context where the totals have a substantive interpretation, there could be a relationship between size and shape.
For example, adult fish of a particular species can have different shapes compared to younger small fish, in terms of the relative values of their morphometric measurements: that is, shape depends on size.

\subsection{Subcompositional coherence}
The main reason for Aitchison's approach using ratios of parts is a consequence of this property.
As an illustration in the context of the Tellus data, suppose geochemist A analyses the full data set, while other geochemists have more specific research objectives that require studying subsets of elements. 
For example, geochemist B could be studying different rock types (e.g., granite, volcanics, sediments) and thus analysing just the 11 major elements. 
Geochemist C could be studying different igneous rock assemblages and thus only be concerned with the 9 rare earth elements. 
Both geochemists B and C will have subcompositions normalized to their respective element totals, and thus have data different from geochemist A who is studying the full composition of 52 elements. 
For example, geochemist A would find a correlation between the two rare earth elements \textit{\textsf{La}} and \textit{\textsf{Y}} equal to $0.479$, whereas geochemist C studying the subcomposition of rare earth elements would find a correlation of $-0.344$ (these are actual correlations computed on the Tellus data).  
If, however, all three geochemists analysed element ratios, they would have identical data and no such paradox. 

\textit{Subcompositional coherence}, or simply \textit{coherence}, is thus the property that is obeyed by the part ratios, ensuring identical results for ratios and their relationship to other ratios whether the parts are included in compositions or subcompositions.
Coherence seems like a perfectly solid property to have, and avoids the problematic situation of changing correlations, for example, when parts are in compositions or subcompositions. 
Invariably, these violations of coherence when not using ratios are illustrated with data having very few parts, but when the number of parts is very large, as in many present-day biological applications, this lack of coherence can become diluted and less important, as will be demonstrated later in Sections 4.4 and 6.5.
A measure of the lack of coherence, called \textit{subcompositional incoherence}, can be defined \cite{Greenacre:11a}, and when this lack of coherence is very small, we can then talk of quasi-coherence, which would be satisfactory for all practical purposes.

\subsection{Pairwise and additive logratios}

The most fundamental data transformation for CoDA is thus the  forming of part ratios, followed by logarithmic transformation --- these are called \textit{pairwise logratios} and abbreviated throughout as LR. 
The log-transformation converts the multiplicative ratio scale to an additive interval scale, also serving the practical purpose of reining in the tail of the right-skewed distributions of the ratios. 
Given a composition ${\bf x} = [ x_1 \ x_2 \ \cdots \ x_D ]$, a general LR is defined as
\begin{eqnarray}
{\rm LR}(j,k) &=& \log(x_j/x_k),\ j,k\in \{1,\ldots,D\}, \label{LR} \\ 
              &\  & \hspace{1.9cm} \ j\neq k \nonumber \\
              &=& \log(x_j) - \log(x_k).\nonumber
\end{eqnarray}
\noindent
In the case of the Tellus data, LRs are of the form log(\textit{\textsf{Si/Al}}), log(\textit{\textsf{Si/Fe}}),..., log(\textit{\textsf{Al/Fe}}), etc..., and there are a total of $\frac{1}{2}\times 52 \times 51 = 1326$ unique LRs.
However, even though there are $\frac{1}{2}D(D-1)$ unique LRs, they are linearly inter-related to the extent that their dimensionality is much lower, equal to $d=D-1$.
Thus, the LRs take the $d$-dimensional compositions in the simplex $\mathbb{S}^d$ to the real space $\mathbb{R}^d$ of the same dimensionality, which for the Tellus data is $d=51$.

There is the important special case of the additive logratio (ALR) transformation, appearing in Aitchison's earliest work. 
This is a subset of $d$ LRs with the same denominator, called the \textit{reference} part.  
The ALR transformation with respect to reference part \textit{ref} is defined as
\begin{eqnarray}
{\rm ALR}(\,j\,|\,{\it ref}) &=& \log(x_j / x_{\it ref}), \ j\in \{1,\ldots,D\}, \label{ALR}\\
	                         &\  & \hspace{2.2cm} \ j\neq {\it ref} \nonumber \\
	                         &=&  \log\left(x_j\right) - \log\left(x_{\it ref}\right)\nonumber .
\end{eqnarray}  
There are $D$ possible ALR transformations, depending on the reference part chosen, so an important step will be to choose the reference to satisfy either a statistical or substantive objective, in the present case a geochemical one. 
For example, it is common in geochemistry to choose  silicon (\textit{\textsf{Si}}), in the form of silicon oxide (silica), as the reference part, since it is the most prolific element with relative values fairly stable across samples.

The use of ALRs will be shown to be beneficial in simplifying CoDA results, both in the Tellus data and in general. 
Furthermore, if the logarithm of the reference part, $\log(x_{\it ref})$, is almost constant (i.e., has very low variance), then the corresponding ALRs are practically the same as the logarithm of the numerator parts, up to an additive  amount that is nearly constant.
When a data set contains very many parts (i.e., it is wider, see Section 6), there is a good chance to find such a near constant reference part, but as will be seen in Section 4, this is also found to be possible for narrower data sets such as the 52-part Tellus data.

\subsection{Centered logratios}
The \textit{centered log ratio} (CLR) transformation is a symmetric transformation of the parts $x_j$ ($j\in \{1,\ldots,D\}$, defined as the logratios of individual parts with respect to their geometric mean $g({\bf x}) = (x_1 x_2 \cdots x_D)^{1/D}$:
\begin{eqnarray}
   {\rm CLR}(j) &=& \log\left(\frac{x_j}{g({\bf x})}\right), \ \  j\in \{1,\ldots,D\}    \label{CLR} \\
                &=& \log(x_j) - \frac{1}{D} \sum_{k=1}^D \log(x_k) \nonumber
\end{eqnarray} 		
The $j$-th CLR, CLR($j$), is also equal to the average of the $D$ LRs \ ${\rm LR}(j,k), k \in \{1,\ldots, D\}$, where one of them, ${\rm LR}(k,k)$, is zero.   

As noted  above, although there are $\frac{1}{2}D(D-1)$ LRs, their dimensionality is only $d$, the same as a set of $d$ ALRs or the set of $D$ CLRs, which sum to zero and hence also have dimensionality $d$.
Furthermore, the usefulness of the CLR is that the Euclidean distances between cases using the $D$ CLRs are identical to the distances using all $\frac{1}{2}D(D-1)$ LRs \cite{AitchisonGreenacre:02}.
In other words, the two geometries are identical and the CLR transformation is termed \textit{isometric} --- see Section 3.8.

How LRs relate to the logratios obtained via CLR can be understood when considering the following embedding of the LR space $R^d$ in $R^D$:
\begin{align}
  \mathbb{T}^d = \{(y_1,\ldots,y_D):& \  y_j \in R, j\in\{1,\ldots,D\}, 
          \label{tangent} \\ 
                                  & \ y_1+\cdots+y_D = 0\} \nonumber
\end{align}
This is the space of the CLR transformed parts, and the isometry discussed in Section 3.8 refers to transformations mapping the simplex to $\mathbb{T}^d$. 

\subsection{Subcompositional dominance}
Whereas subcompositional coherence applies to the analytical properties of the parts, usually the columns of a dataset, \textit{subcompositional dominance} refers to the cases, usually the rows.
Stated briefly, this property asserts that if compositional distances are computed between the cases, and then some parts are eliminated and the data renormalized, the between-case distances based on the subcompositions should be no larger than those based on the compositions.  
Clearly, this would be true if the distances were based on summing positive amounts for each of the LRs, since there would be fewer positive amounts to add up for the smaller subcomposition.

However, there is no pre-ordained reason why distances should be sums, they could just as well be (weighted) averages (see Sections 3.9 and 4.1), in which case the above property is not satisfied.
The benefits of defining distances as averages rather than sums is easily demonstrated.
Summing leads to distances getting increasingly larger as the number of parts increases, even when adding random parts, while averaging over the number of parts would make the distances comparable across studies having different numbers of parts. 
If parts that are compositionally close to being constant were eliminated, then with averaging over the parts in the resultant subcomposition, the distances between cases would tend to increase, since the higher differences in the higher variance parts are being conserved.

Since the distance metric chosen for a given analysis depends on the research question at hand, the conclusion here is that whether subcompositional dominance should or should not hold likewise depends on the research question. 
Thus, it is unnecessary as a required property for the good practice of CoDA.

\subsection{Distributional equivalence, or proportionality}
This is not one of Aitchison's basic properties for CoDA, but is included here as one that could be judged equally as attractive as subcompositional coherence, and at the same time demonstrates the benefit of part weighting \cite{GreenacreLewi:09}.
In fact, this property is one of the foundations of Benzécri's correspondence analysis \cite{Benzecri:73}, which has a very close relationship to CoDA (see Section 4.3).

Two parts in a composition are \textit{distributionally equivalent} if one is a scalar multiple of the other.
This has independently been called the property of \textit{proportionality} \cite{Lovell:15, Erb:16, Quinn:17}, which has been proposed as a basis for measuring correlation between parts.
Proportionality can be measured using the variance of the pairwise logratio LR $\textsf{Var}\left( \log(x_j/x_k) \right)$.
In the Tellus data the LR with the lowest variance, equal to $0.0062$, is $\log(\textit{\textsf{Sm/Yb}})$, so \textit{\textsf{Sm}} and \textit{\textsf{Yb}} are highly proportional.
At the other extreme, the highest variance, equal to $5.06$, is for $\log(\textit{\textsf{Br/Nd}})$, so \textit{\textsf{Br}} and \textit{\textsf{Nd}} are highly non-proportional.
A measure of proportionality can be obtained using the uncentered correlation coefficient on the original compositions, although it is theoretically incoherent.
The highest such correlation, with value 0.997, is found for the same pair \textit{\textsf{Sm}} and \textit{\textsf{Yb}}, which means that these two elements are almost distributionally equivalent.
It turns out that the compositional values for \textit{\textsf{Sm}} are an almost constant multiple of about 2.4 times those of \textit{\textsf{Yb}}.

Distributional equivalence is the property that an analysis of a compositional data set should not be affected if distributionally equivalent components are amalgamated.
Hence the almost exact proportionality of \textit{\textsf{Sm}} and \textit{\textsf{Yb}} means they could be amalgamated with almost no effect on the compositional analysis.
However, with the usual CoDA methods used today this property does not hold.

To take an extreme example, suppose that a column of the compositional data was arbitrarily divided into 10 equal parts.
Then these 10 parts would contribute 10 times the variance of the original column, whereas there is arguably only one variance.
It should not matter whether they are split or amalgamated into one part.
Weighting of the parts would rectify this situation \cite{GreenacreLewi:09} and make the results invariant to such amalgamation of distributionally equivalent (proportional) parts --- see Section 4.1.

Similar scenarios may occur in other real-world applications of CoDA. For example, in microbiome research, all compositional parts are related to one another by a phylogenetic tree, and it is left to the judgement of a researcher to decide where to draw the lines for subdivision. 
Even when bacterial species are well-defined, it may be appropriate to subdivide species further, for example to delineate between human-pathogenic and non-pathogenic strains of heterogeneous bacteria such as \textit{E.~coli.}

From another viewpoint, note that the subcomposition of these distributionally equivalent parts is completely specified and thus contains the same information (Shannon entropy) for each sample.
If parts have a probabilistic interpretation, as in the case of a discrete probability distribution, amalgamating them leads to a new probability of a coarse-grained event. 
The subsequent loss of information is considered in the notion of information monotonicity.
If distributionally equivalent parts are amalgamated, inter-sample distances remain unchanged, while generally amalgamations lead to non-growing distances \cite{Erb:21}.   
Although this notion of monotonicity has also not received much attention in CoDA, the mathematical properties of dissimilarities fulfilling it are well-studied in information geometry \cite{Amari:16}.

A similar idea is used in amalgamation clustering \cite{Greenacre:20}, where the loss of explained logratio variance is minimized at each clustering step when two components are amalgamated, and explained variance reduces monotonically as the agglomerative hierarchical tree is constructed  --- see Fig.~\ref{TellusTree2}.
An alternative clustering algorithm could be developed by amalgamating the most proportional parts at each clustering step.
As with coherence, deviations from distributional equivalence can be defined and measured.
    

\subsection{Log-contrasts}
The three logratio transformations, LR (\ref{LR}), ALR (\ref{ALR}) and CLR (\ref{CLR}), are all linear in the log-transformed components and can thus be expressed as matrix transformations \cite{Greenacre:21c}.
A \textit{log-contrast}  \cite{AitchisonBaconShone:84} is a linear combination of the log-transformed components with coefficients summing to 0, ensuring scale invariance: $\sum_j a_j \log(x_j)$, where $\sum_j a_j = 0$.
For example, LRs and ALRs have all coefficients 0 except for two that are $1$ and $-1$ corresponding to the numerator and denominator components respectively.
A particular CLR with the $j$-th component in the numerator has coefficients equal to $-\frac{1}{J}$ except the $j$-th one which is equal to $1-\frac{1}{J}$.
Log-contrasts are important in the interpretation and choice of logratio transformations, for example \cite{Coenders:20},\cite{ Coenders:22}.

\subsection{Isometry and isometric logratios}
The requirement of isometry, which is also not one of the properties proposed by Aitchison, has become more influential in the later development of CoDA, and pertains to the geometric structure of the cases in the logratio-transformed space of the simplex.
On the one hand, there is defined an exact geometry based on the $(D-1)$-dimensional space of all LRs, referred to here as the \textit{logratio geometry} --- this can be thought of as the set of all logratio distances between cases based on all their LRs.
On the other hand, there is an alternative geometry based on a set of transformed variables of the researcher's choice.
The latter set of transformations are called \textit{isometric} if they reproduce exactly the logratio geometry, up to a possible overall scale factor.

Some authors hold up isometry as a type of gold standard for CoDA (see Section 7.3) but this strict requirement limits their choice of transformations, forcing complex ones to be used.
Whereas the set of $D$ CLRs is indeed isometric, the \textit{isometric logratio} (ILR) transformation was introduced into CoDA \cite{Egozcue:05} specifically to obtain a set of $d$ logratios, the dimensionality of the data, that were both linearly independent and isometric. 

\begin{figure}[h!]
\includegraphics[width=\linewidth]{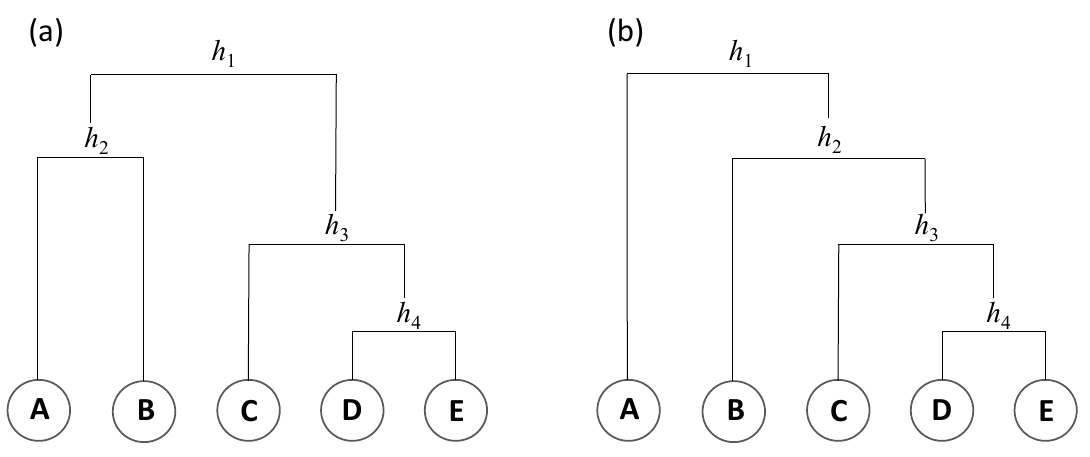}
\caption{Dendrogram graphs representing the groupings for defining four independent ILRs in each case on the five-component composition (heights have no meaning here). (a) A set of isometric logratios. (b) The special case of a set of pivot logratios.}
\label{dendrogram}
\end{figure}

To illustrate the construction of a set of ILRs, consider a 5-component example $\{$A,B,C,D,E$\}$ and the definition of a series of log-contrasts (see Section 3.7).
Define as a first log-contrast the mean of the first two log-transformed components versus the mean of the other three: $h_1 = \frac{1}{2}\big(\log(A)+\log(B)\big) - \frac{1}{3}\big(\log(C)+\log(D)+\log(E)\big)$.
Thus, the coefficients of the log-contrast are $[ \frac{1}{2}\ \ \ \frac{1}{2} \,-\frac{1}{3} \,-\frac{1}{3} \,-\frac{1}{3} ]$, summing to 0.
This is the logratio of the geometric means of the two contrasting subsets: $h_1 = \log\big((AB)^\frac{1}{2}/(CDE)^\frac{1}{3}\big)$ and is also the average of the six LRs combining the numerator and denominator components: $h_1 = \frac{1}{6}\big(\log(A/C)+\log(A/D)+\log(A/E)+\log(B/C)+\log(B/D)+\log(B/E)\big)$. 
The other three ILRs are formed by defining coefficient vectors that are orthogonal to the first one and to one another, so the four ILRs could be defined as follows, although there are other possibilities:
\begin{eqnarray}
    \label{ILR}
   \begin{bmatrix}
     h_1 \\
     h_2 \\
     h_3 \\
     h_4
   \end{bmatrix} =   
   \begin{bmatrix}
   \ \frac{1}{2} &  \ \ \frac{1}{2} & -\frac{1}{3} & -\frac{1}{3} & -\frac{1}{3} \ \\
   \     1       &      -1       &    \ \ \ 0       &   \ \ \ 0    &   \ \ \ 0  \      \\
   \     0       &    \ \ \ 0       & \ \ \  1       & -\frac{1}{2} & -\frac{1}{2} \ \\
   \     0       &    \ \ \ 0       & \ \ \  0       &   \ \ \ 1       &     -1 \ 
  \end{bmatrix}
  \raisebox{-0.6em}{$
  \begin{bmatrix}
    \log(A) \\
    \log(B) \\   
    \log(C) \\
    \log(D) \\
    \log(E)  
   \end{bmatrix}$}
\end{eqnarray}
The pattern for forming these contrasts can be represented in a dendrogram graph --- Fig.~\ref{dendrogram}(a) shows the graph where the four nodes represent the ILRs in (\ref{ILR}).
There is an additional detail to add to the definition of the ILRs, namely that the coefficient vectors are also normalized to have unit sums of squares so that they form an orthonormal set.
Hence, to be precise, the $h_j$s in (\ref{ILR}) each require a scalar multiplier equal to
the inverse square root of the sum of squares of their respective coefficients. 

\begin{figure}[t!]
\includegraphics[width=\linewidth]{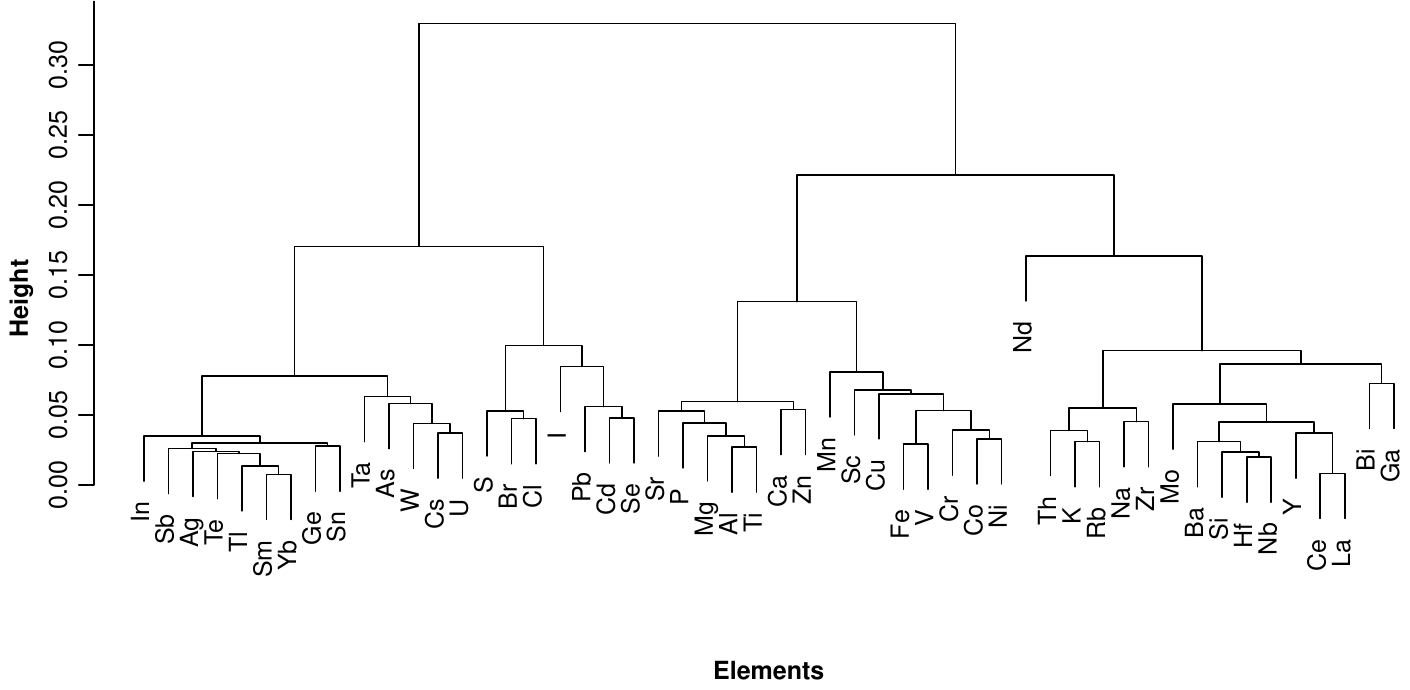}
\caption{Dendrogram of Ward clustering of Tellus parts.}
\label{TellusTree1}
\end{figure}

\begin{figure}[t!]
\includegraphics[width=\linewidth]{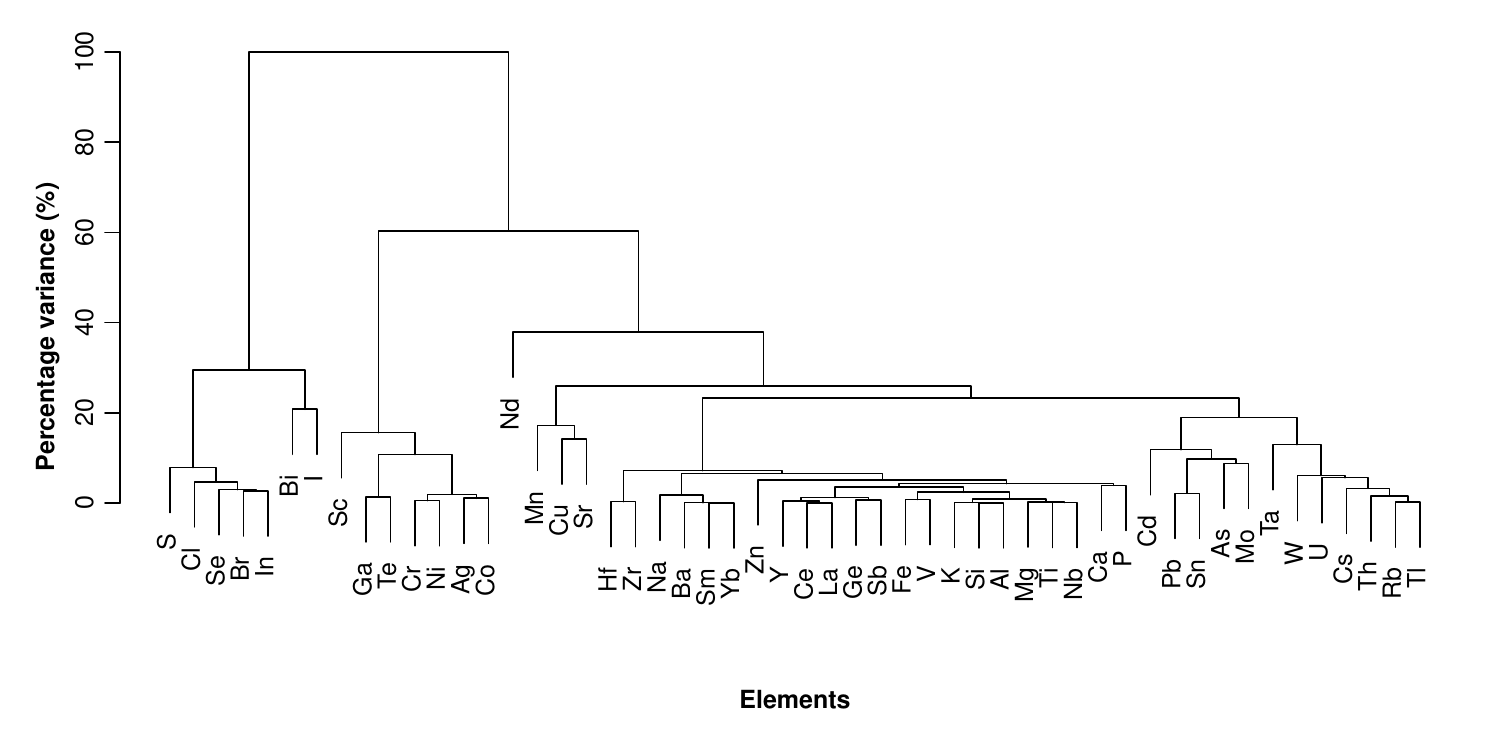}
\caption{Dendrogram of amalgamation clustering of Tellus parts. 
         The vertical scale shows percentage of explained variance lost at each node, 
         which is minimized.}
\label{TellusTree2}
\end{figure}

The question arises how to choose the set of ILRs, since the number of possible ILRs is equal to the number of possible unique dendrograms, of which there are 
$(2D-2)$!$/(2^{D-1}(D-1)$!$)$, according to \cite{Murtagh:84}.
There are already $2^{D-1}-1$ ways of splitting into two groups to fix the first ILR $h_1$.
Fig.~\ref{TellusTree1} shows one dendrogram for the Tellus data, established by Ward clustering of the CLRs of the parts, which is a way often recommended to define a set of ILRs.
The first ILR would be the logratio of the two geometric means of the left hand and right hand sets of elements at the top branch of the dendrogram.
The second would be the logratio of the geometric means of the two subsets splitting the right hand node, and so on.  
An alternative way, unrelated to the ILR, is to cluster by amalgamating \cite{Greenacre:20}, which results in the dendrogram of Fig.~\ref{TellusTree2}.

To reduce the number of possible ILRs to choose from, the special case of the \textit{pivot logratio} (PLR) transformation was then proposed \cite{Fiserova:11,Hron:17}, an example of which is shown in Fig.~\ref{dendrogram}(b).
For a given order of the parts, the first is contrasted with the others, then the second with those from the third onwards, and so on, until the last one which is a pairwise logratio of the last two.
The interpretation of a PLR is easier than that of a more general ILR, since there is only one part in the numerator and so an example such as $\log\big(x_j / (x_{j+1} x_{j+2} \cdots x_D)^{1/(D-j)}\big)$ is, up to a scalar, the average of LRs $\big(1/(D-j)\big)\big( \log(x_j/x_{j+1})+\log(x_j/x_{j+2})+\cdots\log(x_j/x_{D}) \big)$.
However, there is still the problem of how to order the parts, and the 
number of possible PLR transformations is the number of permutations $D!$.
For $D=10$ there are $34\,459\,425$ dendrograms defining ILRs but ``only'' $3\,628\,800$ permutations defining PLRs. 
Clearly, in all these cases, some way of choosing a transformation is needed.
most often based on hierarchical clustering of the parts. 


From a mathematical viewpoint, ILRs appear highly attractive: they are a set of linearly independent variables that are isometric, i.e.~distances between cases based on the ILRs are identical to those based on all the LRs or all the CLRs.
For this reason, their use has proliferated, especially with the availability of software that will choose a ``default'' ILR transformation and thus convert a compositional data set into a set of variables that can be used by regular multivariate statistical methods.
However, their major disadvantage is that their interpretation as single variables is complicated in all practical applications.
It is not clear what ratios of geometric means are actually measuring, whereas ratios of amalgamated parts are very straightforward to understand.


ILRs based on unweighted geometric means have also been termed ``balances'', as if they are balancing the groups of parts.
Most often, an ILR transformation is performed and then the ILRs are used in unsupervised or supervised multivariate analysis, seldom being interpreted as individual variables.
Many users, however, including the originators of balances, often interpret them as balancing the aggregated numerator and denominator parts.
This is clearly an incorrect interpretation, and can lead to mistaken results \cite{GreenacreGrunskyBaconShone:20} --- see Section 7.3. 
Furthermore, since the geometric mean is highly influenced by rare parts, the ILR ``balance" can turn out to be highly non-robust when rare parts are included.

\subsection{Dispersion measures, weighting and distances}
The quantification of total variance in a compositional data set is fundamental, because it is this variance that we want to account for, both in unsupervised learning as well as in supervised learning when the compositions are regarded as response variables.   
Aitchison \cite{Aitchison:86} defined the $D\times D$ \textit{variation matrix} ${\bf T} = [\tau_{jk}]$, where $\tau_{jk} = {\sf Var}\left(\log(x_j/x_k)\right)$, the variance of the $(j,k)$-th LR, ${\rm LR}(j,k)$, in (\ref{LR}).
The diagonal of the symmetric matrix $\bf T$ consists of zeros and the upper and lower triangles each contain the $\frac{1}{2}D(D-1)$ LR variances\footnote{The sample variance would usually be computed by dividing the sum of squared deviations from the mean across the $N$ cases by $N-1$, i.e.~the unbiased estimate. However, the definition preferred here of \textsf{Var} is to divide by $N$, so that each case has a weight of $1/N$ and the weights sum to 1.}.

An interesting result, also shown in \cite{Aitchison:86} and due to the unit-sum constraint, is that the covariances between the LRs can be computed from their variances: 
\begin{equation*}
\textsf{Cov}\left({\rm LR}(j,k), {\rm LR}(u,v)\right) = {\textstyle\frac{1}{2}}\left( \tau_{jk} + \tau_{uv} - \tau_{ju} - \tau_{kv}\right)
\end{equation*}

To measure the total dispersion in a compositional dataset, Aitchison defined the total logratio variability as the sum of the $\frac{1}{2}D(D-1)$ LR variances divided by $D$, $\frac{1}{D}\sum\sum_{j<k} \tau_{jk}$. 
The reason for the division by $D$ was to make the total variability equivalent to 
the more efficient way of computing this same measure, by simply summing the $D$ variances of the CLRs in (\ref{CLR}): $\sum_j {\sf Var}\left( {\rm CLR}(j)\right)$.  

A slightly different measure of total variance, denoted here by {\sf TotVar}, was introduced in \cite{Greenacre:18}, dividing the Aitchison measures by $D$, which assigns weights of $1/D^2$ to every LR variance, and $1/D$ to each CLR variance, thus averaging the CLR variances rather than summing them:
\begin{eqnarray}
  {\sf TotVar} &=& \frac{1}{D^2} \mathop{\sum\sum}\limits_{j<k = 2}^D {\sf Var}\left( {\rm LR}(j,k) \right)
               \label{TotVarLR} \\
               &=& \frac{1}{D} \sum_{j=1}^D {\sf Var}\left( {\rm CLR}(j) \right) 
               \label{TotVarCLR}
\end{eqnarray}
For the Tellus data set, the total logratio variance is equal to 0.3446.
As mentioned before, this definition has the advantage of comparability with measures of total logratio variance across data sets with different numbers of parts, whereas $\sum_j {\sf Var}\left( {\rm CLR}(j)\right)$ increases with the number of parts.
In addition, thinking of each part receiving an equal weight of $1/D$ inspires the general idea of varying the part weights.

The idea of weighting the parts in CoDA originates in the works of Paul Lewi \cite{Lewi:76} --- see also \cite{Lewi:86, Lewi:05, GreenacreLewi:09}.
Lewi's contribution needs to be more widely recognized by the CoDA community, since he anticipated many concepts defined later by Aitchison and other authors, as described in \cite{Smithson:22}.
The more general definitions of (\ref{TotVarLR}) and (\ref{TotVarCLR}), with varying part weights $c_j, j\in \{1,\ldots,J\}$ satisfying $\sum_j c_j = 1$ are then:
\begin{eqnarray}
  {\sf TotVar} &=& \mathop{\sum\sum}\limits_{j<k = 2}^D c_j c_k {\sf Var}\left( {\rm LR}(j,k) \right)
               \label{TotWtVarLR} \\
               &=& \sum_{j=1}^D c_j {\sf Var}\left( {\rm CLR}(j) \right) 
               \label{TotWtVarCLR}
\end{eqnarray}
where the unweighted (i.e., equally weighted) definitions are the special case $c_j = 1/D$ for all $D$ parts.

Differential weighting can be useful when parts with low average proportions create much higher ratios than parts with higher average proportions, and thus contribute excessively to the total logratio variance.  
The higher logratio variances in the rarer parts are often due to high relative measurement error.
In Lewi's conception of CoDA, default part weights were the average proportions of the parts, thus downweighting the parts with low averages compared to the higher ones. 
This system of weights is identical to the one used in correspondence analysis, but any other system that is appropriate to the data and research objective could be chosen.

\section{Application to the Tellus data}
Once a compositional data set is transformed to a set of logratios, regular interval-scale statistical methods can be applied, both for unsupervised and supervised learning.
In this section we describe what we consider the simplest unsupervised analysis of the Tellus data.
First, we show that an additive logratio (ALR) transformation, the simplest of all logratio transformations, satisfactorily represents the geometric structure of these data.
Second, we show that an alternative strategy based on correspondence analysis, which needs no zero replacement, serves the same purpose.  

The data as well as the complete R script for computing these results are provided online --- see the Data Availability statement at the end of the article.

\subsection{To weight or not to weight}
\begin{figure}[h!]
\includegraphics[width=0.7\linewidth]{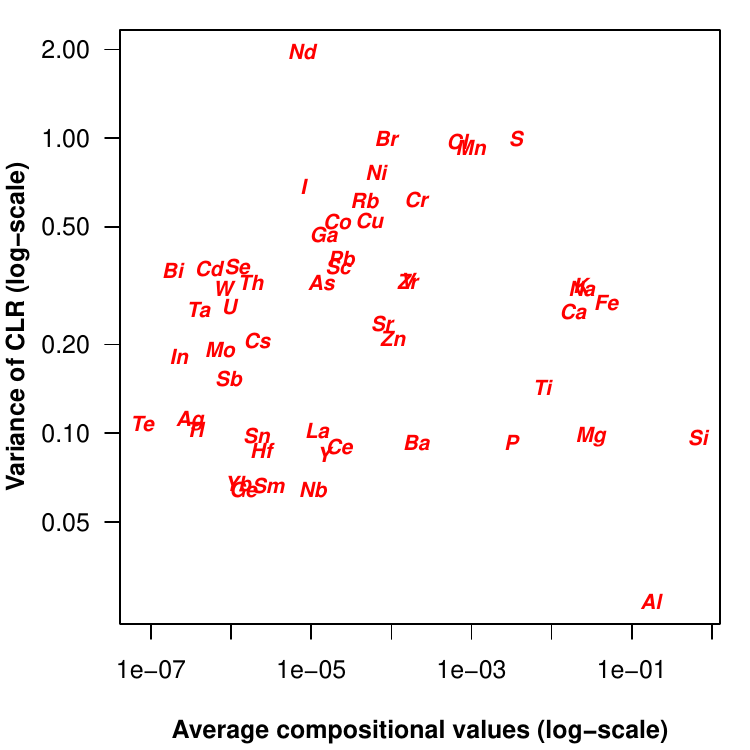}
\vspace{-0.2cm}
\caption{For each of the 52 parts, on the y-axis: the variance of the CLR; and on the x-axis: the average compositional value. Both axes plotted on logarithmic scale.}
\label{weightingplot}
\end{figure}

Since the total logratio variance is the average of the 52 CLR variances, the first step is to decide whether the parts require weighting or not, which is usually required when rare parts engender high variances compared to the others.
This is diagnosed from a plot of the CLR variances versus the average proportions of the parts (Fig.~\ref{weightingplot}).

There seems to be no tendency, as in some other applications (see, for example, the archaeometric example in \cite{Greenacre:18}), for the low mean parts to have excessively high CLR variances. 
Thus the decision is to use unweighted parts and compute the CLRs the usual way and then the total logratio variance (\ref{TotVarCLR}), which is identical to (\ref{TotVarLR}): \textsf{TotVar} = 0.3446.
This total can be broken down into contributions from the 52 elements: the highest contributors are rare earth element \textit{\textsf{Nd}} (10.9\% of the total variance), trace element \textit{\textsf{Br}} (5.5\%) and major element \textit{\textsf{S}} (5.5\%), and the lowest are trace elements \textit{\textsf{Nb}} (0.4\%), \textit{\textsf{Ge}} (0.4\%) and major element \textit{\textsf{Al}} (0.1\%). 
The total logratio variance quantifies the geometric dispersion of the samples in terms of the CLRs (equivalently, all LRs).
We will first show that a well-chosen ALR transformation represents this dispersion of the samples almost exactly, which leads to a considerable simplification of our subsequent results. 

\subsection{Selecting an ALR transformation}
\begin{figure}[h!]
\includegraphics[width=0.7\linewidth]{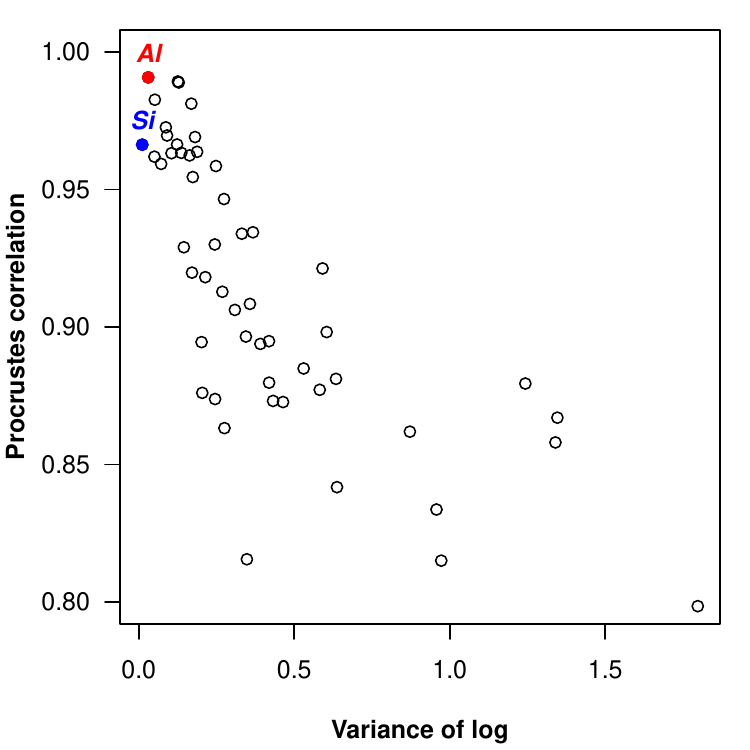}
\vspace{-0.2cm}
\caption{For each of the 52 reference parts, on the y-axis: the Procrustes correlation of the sample geometry of the ALR-transformed data with the geometry based on all LRs (pairwise logratios); and on the x-axis: the variance of the log-transformed reference component $\log(x_{\it ref})$. The element \textit{\textsf{Al}} has the highest correlation, whereas \textit{\textsf{Si}} has the lowest variance, but \textit{\textsf{Al}}'s variance is the second lowest.}
\label{procrustes}
\end{figure}

Following the approach of \cite{GreenacreMartinezBlasco:21}, we search for an ALR transformation that best represents the geometric structure of the Tellus data set. 
Using the function \texttt{FINDALR}, from the \texttt{easyCODA} package \cite{Greenacre:18} in R \cite{R:21}, the 52 different ALR transformations are computed, using in turn the 52 possible reference parts.
For each transformation, the Procrustes correlation \cite{Gower:04}, which measures the concordance of two multidimensional geometries, is computed between the geometry of the 6799 samples in the ALR space and the logratio geometry of the same samples in the CLR space.
These correlations are plotted in Fig.~\ref{procrustes} against the variances of the log-transformed reference parts. 

The highest Procrustes correlation, equal to 0.990, is achieved using reference element \textit{\textsf{Al}}, whereas the lowest variance of the log-transformed reference is for element \textit{\textsf{Si}}. 
Both \textit{\textsf{Al}} and \textit{\textsf{Si}} are useful as references from a geochemical point of view, since they are both significant elements for rock-forming minerals.
\textit{\textsf{Al}} is significant for representing clay minerals and feldspars, while \textit{\textsf{Si}} is significant for representing silicates such as quartz, olivine and feldspars.
For choosing between \textit{\textsf{Al}} and \textit{\textsf{Si}} there is a trade-off between isometry (where \textit{\textsf{Al}} wins) and ease of interpretation (where \textit{\textsf{Si}} wins).

Looking at Fig.~\ref{procrustes} it can be seen that \textit{\textsf{Al}}, which is the second most abundant element, has the second lowest variance in the log, but well above \textit{\textsf{Si}} in terms of Procrustes correlation (for \textit{\textsf{Si}} it is 0.963).
It was thus decided that \textit{\textsf{Al}} was preferable as the reference, and its low variance in the log suggests that the individual ALRs might be similar to the logarithm of the numerator part, according to the definition (\ref{ALR}), which makes their interpretation even easier.
In the Supplementary Material A1\cite{GreenacreEtAl:23}, each of these ALRs is plotted against the logarithm of the numerator part, where it is shown that this conclusion is generally confirmed, especially for those elements that are crucial for the interpretation. 

\begin{figure}[h!]
\includegraphics[width=0.6\linewidth]{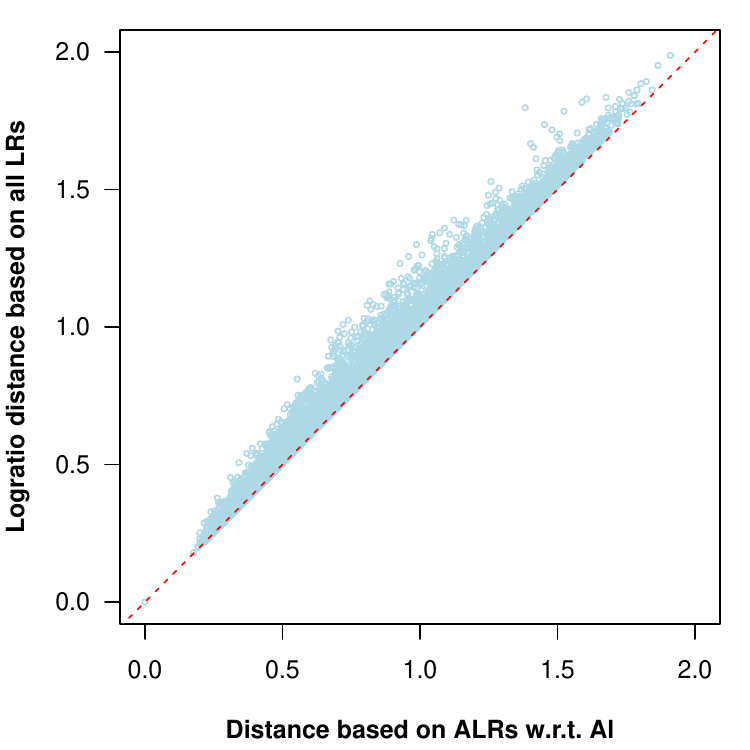}
\vspace{-0.2cm}
\caption{A scatterplot of $10\,000$ inter-sample distances computed on the ALR-transformed data (horizontal axis) and CLR-transformed data (vertical axis, the latter distances being equivalent to the distances based on all LRs). The Procrustes correlation between the two geometries is 0.990.}
\label{distances}
\end{figure}

To demonstrate the closeness of the ALR geometry to the logratio geometry, that is, its closeness to isometry (see Section 3.8), the \textit{logratio distances} between the samples (i.e., the Euclidean distances computed on the CLR-transformed data) can be plotted against the Euclidean distances between the samples using the ALRs with respect to the chosen reference \textit{\textsf{Al}}.
Since there are $6799\times6798/2 = 23\,109\,801$ between-sample distances, Fig.~\ref{distances} shows only a random sample of $10\,000$ of them, to illustrate the concordance of the two geometries.
Notice that the ALR distances approximate the logratio distances from below, so all points are above the diagonal line of equality shown in Fig.~\ref{distances}.

The high Procrustes correlation of 0.990 and the concordance of the distances seen in  Fig.~\ref{distances}, point to the fact that the ALR transformation with respect to element \textit{\textsf{Al}}, using 51 LRs, accurately reflects the exact logratio structure of the data, which is based on all 1326 LRs.
In other words, this ALR transformation is \textit{quasi-isometric} and the Tellus compositions can be effectively replaced by this set of ALRs.

\begin{figure*}[h!]
\centering
\includegraphics[width=0.95\linewidth]{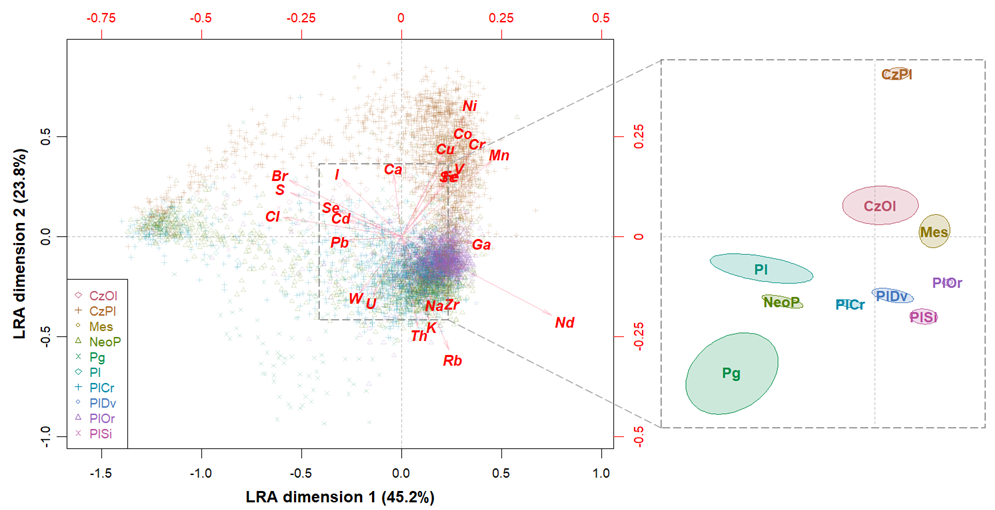}
\vspace{-0.2cm}
\caption{Logratio analysis (LRA) of the Tellus data (i.e., PCA of the CLR-transformed data), showing the optimal two-dimensional solution. This is a contribution biplot showing only the components with more than average contribution to the solution \cite{Greenacre:13}. An enlargement of the indicated rectangle is shown on the right, with 99\% confidence ellipses of the 10 age bracket groups of samples. Abbreviations of the age brackets are: CzOl (Cenozoic, Palaeogene, Oligocene , CzPl (Cenozoic, Palaeogene, Palaeocene), Mes (Mesozoic), NeoP (Neoproterozoic), Pg (Palaeogene), Pl (Lower Palaeozoic), PlCr (Palaeozoic, Carboniferous), PlDv (Palaeozoic, Devonian), PlOr (Palaeozoic, Ordovician), PlSi (Palaeozoic, Caledonian, Silurian).}
\label{LRAbiplot}
\end{figure*}

\begin{figure*}[h]
\centering
\includegraphics[width=0.95\linewidth]{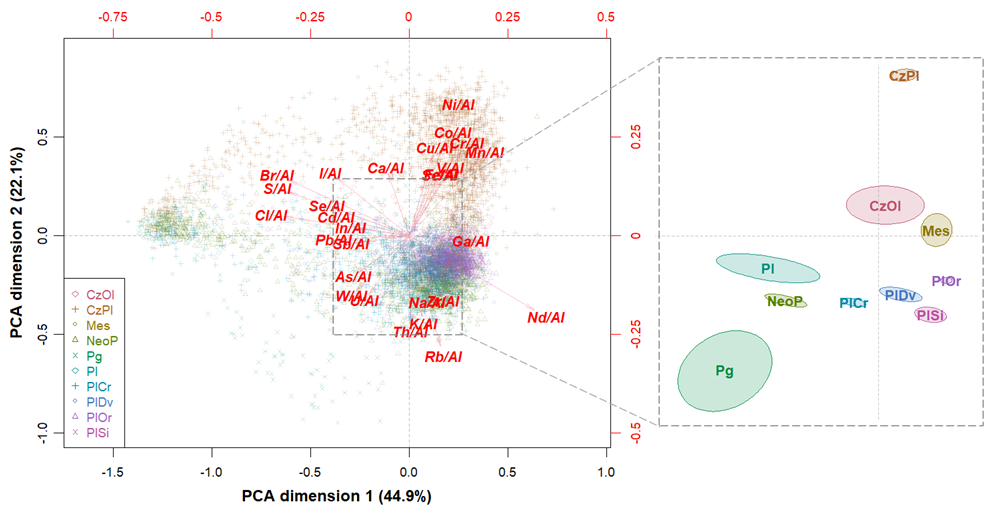}
\vspace{-0.2cm}
\caption{Principal component analysis (PCA) of the ALR-transformed Tellus data, showing the optimal two-dimensional solution. As in Fig.~\ref{LRAbiplot} the components with more than average contribution to the solution are shown. An enlargement of the indicated rectangle shows the 99\% confidence ellipses of the 10 age bracket groups of samples.}
\label{ALRbiplot}
\end{figure*}

Another convincing demonstration of this property of the ALRs is to show the two-dimensional PCA of the CLRs in Fig.~\ref{LRAbiplot} along with the two-dimensional PCA of the ALRs, shown in Fig.~\ref{ALRbiplot}. 

The PCA of the CLRs is called \textit{logratio analysis} (LRA)\cite{Greenacre:18}, since all 1326 LRs are being analysed simultaneously, whereas the PCA of the ALRs involves only 51 LRs. 
In both cases only the highly contributing variables, that is, those making more than average contribution to the two-dimensional solution, are indicated.
It is clear that the two biplots are practically identical. 
The percentages of variance explained on the first two dimensions are also quite similar, with $45.2 + 23.8 = 69.0\%$ and $44.9 + 22.1 = 67.0\%$ explained respectively.
Furthermore, the Procrustes correlation between the configurations in the two-dimensional configurations of Figs.~\ref{LRAbiplot} and \ref{ALRbiplot} is 0.997, even higher than the full-space concordance. 



The 6799 samples are classified into 10 ``age brackets'' (see the caption of Fig.~\ref{LRAbiplot}), which make it easier to compare the two results --- these groups are coloured differently and 99\% confidence ellipses for each group mean are added, but shown separately in the expanded right hand plots for legibility.
The confidence ellipses are achieved by bootstrapping each group's set of coordinates 1000 times, giving 1000 replications of its group mean, from which a 99\% concentration ellipse of the mean can be computed \cite{Greenacre:16b}, based on bivariate normal theory.
The sizes of the ellipses are determined both by the spatial variation of the sample points within each group, seen in the left hand plots of Figs~\ref{LRAbiplot} and \ref{ALRbiplot}, as well as the respective sample sizes.
The groups \textsf{CzPl} and \textsf{Pg}, for example, at top right and bottom left of the confidence ellipse plot, are the most and least frequent samples in the data set (1691 and 124 out of the 6799 samples, respectively).
The only noticeable difference between the ellipses in Figs~\ref{LRAbiplot} and \ref{ALRbiplot} is that the latter ones are very slightly larger.
The separation of all the confidence ellipses, with no overlapping, suggests that all the age bracket groups are significantly different from one another \cite{Greenacre:16b}.

The directions of the ALRs from the origin in Fig.~\ref{ALRbiplot} match very closely those of the CLRs in Fig.~\ref{LRAbiplot}, but here the ALRs are interpreted as regular logratio variables.
On the other hand, in Fig.~\ref{LRAbiplot} the CLRs  are not interpretable \textit{per se}, it is rather the directions of the links between them that should be interpreted \cite{Greenacre:03}, since these represent the differences between the CLRs, which are the LRs: $\log\left(x_j/g({\bf x})\right) - \log\left((x_k/g({\bf x})\right) = \log(x_j/x_k)$.
For example, the link from \textit{\textsf{Br}} to \textit{\textsf{Nd}} represents the direction of the logratio log(\textit{\textsf{Nd/Br}}) and is discriminating between the age brackets \textsf{PlSl} and \textsf{PlOr} on the right and \textsf{Pl} on the left. 
Another example is the link from \textit{\textsf{W}} to \textit{\textsf{Ni}}, which displays the direction of the logratio log(\textit{\textsf{Ni/W}}), distinguishing between groups \textsf{CzPl} and \textsf{Pg}.
If PCA had been applied to the complete set of 1326 LRs, the logratios log(\textit{\textsf{Nd/Br}}) and log(\textit{\textsf{Ni/W}})  would have exactly the same orientations as the corresponding links in Fig.~\ref{LRAbiplot} but anchored at the origin.
In this way the LRA shows the optimum biplot of all 1326 LRs.

Notice that the simultaneous optimization of the $D$ CLRs and the $\frac{1}{2}D(D-1)$ LRs is a result of the double-centering of the matrix of log-transformed parts: that is, the preliminary row centering to compute the CLRs and then the column centering inherent in the PCA. 
The interpretation of differences between variables in biplots was already known by linking pairs of variables \cite{Gabriel:72}, but these differences were not optimally displayed \cite{Greenacre:03}.

A full geochemical interpretation of Figs~\ref{LRAbiplot} and \ref{ALRbiplot} is beyond the scope of this article, except to note the most obvious feature of the separation of the points on the negative side of the first dimension. 
The cluster of points on the left represents peat, that is, soils that are highly enriched in organic material, chalcophile elements (\textit{\textsf{S, Se, Cd, Pb}}), and lithophile elements (\textit{\textsf{Cl, Br, I}}). 
The relative enrichment of \textit{\textsf{I}} also reflects coastal effects associated with proximity to sea water \cite{McKinley:17}.


\subsection{The correspondence analysis alternative}

Correspondence analysis (CA) \cite{Benzecri:73, Greenacre:16a} has a long history of being used to analyse geochemical compositional data, for example \cite{David:77, Grunsky:85, Jackson:97}.
More recently, it has been shown that CA is theoretically related to LRA, thanks to the Box-Cox power transformation \cite{Greenacre:09, Greenacre:10a}.
Both CA and LRA are variants of PCA, both double-center the data (relative abundances in the case of CA, log-transformed relative abundances in the case of LRA), and the Box-Cox power transformation $f(x) = (1/\alpha) (x^\alpha - 1)$ links the two since it tends to the log-transformation as the power $\alpha$ tends to 0.
(Notice that the term $-1/\alpha$ of the transformation is eliminated by the double-centering, so can be omitted, whereas the multiplicative factor $1/\alpha$ is important to retain.)

The distance used in CA is the chi-square distance, which similarly tends to the logratio distance when the raw compositions are Box-Cox transformed and $\alpha$ tends to 0.
The chi-square distance between two compositional samples $\bf x$ and $\bf y$ is defined as:
\begin{equation}
 \sqrt{\sum_{j=1}^D \frac{(x_j-y_j)^2}{c_j}}
 \label{DistChi}
\end{equation}
where $c_j$ is the average (i.e., expected value) of the $j$-th part.
In other words, the chi-square distance is a Euclidean-type distance of the original compositional parts standardized by dividing by the square roots of their expected marginal values.
There is a  symmetric definition of the chi-square distance between parts, where the columns of the compositional data matrix are first closed and the columns then standardized by the square roots of their respective marginal averages.

\begin{figure}[b!]
\centering
\includegraphics[width=0.6\linewidth]{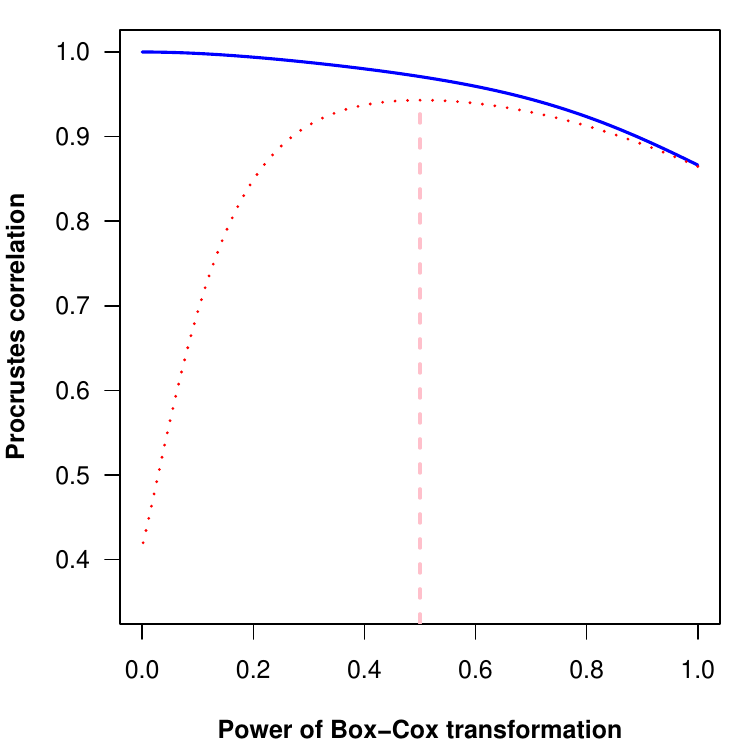}
\vspace{-0.2cm}
\caption{The blue curve shows that convergence, from right to left, of the correspondence analysis (CA) solution as the power of the Box-Cox transform decreases from 1 and tends to 0 (this is for the data set with zeros replaced). The agreement is measured by the Procrustes correlation. The red dotted curve shows the situation when the data zeros have not been replaced, where the power transformation starts to break down starting at exactly 0.5 (the square root transformation), where the Procrustes correlation is at its maximum.}
\label{BoxCox}
\end{figure}

The connection between CA's chi-square distance and LRA's logratio distance can be easily demonstrated using the Procrustes correlation.
This time it is the geometry of the 52 element parts that is of interest, but it could also be done for the samples.  
In Fig.~\ref{BoxCox} the blue curve shows, for decreasing values of the power $\alpha$ in the Box-Cox transformation (the horizontal axis, read from right to left), the Procrustes correlation between the logratio geometry of the elements, and the corresponding CA geometry.
This shows how the CA converges to the LRA, tending to a value of 1 as $\alpha$ tends to 0 --- the smallest $\alpha$ used is $0.001$.
Of course, this convergence is demonstrated on the Tellus data set with the zeros replaced.
The red dotted curve, by contrast, shows the Procrustes correlations when the original data including the zero values are analysed by CA and correlated with the logratio geometry. 
Here it can be seen that the correlation curve rises just below the blue one (again, following the curve from right to left) to a maximum when $\alpha = 0.5$ (a square root transformation) and then starts to descend rapidly.
This is where the approach to the log-transform is starting to kick in and the geometry starts to deteriorate due to the zero values.
At the maximum of the red dotted curve at $\alpha=0.50$, the Procrustes correlation equals $0.961$.

The conclusion is that a CA of power-transformed data, using $\alpha=0.50$, comes very close to the logratio geometry.
It should be remembered that the logratio geometry had to have its 3883 zero values substituted, otherwise logratios could not be computed.
Thus, it is a mute point whether it is valid to consider the logratio geometry as the reference standard of the two being compared here, since it contains imputed data.

\begin{figure*}[t]
\centering
\includegraphics[width=0.99\linewidth]{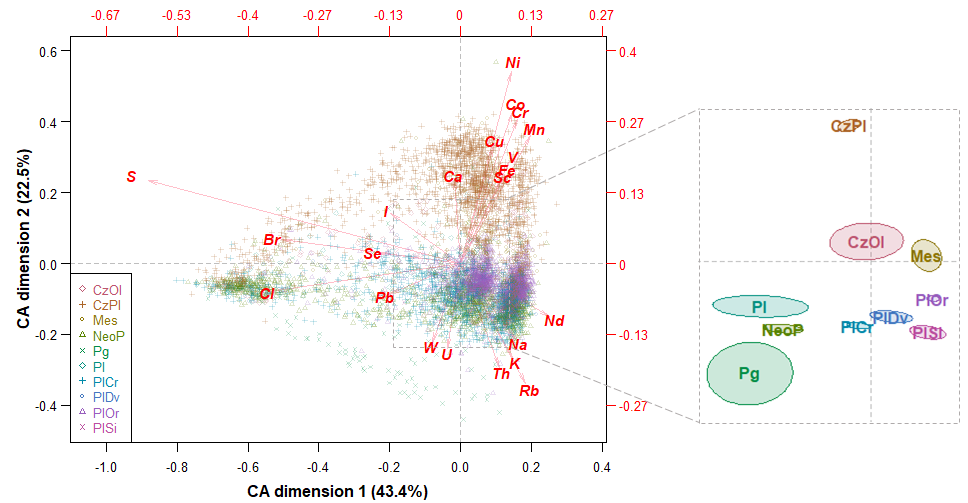}
\vspace{-0.2cm}
\caption{Correspondence analysis (CA) of the Box-Cox power-transformed Tellus data, showing the optimal two-dimensional solution. As in Figs~\ref{LRAbiplot} and \ref{ALRbiplot}, the components with more than average contribution to the solution are shown. An enlargement of the indicated rectangle is shown on the right, containing the 99\% confidence ellipses of the 10 age bracket groups of samples.}
\label{CAbiplot}
\end{figure*}

The two-dimensional CA solution is shown in Fig.~\ref{CAbiplot}, which again --- as expected --- closely resembles the previous figures, Figs \ref{LRAbiplot} and \ref{ALRbiplot}, the only noticeable difference being a separation on the right of two similar dense clouds of points where the right hand cloud corresponds to the zeros of element \textit{\textsf{S}} (which has the majority, 2535, of the 3883 zeros) --- notice the direction of element \textit{\textsf{S}}, which is an important contributor to this solution. 
There is also a slight counter-clockwise rotation of the solution.
The percentages of variance on the two principal dimensions have a total of $41.5+23.4=64.9\%$, very similar to the totals computed before. 

The quasi-isometries of both the ALR transformation and the power-transformed CA imply that any other multivariate analysis on the samples, for example clustering, would give similar results.
As an illustration of this conjecture, non-hierarchical k-means clustering was applied to the logratio distances, to the distances based on the ALRs, and to the chi-square distances after the square-root power transformation.
Three clusters were specified in each run of the \textsf{R} function \texttt{kmeans}.
The cross-tabulations of the two alternative solutions with the one based on all logratios, shows a very close agreement, 99.0\% and 97.6\% respectively, as shown in the cross-tabulations in Table \ref{kmeans.tabs}.
The respective values of the \textit{adjusted Rand index} \cite{Rand:71, Hubert:85}, which measures similarity between two clusterings, are 0.971 and 0.914.
It should be emphasized again that CA is applied to the original data set with zeros, whereas the logratio distances and those based on the ARL transformation have the zeros replaced.

 \begin{table}[h]
    \begin{minipage}{.4\textwidth}
      \centering
      \begin{tabular}{lrrr}
        \hline 
        (a)  & ALR1 & ALR2 & ALR3 \\
        \hline 
        LR1 & 4467 & 9 & 37\\
        LR2 & 7 & 1470 & 11\\
        LR3 & 4 & 0 & 794\\
        \hline
      \end{tabular}
    \end{minipage}
    \begin{minipage}{.4\textwidth}
      \centering
      \begin{tabular}{lrrr}
        \hline 
        (b)   & CA1 & CA2 & CA3 \\
        \hline
        LR1 & 4369 & 45 & 99\\
        LR2 & 17 & 1444 & 27\\
        LR3 & 10 & 0 & 788\\
        \hline
      \end{tabular}
    \end{minipage}
    \caption{Comparison of three-cluster solution obtained using the all pairwise logratio (LR) transformation and (a) the ALR-transformed data, with Al as the reference (99.0\% agreement in the clusters, adjusted Rand index = 0.971}) and (b) the CA-transformed data, using Box-Cox power transformation with power 0.5 (square root), and requiring no zero replacements (97.1\% agreement, adjusted Rand index = 0.914).
    \label{kmeans.tabs}
  \end{table}

The CA solution shows slightly less agreement than the ALR solution, but it should again be emphasized that the CA solution needs no zero replacement, so it is being compared with a solution where the data zeros have been replaced, hence a different data matrix.
It might well be that in the final analysis the CA strategy is more acceptable for handling compositional data with zeros, as long as it can be checked against the logratio approach to support its closeness to the strictly coherent approach, always remembering that the logratio approach requires imputation of the zeros.

It is coincidental but also remarkable in this example that the optimal power transformation for the CA is 0.5, the square root, since this is one of the accepted transformations of data in the form of proportions \cite{Stephens:82, Atkinson:82}.
In this context it is interesting to note that such an approximation (in the form of the square root transformation) is isometric with respect to a non-Euclidean geometry based on the Fisher-Rao metric, which is well known for probability distributions \cite{Erb:21}.
The square root transformation maps the proportions onto the multidimensional unit sphere in the positive orthant \cite{ScealyWelsh:11}, and the Euclidean distances are then the same as the Hellinger distances, named after Ernst Hellinger \cite{Hellinger:09}.  
In the case of CA, however, the sphere will be deformed by the chi-square metric, just like the untransformed proportions in CA are mapped into an irregular multidimensional simplex.
As the power tends to zero the ``sphere'' in CA becomes less deformed as the margins tend to uniform ones, and then the chi-square distance and the Euclidean distance on the power-transformed compositions become the same.

\subsection{Quasi-coherence}
Subcompositional coherence is concerned with the difference in relationships between a subset of parts in a full composition and those in the same subset treated as a subcomposition with renormalization of the parts.  
A measure of incoherence was proposed between the logratio distances between the parts in the subcomposition and the distances between the same parts in the full composition, using the stress that measures the error between two sets of distances in multidimensional scaling \cite{Greenacre:11a}.
It is clear that the LRs are coherent, so the proposed stress measure of incoherence will be zero for logratio methods.
Similarly, the Procrustes correlation between the geometry of the parts in an LRA of the full composition and the LRA of the subcomposition will be 1.
Using the Procrustes correlation as an alternative to stress unifies the treatment of isometry (of the cases) and coherence (of the parts).
However, in CA, for example, we know the chi-square distances will not be coherent, but the question is: how incoherent?

The level of incoherence can be investigated by generating random subcompositions of the Tellus data, in this case with zeros not replaced.
For each subcomposition, the parts were renormalized and the chi-square distances between the parts were computed. 
These are compared with the chi-square distances between the parts using the original compositional values in the full composition. 
In each case the full set of principal coordinates is computed, thus defining the geometry of the compositional parts and of the subcompositional parts, and then the Procrustes correlation is used to measure the degree of concordance. 

\begin{figure}[h!]
\includegraphics[width=0.8\linewidth]{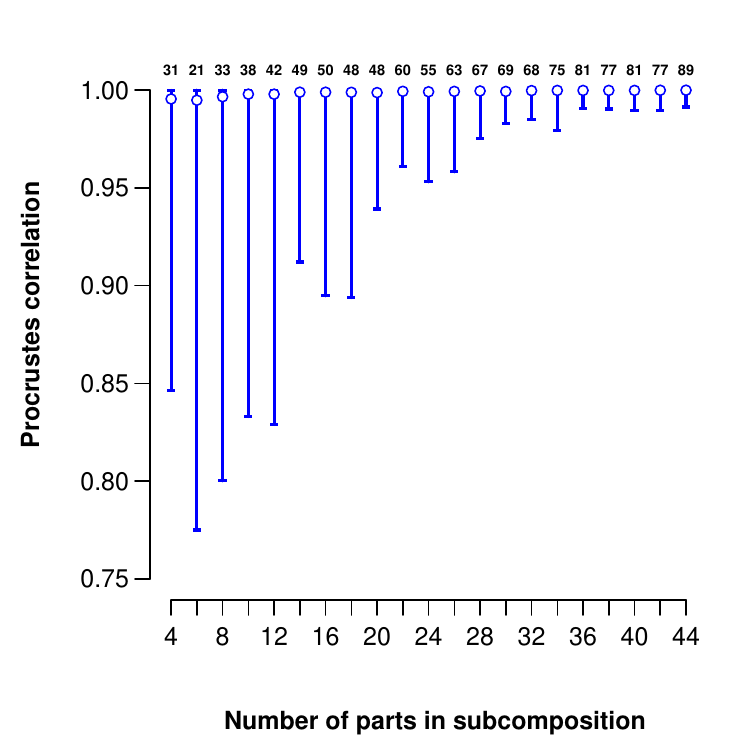}
\vspace{-0.2cm}
\caption{Procrustes correlations between the chi-square geometry of parts in random subcompositions and their respective chi-square geometry in the original full composition of the Tellus data. A 100 random subcompositions are generated of sizes 44 (out of the 52 Tellus parts) down to 4. The extents of the 2.5\% and 97.5\% percentiles of each set of 100 correlations are indicated by vertical lines, and their medians by dots. The median correlations are all very close to 1, and many subcompositions exhibit almost perfect coherence. The numbers at the top show the number of times out of the 100 subcompositions the Procrustes correlations were $>0.999$.}
\label{Tellus_coherence_CA}
\end{figure}

The results are shown in Fig.~\ref{Tellus_coherence_CA}, and it can be seen that the median Procrustes correlations are very close to 1, with some long tails of lower correlations below the medians for the smaller subcompositions where the renormalization can have the most effect. 
Above each bar is the number out of 100 (i.e., the percentage) of correlations greater than 0.999, indicating near perfect coherence. 

\begin{figure}[h!]
\includegraphics[width=0.8\linewidth]{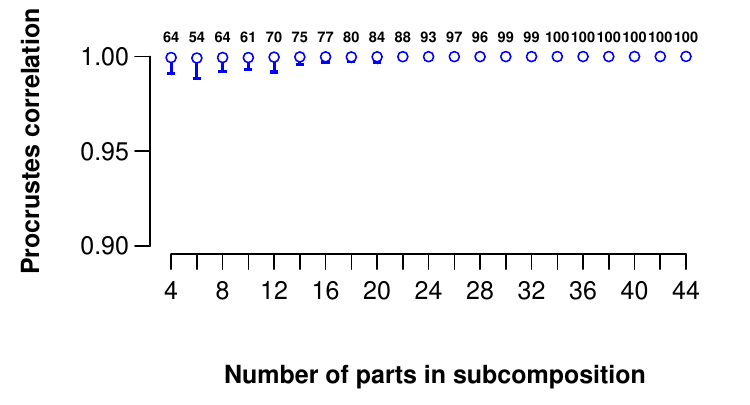}
\vspace{-0.2cm}
\caption{As in Fig.~\ref{Tellus_coherence_CA} but for compositional data transformed by square root. Almost all the subcompositions have Procrustes correlations very close to 1 (notice the shorter range of the vertical scale), showing that the chi-square distance on square-rooted compositions is quasi-coherent.}
\label{Tellus_coherence_CAsqrt}
\end{figure}

As a more extreme example, the chi-square distances between the 9 rare earth elements, in a specific renormalized subcomposition, were compared to those computed in the full composition.
The Procrustes correlation for this single comparison was 0.972.
If the Tellus compositions are square-rooted, this being the optimal power transformation identified in Section 4.3, the resulting Procrustes correlation for the same comparison rises to 0.998. 
This suggested performing the same coherence exercise on the Tellus compositions after square-root transformation.
The results in Fig.~\ref{Tellus_coherence_CAsqrt} show the dramatic improvement in coherence, right down to the smallest subcompositions.
In other words, these Tellus compositional data can be square-root transformed, renormalized, and then each part chi-square standardized by dividing by the square root of its (transformed) part mean, giving a quasi-coherent set of transformed data for further analysis, with no need for replacement of data zeros.

\section{Variable selection}
The issue of variable selection is a crucial one in all aspects of CoDA, since there are very many LRs available and many possible ALR transformations
The logratio space has dimensionality $d=D-1$, which means that a subset consisting of only $D-1$ linearly independent LRs is needed to generate all the other $\frac{1}{2}D(D-1) - (D-1) = \frac{1}{2}(D-1)(D-2)$ LRs.

\begin{figure}[t]
\includegraphics[width=0.95\linewidth]{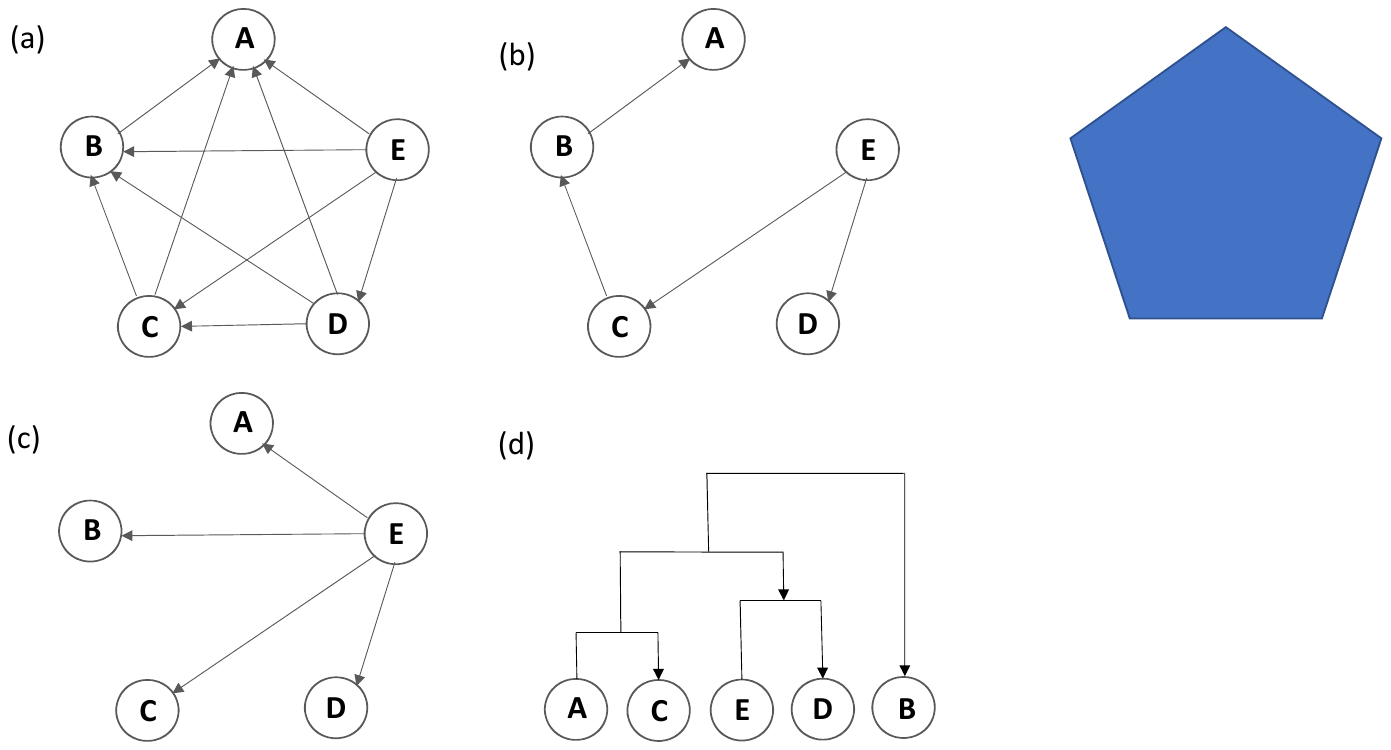}
\caption{(a) Graph of all ten pairwise logratios (LRs) for a 5-component composition, shown as edges between the component vertices, and (b, c) two directed acyclic graphs (DAGs) of four linearly independent logratios, which are sufficient to generate all LRs. Graph (b) shows logratios based on A/B, B/C, C/E and D/E (the arrow points to the numerator component). Graph (c) is an ALR transformation with respect to reference component E, i.e. logratios based on A/E, B/E, C/E and D/E. Graph (d) is a dendrogram which combines parts at each node, which is only one of the different possibilities for combining in this way.}
\label{DAG}
\end{figure}

For a 5-component composition, for example, Fig.~\ref{DAG}(a) shows a graph of all $\frac{1}{2}\times 5\times 4 = 10$ LRs, where the arrows point to the denominators of the respective ratios. 
Fig.~\ref{DAG}(b) and (c) show two different linearly independent subsets of $D-1=4$ LRs that generate all the other 6 LRs. 
Fig.~\ref{DAG}(b) graphs the ratios \textsf{A/B, B/C, C/E} and \textsf{D/E}.
Then, to get the ratio \textsf{A/E}, the arrows can be followed to obtain \textsf{C/E} $\cdot$ \textsf{B/C} $\cdot$ \textsf{A/B} = \textsf{A/E} (any arrow direction can be reversed to give the inverse ratio).
Linearly in the logarithms this equality is $\log(\textsf{C/E}) + \log(\textsf{B/C}) + \log(\textsf{A/B}) = \log(\textsf{A/E})$.
The five parts form a directed acyclic graph (DAG) \cite{Greenacre:18, Greenacre:19} --- all parts are connected and there is no cycle amongst them. 

Fig.~\ref{DAG}(c) shows another DAG with four arrows, which is the ALR transformation with \textsf{E} as the reference part. 
In Section 4.2 a procedure was described, in an unsupervised context, of finding an optimal ALR transformation for the Tellus data, which was using the reference element \textit{\textsf{Al}}.

When all LRs are candidates for forming a subset of $D-1$ independent ones for a $D$-component composition, the number of possible DAGs is the Cayley number $D^{D-2}$ \cite{Bona:06}, equal to 125 for $D=5$, but increasing more than exponentially as $D$ increases.
For the medium-sized Tellus data set with $D=52$, the number of DAGs is $52^{50}$, beyond any possibility to be explored. 

Finally, Fig.~\ref{DAG}(d) shows a different way of defining the four logratios, where parts are combined at each node of a dendrogram. If they are combined using geometric means, this is equivalent to defining a set of ILRs --- see Fig.~\ref{TellusTree1}.
But they can also be combined more simply as amalgamations --- see Fig.~\ref{TellusTree2} --- in which case the variance explained is theoretically not 100\% but in practice very close to 100\% (for Fig.~\ref{TellusTree2} it is 99.99\%).


Rather than doing a dimension reduction, where all parts participate, as in Figs~\ref{LRAbiplot} and \ref{ALRbiplot}, the idea of reducing the number of variables is an alternative \cite{Krzanowski:87}.
The question of selecting a small set of LRs that satisfy a practical objective has been addressed in \cite{Greenacre:19, Graeve:20, Wood:21}.
The idea is very simple and involves choosing LRs in a stepwise fashion, with the objective of explaining the maximum amount of variance in the compositional data set.
A complete set of $D-1$ independent LRs forming a DAG explains 100\% of the logratio variance (assuming $D$ is less than the number of observations $N$, otherwise for wide matrices $N-1$ LRs are needed). 
But perhaps much fewer LRs would be satisfactory, since all data have inherent error and hence explaining 100\% of the variance is unnecessary.

The stepwise algorithm starts by selecting the LR that explains the most logratio variance. 
It is then retained and the next logratio which, together with the first, explains the maximum variance, is retained as the second LR, and so on until a large percentage of variance, say around 95\%, is explained.
Alternatively, a stopping rule can impose a strong penalty on the number of logratios chosen \cite{Coenders:22, Gordon-Rodriguez:21, Combettes:21}. 

\begin{figure}[t]
\includegraphics[width=\linewidth]{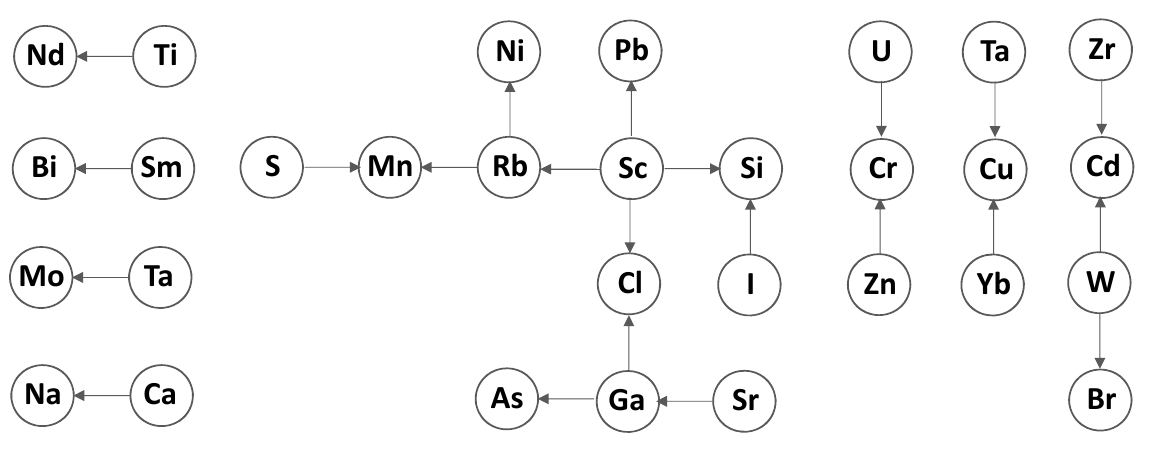}
\caption{Graph of the 30 elements that form a graph with 22 edges representing the 22 logratios chosen stepwise to maximize variance explained.}
\label{TellusGraph}
\end{figure}

The computations for LR selection can be performed using redundancy analysis (RDA) \cite{Wollenberg:77}, where the full set of CLRs serve as the response set (remembering that this set contains all the variance of the LRs), and the selected LRs are the explanatory variables chosen stepwise.
Thus, the process aims to find a small set of LRs that predicts the full set very closely.  
In parallel, the Procrustes correlation of the geometry of this reduced set of LRs with the exact geometry can be monitored. 
Using the function \texttt{STEP} in the \texttt{easyCODA} package, 22 LRs are selected to bring the explained variance over 95\% as well as the Procrustes over 0.95 --- the values obtained are 95.2\% and 0.957 respectively (see Supplementary Material A2\cite{GreenacreEtAl:23}).
The number of parts used in the 22 ratios is 30, thus omitting 22 parts.
The stepwise order of the LRs is given in the Supplementary Material A2\cite{GreenacreEtAl:23}, and the graph of the 22 ratios is drawn in 
Fig.~\ref{TellusGraph}.

Similarly, if a set of near-isometric ALRs has been identified, as in \cite{GreenacreMartinezBlasco:21}, they explain 100\% of the logratio variance, but a smaller set can be chosen, this time by stepwise backward elimination. At each step the ALR is eliminated that reduces the explained variance and/or the Procrustes correlation the least.
For example, in the ALR analysis reported in Section 4.2, 100\% of the total logratio variance was explained by the full set of 51 ALRs with reference element Al, and the Procrustes correlation with the logratio geometry was 0.991. It turns out that 13 of the elements, one quarter of the total of 52, can be eliminated with only 1\% loss in explained variance, and the Procrustes correlation reduced slightly by 0.005 to 0.986.
To compàre with the LR selection reported above, as many as 27 ALRs can be eliminated to leave the explained variance above 95\% at 95.3\% and the Procrustes correlation above 0.95 at 0.967.
Thus there are 24 ALRs retained, two more logratios than  the 22 LRs, but they only involve 25 parts, 5 less than the 29 for the LR selection (see Supplementary Material A3\cite{GreenacreEtAl:23}).

This backward elimination process of the ALRs is effectively choosing a subcomposition of the components, so for computing the Procrustes correlation an alternative would be to use the logratio geometry of the subcomposition, including all the LRs in the subcomposition, not just the ALRs.
Using the 25-part subcomposition resulting from the ALR analysis, after eliminating 27 elements, gives the same variance explained (95.3\%), but the Procrustes correlation is improved slightly to 0.972.

The above approaches suit an unsupervised context where a data reduction is sought, but could just as well be performed in a supervised context, when there is an observed response variable, in which case the LRs are chosen stepwise to maximally explain or predict the response, for example in the context of general linear models \cite{Coenders:20,Coenders:22}, or classification and regression trees. 
In this case the use of the Procrustes correlation measure of isometry is not necessary, since there is no reason here why the LRs should be close to isometry.

In all pairwise logratio selection strategies the statistical optimality criteria can be juxtaposed with domain knowledge for the choice of a set of logratios that satisfies both statistical criteria and  substantive relevance to the research question.
This approach has already been successfully implemented in three studies, two in biochemistry and one in archaeology \cite{Graeve:20, Rey:21, Wood:21}, where at each step a list of the top 20, say, LRs according to the statistical criteria were consulted by the researcher, who then selected an LR according to expert knowledge. 
Usually the top LRs are very close to one another in terms of statistical optimality, so very little is sacrificed by choosing a slightly less optimal LR that has a clearer interpretation.


\section{C\NoCaseChange{o}DA for wide data matrices}
\subsection{Genomic sequencing data}

In this section we look at an example of CoDA for data matrices with hundreds or thousands of parts, as often found in the various `omics' research fields today.
Count data coming from genomic sequencing experiments are known to be relative. 
Their size is essentially determined by features of the experimental protocol, like the sequencing depth, not the number of molecules in the input material \cite{Quinn:18}. 

There are various ``normalisation" strategies that analysts use to circumvent this problem. As an example, the ``counts per million" (CPM) normalisation is a simple count depth scaling and equivalent to a closure operation. More sophisticated methods try to estimate the size factors that would make the counts proportional to the true molecular counts before the constraints  imposed by the experimental procedures take effect. This can be achieved by identifying variables or suitable aggregations of them that remain unchanged across samples \emph{before} constraints take effect. If assumptions concerning the unchanged reference hold true, dividing all the variables by such a reference (i.e., multiplying each sample by a suitable size factor) will put samples on the same scale again. This so-called ``effective library size” normalisation can in principle recover absolute abundances, but suitable references may not exist or are difficult to identify.

The increasing complexity of some experimental protocols has led to the development of specific and ever more complex normalisation schemes that account for various sources of technical variability \cite{Booeshaghi:22}. As an example, for single-cell RNA sequencing there are nonlinear procedures that normalise genes individually depending on their relative abundance \cite{Hafemeister:19}.

Regardless of the method used, it should be kept in mind that there are essentially two possibilities: counts either remain relative or they are put on the same scale. We either ignore sample sizes in the analysis or we assume that they represent true sizes. While the former leaves us with scale-free analysis, the latter means abundance can be directly compared  between samples (and assumptions are needed for this). These two strategies are not always explicitly distinguished from each other in the literature.

The CoDA approach is scale-free, and in order to obtain compositions from relative counts, these can be considered random variables that are conditioned on their totals via a multinomial model \cite{Townes:19}. The resulting count probabilities are compositions that are estimated as parameters.

An approach working on such modelled compositions can then be compared again with the correspondence analysis approach, which works on the relative counts directly. In the following, we sketch how an exploratory analysis of a single-cell gene expression (scRNA-seq) data set could proceed from these two related perspectives.

\subsection{An example from single-cell sequencing}

The identity of a cell’s mRNA molecules expressed by thousands of genes simultaneously can be interrogated using sequencing protocols adapted to single cells. (In the more traditional, bulk RNA-seq approach,  mRNA of mixtures of many cells are sequenced.) Much work has been done in recent years to identify suitable analysis practices for single-cell data, see e.g. \cite{Luecken:19}.  Presence of compositional bias in single-cell data was reported in \cite{Buettner:21}. Independently from the compositional approach, CA has recently been proposed as a suitable analysis tool for single-cell RNA-seq data \cite{Gralinska:22,Hsu:22}.

In Supplementary Material A4\cite{GreenacreEtAl:23}, the  pre-processing of a large data set of the raw counts of 29,015 genes in 12,611 cells, with 98.7\% sparsity (i.e., data zeros) is described. Most cells are annotated with one of five cell types, and the data are aggregated into a matrix with 45 replicate samples of the cell types and a reduced set of 6147 genes, with a sparsity still high at 64.1\%. The number of replicates is seven or eight for two of the cell types and ten replicates for the other three cell types. This $45\times 6147$ matrix of counts is of primary interest, a size that is typical in this research field: very few rows and very many columns.

In order to follow the logratio approach described in the previous sections, we estimate the count probabilities, thereby effectively imputing zero counts. We do so by choosing a quasi empirical-Bayes modelling similar to what is done for bulk-RNA seq data (see Supplementary Material A5\cite{GreenacreEtAl:23}). Our proposal here is simple and only intended for exploratory analysis and visual comparison with the CA biplots.


\subsection{Visualization comparison of modeled compositions and relative counts}
\begin{figure*}[t]
\centering
\includegraphics[width=0.99\linewidth]{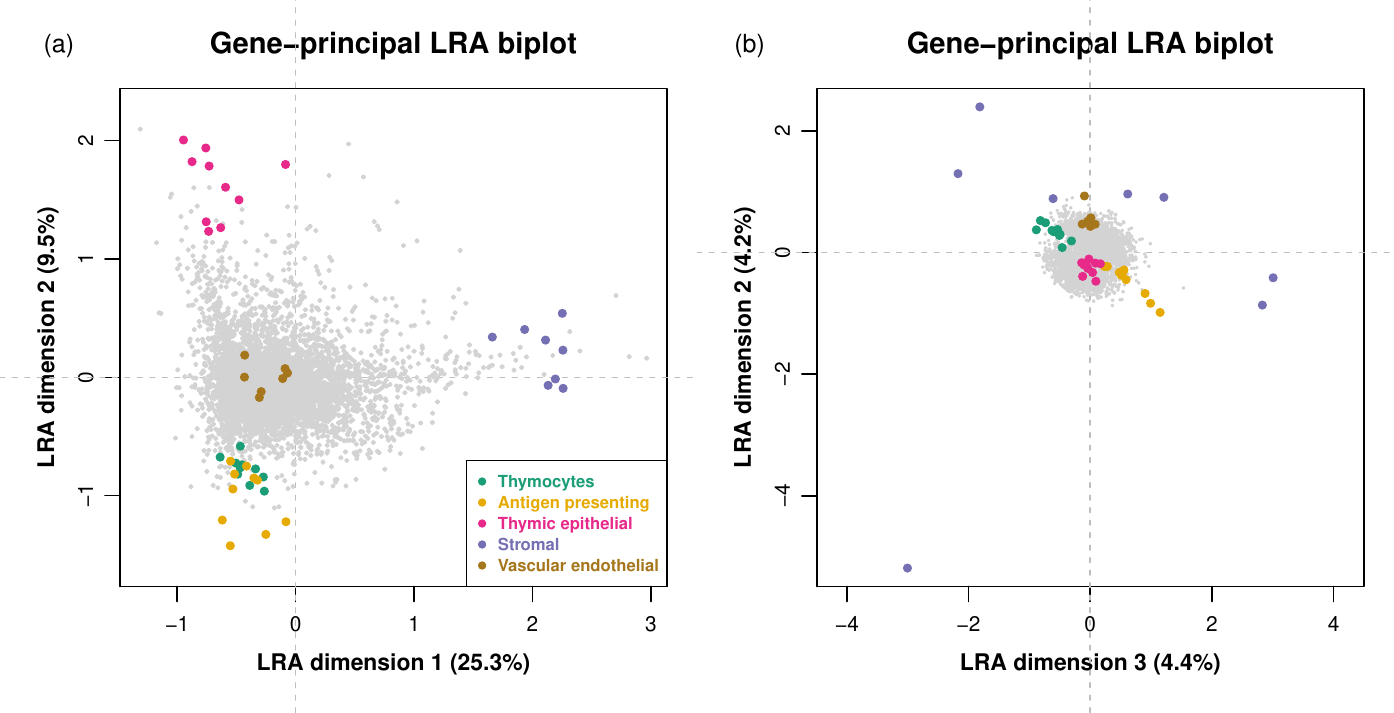}
\vspace{-0.2cm}
\caption{(a) LRA gene-principal biplot obtained using 45 pooled samples from 5 cell types. Genes are shown as small grey dots. 
Many of them show clear} expression preferences for the various cell types. (b) As before but showing dimensions 3 and 4.
\label{FigIonas1}
\end{figure*}

For visualisation, we proceed with an LRA (i.e., PCA of the log-transformed and double-centred matrix) of the 45 estimated parameter compositions. 
Fig.~\ref{FigIonas1} shows the LRA gene-principal biplots for dimension pairs 1--2 and 3--4.

\begin{figure*}[t]
\centering
\includegraphics[width=0.99\linewidth]{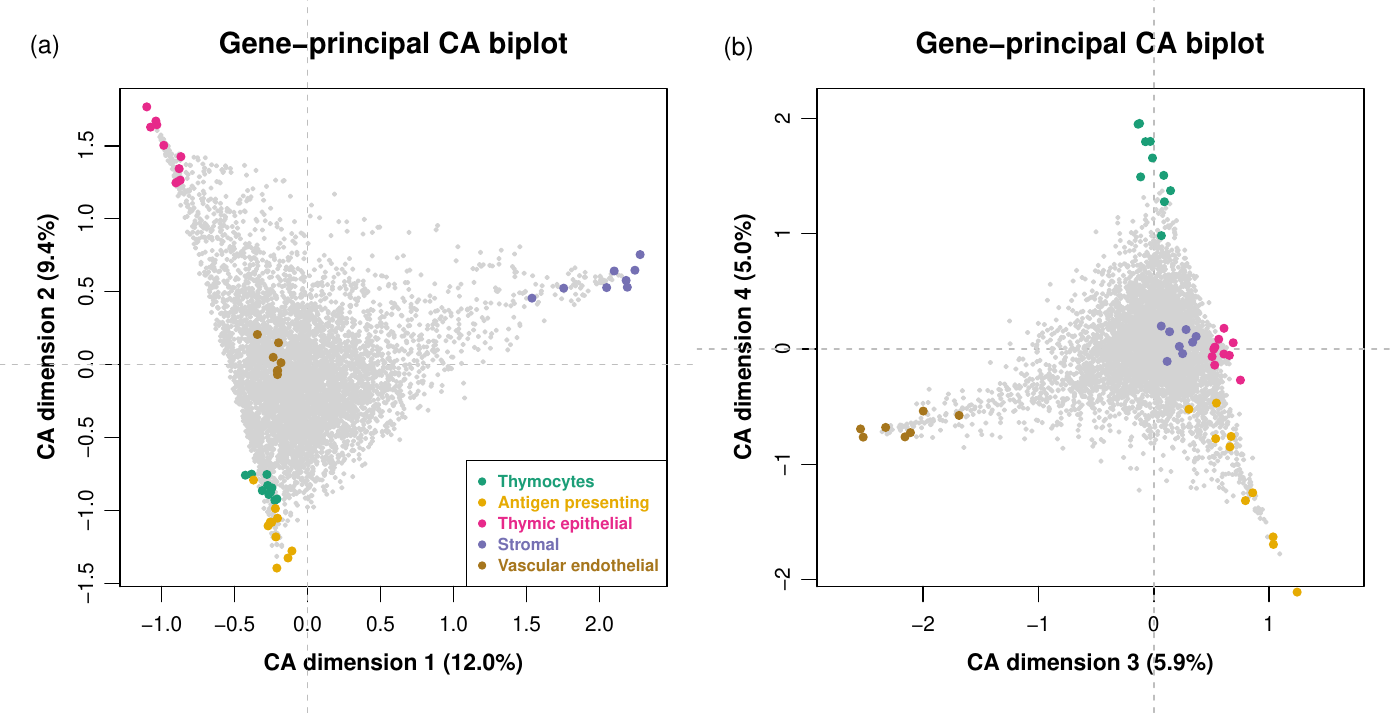}
\vspace{-0.2cm}
\caption{CA gene-principal biplots of 45 pooled samples from 5 cell types, as in Fig.~\ref{FigIonas1}. (a) Dimensions 1 and 2. (b) Dimensions 3 and 4. The pooled samples are more clustered than in Fig.~15 and their gene associations clearer.}
\label{FigIonas2}
\end{figure*}

For comparison, we also show the CA gene-principal biplots obtained from the data matrix of the same replicates, but without zero imputation (Fig.~\ref{FigIonas2}). 
Aiding intuition, CA’s projection of the weighted simplex becomes apparent: the large number of diversely expressed genes are spread out between the corners formed by the cell types  and genes that are expressed exclusively in one of the cell types fall in the respective corners. 
In the first two dimensions the overall spatial arrangement is quite similar to that of the LRA.
There are differences in dimensions 3 and 4 --- for LRA (Fig.~15b) the variance of the genes in principal coordinates is low and the sample pattern unclear, whereas for CA (Fig.~16b) the sample structure is clearer, and all sample groups and corresponding genes are effectively separated in the four-dimensional result.
While for LRA we estimate the parameters, CA is performed on the data themselves, with the zeros contributing to defining the multivariate structure. 
CA's chi-square standardization results in Mahalanobis distances for multinomial counts \cite{Greenacre:16a}, so from this perspective the approach is justified in its own right.

An advantage of CA becomes apparent when we set out to plot the original data consisting of 12,611 cells, including the ones that were not included in the 45 samples or that could not be pooled due to the missing annotation. 
It seems infeasible to meaningfully impute zeros in a matrix with 92\% zero elements (the sparsity of the unpooled data). 
While LRA appears out of the question, CA handles these cells easily as supplementary points of the plot presented before. 
In Supplementary Material A6\cite{GreenacreEtAl:23}, Fig.~A2, they are visualised without showing pooled samples and genes. 
Beyond mere visualisation, the coordinates of the supplementary CA points could also be used to predict cell types of cells without annotation (shown in yellow), e.g., using nearest annotated neighbours.

Similar to the exercise for the 52-part Tellus data in  Fig.~\ref{Tellus_coherence_CA}, Supplementary Material A7\cite{GreenacreEtAl:23} shows for the single-cell data how close to coherence random subcompositions are in CA. 
In fact, it is demonstrated that the coherence is extremely high for data with a very large number of parts.

\section{Are exact compositional coherence and isometry necessary for C\NoCaseChange{o}DA?}
To recap, coherence is the property that the relationships between a set of compositional parts should not be affected by the particular composition they find themselves in --- the composition can be extended, for example, and the relative values changed, without affecting the results for the original set of parts. 
Strict adherence to the coherence property imposes the use of ratios of parts.
Isometry is the property that the logratio geometric structure of the samples, determined in the $d$-dimensional space of all pairwise logratios, is preserved by the particular data transformations applied to the compositional data.
Strict adherence to the isometry property imposes the use of logratios transformations that are isometric.
 
For more than 20 years after the advent of logratio-style C\NoCaseChange{o}DA at the time of Aitchison's 1982 paper, researchers following the logratio approach were more or less happy with LRs, ALRs and CLRs, but each of these had their own apparently problematic aspects.
The LRs, the most fundamental concept, underpinned the approach to CoDA, but they were very many of them whereas their rank was only $d = D-1$, so a method for selecting them was required. 
A chosen subset of LRs was clearly coherent, but not isometric.
The ALRs had exactly the right number of them ($d$), were linearly independent and explained all the logratio variance, but the reference had to be chosen and no ALR transformation reproduced the logratio geometry exactly. 
A particular set of ALRs was again clearly coherent (if the composition were to be extended, for example), but again not isometric.
And the CLRs, of which there were $D$, had rank $d=D-1$, reproduced the logratio geometry exactly, hence were isometric, but were not coherent because each involved all the components.
Thus, CLRs could not be interpreted as their numerator components, although it was tempting to do so.

The ILR transformation was intended to solve the above ``deficiencies", but it came with the burden of a very complicated interpretation of the individual ILR variables in practice, unless one considered them a ``black box".
In response to the question posed by the above section heading, and in the light of the examples given previously, our answer about strict, mathematically exact, adherence to these two properties is in the negative: we believe that approximate coherence (quasi-coherence) and approximate isometry (quasi-isometry) are sufficient for good practice in CoDA, not unlike well-accepted statistical practice where approximate normality is acceptable for performing certain tests that rely theoretically on exactly normal-distributed data.

\subsection{Measuring lack of coherence (incoherence)}
If coherence is deemed an important property, any way of transforming compositional data, even non-transformation, can be applied and the closeness to exact coherence can be measured, as in the examples of Sections 4.4 and Supplementary Material A7\cite{GreenacreEtAl:23}.
The general idea is to study the behavior of random subcompositions of different sizes and see the effect on either the geometry of the subcompositional parts or on results such as effect sizes and p-values, if the parts are used in model-fitting (an example of the latter in a regression context is given in \cite{Greenacre:21}).
The measure of incoherence can be the stress measure from multidimensional scaling \citep{Greenacre:13}, or the Procrustes correlation, on the geometry of the subcompositional parts, as used in Sections 4.4 and 6.5.
Low stress near 0\% or high Procrustes correlation near 1 would imply high coherence.
Based on the results of this exercise, it is possible that mathematically strict coherence can be dispensed with, opening up the field to approaches that are easy to understand and interpret, as well as coping with zeros without the need to replace them.
In addition, given the stochasticity and noise in statistical data, in the same way we do not retain all dimensions in component methods such as PCA or CA, one could also imagine that it is acceptable to use methods that only approximately preserve coherence.

\subsection{Measuring lack of isometry (anisometry)}
As mentioned before, the property of isometry is not due to Aitchison, and its necessity is questionable.
When an LRA or any other dimension-reduction technique or variable selection is performed, the resultant small set of dimensions or variables do not constitute an isometric version of the original logratio distances, only an approximate one.
Thus, it seems illogical to prohibit non-isometric transformations from the start, especially those that might be very close to isometry --- what is needed is  a measure of deviation from isometry..
Such a measure is useful, for example, to choose between alternative ALR transformations, as was done in Section 4.2.
We have shown that the Procrustes correlation can again be used for this purpose, but this time on the geometry of the cases.
If an alternative transformation produces a geometry of the cases with a very high Procrustes correlation with their logratio geometry, then this alternative can be validly used, if isometry is deemed a useful property in the context of the research objective. 
This transformation might not even involve logratios, but something simpler such as the original compositional data that have been optionally power transformed and standardized in some way, for example the chi-square standardization. 

\subsection{Insistence on the use of isometric logratios}
A problem with the advent of ILR ``balances" was that they were heavily promoted in the literature as being the ``correct" way to transform compositional data. 
Several authors have been responsible for unreasonably insisting on their mandatory use.
For example:
\begin{itemize}
  \item ``Compositional data analysis requires selecting an orthonormal basis with
which to work on coordinates.'' (first sentence of abstract of \cite{Martin:18})
  \item ``Compositional vectors need to be expressed in orthonormal coordinates, thereby allowing further processing using standard statistical tools'' \cite{Fiserova:11}; 
  \item ``Compositional data should be expressed with respect to orthonormal coordinates\footnote{Notice that the term ``orthonormal coordinates'' conveys the mistaken impression that the ILRs are themselves orthonormal vectors, but the term refers rather to the fact that the ILRs are linear combinations of the log-transformed components with orthonormal coefficients -- see the orthogonal coefficients in \eqref{ILR} before rescaling to be orthonormal). Recently it has been proposed that ILRs be renamed ``orthonormal logratio coordinates'' \cite{Hron:21}, which unfortunately further propagates this false impression and creates confusion.} that ``guarantee isometry between the Aitchison geometry [of the simplex] and the real space" \cite{Kynclova:17}; 
  \item ``isometric logratio coordinates, real coordinates with respect to an orthonormal basis in the Aitchison geometry, are preferable" ... ``The overall preferred option is to display compositional data in any interpretable orthonormal (ilr) coordinates, free of possible caveats resulting from an inappropriate treatment of observations carrying relative information.'' ... ``one should be aware that the real data structure can hardly be recognized without expressing the compositions in orthonormal coordinates'' ... ``The only way out is to express the data in interpretable orthonormal coordinates.'' ... ``Fortunately, almost no peculiarities can be expected when performing statistical inference for compositional data in orthonormal coordinates.'' \cite{Filzmoser:18};
  \item ``It is not possible to apply statistical analysis correctly" (referring to using a well-known ratio in water chemistry, so the author `revises' the ratio as an ILR and then states that the analysis is now) ``coherent with the nature of compositional data, thus obtaining a simple tool to be used in a statistical sense, going beyond the descriptive approach" \cite{Buccianti:15}. 
\end{itemize}

\noindent
Notice the implication, especially in the last quotations above, that to not use the ILR transformation will be an erroneous treatment of the data.

By contrast, it is remarkable that Aitchison himself expressed disagreement about this
development. 
He refers to the ``fallacy" of using orthonormal coordinates, saying that 
\begin{quotation}
``given the elegance of the algebraic geometric (Hilbert space) structure of the simplex it is easy to fall into the pure-mathematical trap that all compositional problems must depend on this structure, that all statistical problems should be addressed in terms of coordinates
associated with orthonormal, isometric bases" (page 20 of \cite{Aitchison:08}).
\end{quotation}
\begin{quotation}
``Where things go wrong is an implicit belief that [the isometric logratio transformation] is the only safe way to tackle compositional problems. 
In my view such an approach to a practical problem is fraught with difficulties."
\end{quotation}
The ILRs guaranteee perfect isometry, but generally no interpretation of the ILRs is ventured, even though sometimes they are surprisingly referred to as ``interpretable'' \cite{PawlowskyEtAl:15}.
A truer description is that they are a ``perfect black box'' \cite{BoogaartTolosana:13}.
Further criticisms of ILRs in the literature can be found in \cite{Cortes:09, ScealyWelsh:14}.
As shown by the Tellus application, as well as several other publications \cite{GreenacreMartinezBlasco:21, Martinez:21, Martinez2:21}, a carefully selected ALR transformation or the standardization inherent in correspondence analysis, with optional power transformation, can be close enough to isometry for all practical purposes.
One of these papers \cite{GreenacreMartinezBlasco:21} gives three in-depth applications of successfully using the ALR transformation as well as for 30 more simulated data sets to show its good performance.

\subsection{The geometric mean for grouping parts}
The use of the geometric mean to combine parts to define ``balances", even in spite of domain knowledge, is promoted by the proponents of ILRs, who say they are "easily interpreted in terms of grouped parts of a composition"  (page 38 of \cite{PawlowskyEtAl:15}). 
The ratio of two geometric means in an ILR ``balance'' is often misinterpreted as the ratio of two compositional sums, as in page 41 of  \cite{PawlowskyEtAl:15}, an example which is glaringly incorrect.
Fortunately, the difficulties with using the geometric mean in CoDA for grouping components is gaining some traction in the literature, for example these recent quotes from \cite{Smithson:22}:
\begin{quotation}
  ``Thus, a geometric mean of parts cannot be regarded as equivalent to combining the parts'' ... ``The geometric mean of compositional parts cannot be interpreted as simply combining the parts, whereas their sum can.''
\end{quotation}
Even some of the main protagonists of ILR ``balances" are expressing some doubts about their usefulness; for example \cite{Hron:21}: 
\begin{quotation}
``Even if balances can be successfully constructed, the resulting (orthonormal logratio) coordinates usually include some which do not have a straightforward interpretation (if any at all) in the context of the particular application. In such cases their practical usability is questionable''.
\end{quotation}
One of the most complicated logratio transformations involving geometric means, the ``symmetric balance'' \cite{Kynclova:17}, has been conceived in an attempt to define what could questionably be called a correlation between compositional parts.

Several authors have actually proposed ILRs as single
variables, and have often substituted existing measures that involve sums with geometric means for the sake of creating an ILR.
An example already cited \cite{Buccianti:15} is in water chemistry, where the well-known Gibbs diagram involves a ratio with respect to the total dissolved solids (i.e., an amalgamation), which is then substituted by the geometric mean.
Another example of using an ILR for its own sake is \cite{Egozcue:19}, 
where an existing measure of economic inequality involving a sum is replaced by one using a geometric mean.
The interpretation of this alternative index is now more complicated and there seems to be no substantive benefit 
at all, except to satisfy the definition of an ILR \cite{Greenacre:19d}.

\subsection{The role of amalgamated parts}
As a reaction to the problems of using ILRs involving ratios of geometric means, there has been a return to defending the use of Aitchison's simpler summed (amalgamated) components to combine parts in logratios \cite{Greenacre:20, GreenacreGrunskyBaconShone:20}.
The \textit{summed}, or \textit{amalgamated, logratio} (SLR), uses amalgamations as originally proposed by Aitchison \cite{Aitchison:86}:
\begin{eqnarray}
  {\rm SLR}(D_1,D_2) = \log\left( \frac{\sum_{j\in D_1} x_j}{\sum_{j\in D_2} x_j} \right)
  \label{SLR}
\end{eqnarray}
where $D_1$ and $D_2$ are two non-intersecting subsets of parts.
Rather than basing the amalgamations on an algorithm, as is commonly done for ILRs, one could instead define combinations of parts based on domain knowledge.
For example, biochemists studying fatty acids already do it as a matter of course when they sum all saturated fatty acids into a single amalgamated part and create a ratio with a similar amalgamation of monounsaturated fatty acids.
When a data-driven aggregation of parts, rather than a knowledge-driven one, is desired, it is possible to perform a clustering of the parts in a hierarchical clustering by amalgamation rather than by geometric means, as depicted in Fig.~\ref{TellusTree2}, where at each node parts are amalgamated to minimize the loss of explained logratio variance \cite{Greenacre:20}. 
Data-driven SLRs can also be used in place of data-driven ILRs in the context of supervised machine learning, that is, to maximize the predictive accuracy of a dependent variable \cite{Gordon-Rodriguez:21}.

SLRs do not decompose into combinations of LRs, and thus do not fit into the neat linear theory of log-contrasts like the other logratio transformations.
But they have many advantages: they are simple to interpret, are usually based on substantive issues that are relevant to the research question, are subcompositionally coherent with respect to adding new components and explain logratio variance or response variable variance just as any other statistical variable can.
Furthermore, they assist with the zeros problem in CoDA, since some components with data zeros can be incorporated into a summed component, whereas zeros are still problematic in a geometric mean.
Fortunately, researchers who use logratio transformations are starting to use amalgamations again in ratios, for example \cite{QuinnErb:20, Greenacre:20, GreenacreGrunskyBaconShone:20, Wood:21}.

Although amalgamating components is the most natural thing to do in practice, it is eliminated from consideration by the proponents of combining parts using geometric means with the simple dismissal: ``Amalgamation is incompatible with the techniques presented in this book" \cite{PawlowskyEtAl:15}.
In fields such as bio- and geochemistry it is regular practice to form ratios of amalgamations, even without log-transforming, for example \cite{Stanley:19}. 
So it seems unscientific to disallow amalgamation of parts based solely on a mathematical argument that does not have substantive scientific underpinning.

\section{Reappraisal and conclusion}
No practitioner should be forced into any particular approach based on a pure mathematical argument that is disconnected with the application field.
Nevertheless, practitioners can take Aitchison's logratio approach as a type of reference, since it is so widely used today, but not necessarily as an ideal or a gold standard.
The measurements of incoherence and deviation from isometry, if these properties are deemed important, provide a straightforward empirical solution for future applications.
First, for those using the logratio approach, these measures are good diagnostics for  variable selection and thus lead to more parsimonious results.
Second, for those using other approaches, they can check their particular choice of transformation and method against that of the logratio approach, should they choose to do so.

The problem of data zeros, especially structural zeros, is probably the thorniest and least-resolved issue in the logratio approach, and has been called the ``Achilles heel of CoDA'' \cite{GreenacreGrunskyBaconShone:20} .
The definition of a composition should allow for zero values --- in a study of daily time use, for example, if someone does not work, this is a true zero for percentage of working time and replacing it with a small value is creating a fiction for the sake of being able to use the chosen method, whereas it should be the other way round, namely that the method has to cope with the true data zero.
Likewise, if there are structural zeros in a data set, then replacing them with positive numbers merely to allow logratios to be computed seems senseless, if not incorrect.
In reported applications, the issue of data zeros tends to be minimized, with replacements made using one of the available algorithms, and usually no attention paid to the possible effects of these replacements, especially in data sets with very many zeros.
One study of genetic data with many zeros shows that the number of zeros is related to the total count before normalization, and this has consequences on the results of the statistical analysis \cite{teBeest:21}. 

The many other approaches to CoDA which do not rely on logratios are not surveyed here, for example \cite{Stephens:82, Butler:08, ScealyWelsh:11, Ren:17, Stanley:19}, to mention only a few. 
For a strong critique of the logratio approach, the concept of subcompositional coherence and a discussion of alternative approaches, see \cite{ScealyWelsh:14}.

In conclusion, if one accepts the logratio approach to compositional data analysis then Aitchison's original formulation is, as he said himself, simple \cite{Aitchison:97}.
Once the compositional data have been transformed to logratios, everything that follows is standard statistical practice, with some care taken in the interpretation of these transformed components.  
Aitchison's basic concepts of pairwise, additive and centered logratios, with the possibility of amalgamating components based on domain knowledge, remain the most useful in practice today.
The more complex isometric or pivot logratios are not a prerequisite for good practice, in fact they may be detracting from good practice, so researchers should not feel pressured into using them by certain dogmatic statements in the literature.
While it is true that the ILR transformation ``is increasing in popularity'', it is not clear whether all users fully understand what ILRs are measuring or how they are chosen, and they remain a ``black box".
Black boxes cause many problems, in that we cannot ensure that the models are scientifically meaningful, so we should always prefer simpler models that we can explain, unless they are clearly inferior.

There is strong evidence that simple logratio transformations such as ALR as well as even simpler transformations, not even logratios, provide solutions that are measurably close to being subcompositionally coherent and isometric, allowing a more open approach to CoDA.
For example, the potential of correspondence analysis to handle compositional data has been proposed as early as 1977 \cite{David:77}, see also \cite{Grunsky:85, Jackson:97} and has been used in several fatty acid studies where the data contain many zeros \cite{Meier:16, Kraft:17, Haug:17}, as was the case for the wide data set with many zeros in Section 6.

This article has been concerned mainly with unsupervised learning in CoDA and how to validly transform compositional data.
If the compositions serve as predictors or as responses, the distinction between the different logratio transformations is blurred by the fact that they generate the same log-contrasts \cite{Coenders:20, Coenders:22, Yoo:22}.
In this case, the decision on which to use is entirely guided by interpretability, with pairwise and additive logratios being the most attractive options.

The last word is left to John Aitchison himself, who in his keynote address \cite{Aitchison:08} at the Compositional Data Analysis Workshop in 2008, stated the following:
\begin{quotation}
``The ALR transformation methodology has, in my view, withstood all attacks on its validity as a statistical modelling tool. Indeed, it is an approach to practical compositional data analysis which I recommend particularly for non-mathematicians. The advantage of its logratios involving only two components, in contrast to clr and ilr, which use logratios involving more than two and often many components, makes for simple interpretation and far outweighs any criticism, more imagined than real, that the transformation is not isometric."
\end{quotation}
\begin{quotation}
``Where things go wrong is an implicit belief that [the isometric logratio transformation] is the only safe way to tackle compositional problems. 
In my view such an approach to a practical problem is fraught with difficulties. 
The coordinates in any ilr transformation are necessarily a set of orthogonal logcontrasts. 
So imagine a consultation with a urologist who has analysed the 4-part composition of renal calculi extracted from male patients at operation. 
The patients subsequently have been classified as of type R (repeater) or type S (single episode patient). 
The urologist now consults an ilr compositional data analyst to ask if there is any way of deciding whether a new patient, whose extracted kidney stone has composition $[x_1 , x_2 , x_3 , x_4 ]$, is more likely to be R rather than S. 
So the ilr analyst explains to the urologist that it is necessary to examine an ilr transformation, say:
\begin{equation*}
\begin{aligned}
& (\log x_1 - \log x_2 ) / \sqrt{2}, \\[-0.5em] 
&(\log x_1 + \log x_2 - 2 \log x_3 ) / \sqrt{6}, \\[-0.5em]
&(\log x_1 + \log x_2 + \log x_3 - 3\log x_4 ) / \sqrt{12}
\end{aligned}
\end{equation*}
in order to answer his problem. I know a number of urologists and I cannot imagine any of them understanding why such an elaborate transformation would be necessary. 
And why, they would sensibly ask, is this particular ilr transformation relevant? 
Why not another one? 
And an even greater stretch of the imagination would be required to envisage them bringing further compositional problems for analysis.
Indeed, ensuring isometry has little to do with this compositional problem."
\end{quotation}
(see our previous remark about interpretability of logratios when used in models).
And in the same 2008 keynote address, referring back to earlier workshops, Aitchison said:
\begin{quotation}
``At CodaWork’03 and CoDaWork’05, when countering this insistence on the use of ilr transformations, I said I would look forward to a convincing practical use of the method. 
As far as I know there has been no progress in demonstrating its applicability."
\end{quotation}
This view expressed 14 years ago, and vindicated by the present paper, remains relevant today --- still there has been no evidence that demonstrates the obligatory use of the ILR transformation in the practice of CoDA.
Furthermore, there is ample evidence that simpler logratio transformations, or even transformations not based on logratios, such as in correspondence analysis, are feasible.

\section*{Data availability and software}
The data sets as well as R scripts for reproducing the analyses in this paper are provided on GitHub at 

\smallskip

https://github.com/michaelgreenacre/CODAinPractice

\smallskip
\noindent
The function \texttt{FINDALR}  will be available in the new release of the \texttt{easyCODA} package \cite{Greenacre:18} in \textsf{R} \cite{R:21}.
It can also be found in the latest version on RForge, which is installed as follows:

\smallskip
\noindent
install.packages("easyCODA", repos="http://R-Forge.R-project.org").


\begin{supplement}
\noindent
\stitle{A. Additional analyses, figures and tables}
\sdescription{Seven short sections, comprising  supporting analyses, figures and tables:
\begin{itemize}
\item[A1] 
ALRs compared to log of numerator parts
\item[A2]
Stepwise entry of LRs
\item[A3]
Backward elimination of ALRs
\item[A4]
Pre-processing of single-cell genomic data
\item[A5]
Modeling pooled single-cell data
\item[A6]
Visualizing thousands of samples with many zeros
\item[A7]
Dilution of incoherence for wide matrices
\end{itemize}}
\end{supplement}
\begin{supplement}
\noindent
\stitle{B. Tellus geochemical cation data}
\sdescription{The $6799 \times 52$ matrix of geochemical compositional data, in a text file.}
\end{supplement}
\begin{supplement}
\noindent
\stitle{C. R script for analysing Tellus data}
\sdescription{The code for producing most of the analyses and figures in Sections 3 to 5, and in the Supplementary Material.}
\end{supplement}
\begin{supplement}
\noindent
\stitle{D. Single-cell genomic data}
\sdescription{The $12611 \times 29015$ matrix of single-cell compositional data, in an \textsf{R} workspace.}
\end{supplement}
\begin{supplement}
\noindent
\stitle{E. R script for analysing the single-cell data}
\sdescription{The code for producing most of the analyses and figures in Section 6 and in the Supplementary Material.}
\end{supplement}

\clearpage

\section*{Supplementary material}
\renewcommand{\thefigure}{A\arabic{figure}}
\setcounter{figure}{0}
\setcounter{table}{0} 

\subsection*{A1. ALRs compared to log of numerator parts}
\begin{figure}[h!]
\includegraphics[width=0.9\linewidth]{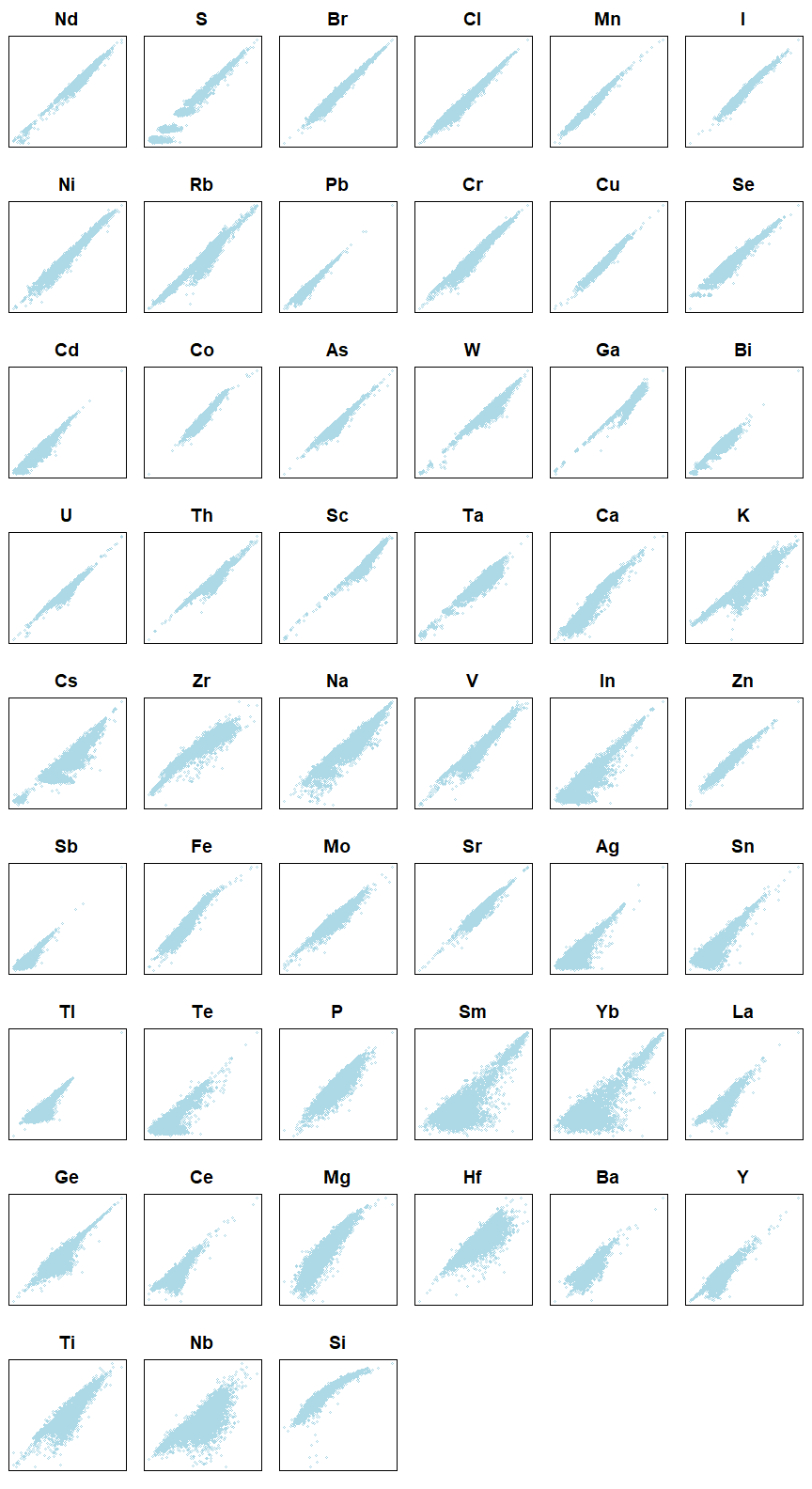}
\vspace{-0.2cm}
\caption{For each of the 52 numerator parts in the ALR transformationwith Al as the reference parts, the scatterplot of the log-transformed numerator (on the y-axis) versus the corresponding ALR (on the x-axis).}
\label{A1}
\end{figure}

\newpage

\subsection*{A2. Stepwise entry of LRs}
\begin{centering}
\begin{table}[h!]
{\large
\begin{verbatim}
    ratio row col R2cum  Procr           
 1  Cl/Ga  19  24  44.7  0.669
 2  Ni/Rb  33  35  66.0  0.807
 3  Si/I    1  27  70.5  0.829
 4  Nd/Tl  32  45  74.2  0.852
 5  Mn/S    5  11  77.0  0.869
 6  Pb/Sc  34  37  79.4  0.883
 7  Bi/Sm  15  39  81.2  0.893
 8  As/Ga  13  24  83.0  0.900
 9  Ga/Sr  24  41  84.5  0.907
10  Cu/Ta  23  42  85.9  0.917
11  Mn/Rb   5  35  87.3  0.924
12  Cd/Zr  17  52  88.4  0.931
13  Mo/Ta  30  42  89.4  0.934
14  Si/Sc   1  37  90.4  0.938
15  Cu/Yb  23  50  91.3  0.941
16  Rb/Sc  35  37  92.0  0.942
17  Cl/Sc  19  37  92.7  0.943
18  Cr/U   21  46  93.3  0.946
19  Cd/W   17  48  93.9  0.948
20  Na/Cs   7  22  94.4  0.952
21  Br/W   16  48  94.8  0.955
22  Cr/Zn  21  51  95.2  0.957
\end{verbatim}
}
    \caption{Stepwise selection of LRs. The column \texttt{R2cum} is the cumulative percentage of explained variance.
    The column \texttt{Procr} is the Procrustes correlation of the set of LRs chosen up to that step.}
    \label{TellusStepwise}
\end{table}
\end{centering}

\newpage

\newpage

\subsection*{A3. Backward elimination of ALRs}
\begin{table}[h!]
\begin{centering}
{\large
\begin{verbatim}
    ratio  row col  R2cum  Procr           
 0  ALL      0   0  100.0  0.991
 1  Sm/Al   39   2  100.0  0.991
 2  Ce/Al   18   2  100.0  0.991
 3  Yb/Al   50   2   99.9  0.991
 4  Nb/Al   31   2   99.9  0.990
 5  Ti/Al   10   2   99.9  0.990
 6  Tl/Al   45   2   99.8  0.990
 7  Si/Al    1   2   99.7  0.990
 8  Hf/Al   26   2   99.7  0.989
 9  Ba/Al   14   2   99.6  0.989
10  Y/Al    49   2   99.4  0.988
11  Sn/Al   40   2   99.3  0.988
12  Mg/Al    4   2   99.2  0.987
13  La/Al   29   2   99.0  0.986
14  Ge/Al   25   2   98.8  0.986
15  Ag/Al   12   2   98.7  0.985
16  Sb/Al   36   2   98.5  0.984
17  Te/Al   43   2   98.2  0.983
18  K/Al     8   2   98.1  0.982
19  Fe/Al    3   2   97.9  0.981
20  P/Al     9   2   97.6  0.979
21  V/Al    47   2   97.4  0.978
22  Se/Al   38   2   97.1  0.977
23  In/Al   28   2   96.8  0.975
24  Th/Al   44   2   96.6  0.974
25  Zr/Al   52   2   96.1  0.972
26  Cs/Al   22   2   95.8  0.969
27  Ca/Al    6   2   95.3  0.967
..   ...    ..  ..    ..    ..
\end{verbatim}
}
\end{centering}
    \caption{Backward elimination of ALRs. The column \texttt{R2cum} is the percentage of explained variance of the remaining ALRs at that step..
    The column \texttt{Procr} is the Procrustes correlation of the set of the remaining ALRs.
    The first row is when all 51 ALRs are included. 
    The next row is when the ALR of Sm/Al is eliminated, causing the least loss of explained variance.
    The next row is when, in addition, the ALR of Ce/Al is eliminated, and so on. 
    By the 27th step, 27 ALRs have been eliminated, leaving 24 ALRs that include the 24 numerator parts and the reference part Al.}
    \label{TellusBackward}
\end{table}

\newpage
\subsection*{A4. Pre-processing of single-cell genomic data}
As a relatively small example data set, we choose cells from the Thymus tissue included in the human foetal gene expression atlas (Cao et al, 2019) and downloaded from

\noindent
\texttt{https://descartes.brotmanbaty.org/bbi/}

\noindent
\texttt{human-gene-expression-during-development/}. 

\noindent
The processed data comprise 12,611 cells with raw counts for 29,015 genes that are not zero in all of the cells. The sequencing depth is low; the median number of counts for a cell is 365. Overall, 98.7\% of the matrix elements are 0. The data come along with an annotation of about 70\% of the cells into 5 cell types (obtained via marker gene expression). 
The cells are obtained from 8 individuals.

The annotation is crucial for us to handle the extreme sparsity. 
It allows us to add the gene counts of cells of the same type and from the same individual (assumed to be measured on the same scale) into pooled replicate samples. This strategy reduces sparsity while retaining some of the variation coming from different cells, individuals, and experimental batches within cell types. 

In one of the individuals, only 12 and 13 cells are annotated as stromal and vascular endothelial, respectively. 
Based on these numbers, we created pooled samples of 12 cells each coming from the same individual and assigned to the same cell type. In this way, 724 pooled samples were created for the 5 cell types. 
If there were less than 12 cells remaining from the same individual and type, they were added to one of the pooled samples of the same individual and type.
To reduce the imbalance with respect to the number of replicates per cell type 
we create a data matrix with no more than 10 replicates for any cell type. 
This leaves us with a matrix containing 45 samples, with seven and eight replicates for two of the cell types and 10 replicates for three of the cell types. 
For a meaningful statistical analysis, we keep only those genes having sufficiently many counts. Here we keep all genes having at least 10 counts per gene in the subset of 45 samples.
This reduces the number of genes to 6147. 
The number of zeros in the resulting $45\times 6147$ matrix still amounts to 64.1\%.

\newpage

\subsection*{A5. Modelling pooled single-cell data}
As argued by a number of authors, for example \cite{Townes:19,Breda:21}, the counts of the unique molecular identifiers (UMI) in single-cell expression data can be considered multinomial. 
A complication in addition to compositionality is sparsity, which is often excessive. There are currently no generally accepted standards for handling zeros \cite{Svensson:20}.

To infer the multinomial parameters for a cell type, it would be ideal to choose a prior distribution for the parameters that takes their intrinsic interdependence into account. The additive logistic-normal distribution (introduced by \cite{AitchisonShen:80} and discussed by Aitchison in his 1982 paper) would be ideally suited for this purpose and has been used in microbiome analysis, e.g., \cite{McGregor:20}. A more generic option that has often been adopted in the genomics community is modelling via a negative binomial distribution, e.g. \cite{Hafemeister:19}.\\
Here we adopt a simple strategy that does not take the covariance structure into account when inferring multinomial probabilities for the counts. This maximum-entropy shrinkage approach has originally been proposed to estimate mutual information in bulk RNA-seq data.
  It consists in shrinking the empirical distribution (i.e., the relative counts) towards the equidistribution. The shrinkage intensity can be determined in a data-driven way by minimising the mean-squared error with respect to the unknown multinomial parameters \cite{Hausser:09}. This approach is equivalent to an empirical Bayes procedure with data-driven inference of the weight of the conjugate prior. While this Dirichlet prior cannot capture dependencies between the parts beyond those introduced by the closure, data-driven flattening constants work much better \cite{Hausser:09} than arbitrarily chosen pseudocounts, which are still widely used in genomics. The R package ``entropy” \cite{Hausser:21} allows us to obtain the shrinkage estimates for individual samples in a computationally efficient way. Note that this procedure imputes the zeros, and the resulting gene frequencies are now suited for logratio analysis.
\newpage
%
\subsection*{A6. Visualizing thousands of samples with many zeros}
\begin{figure}[h!]
\includegraphics[width=\linewidth]{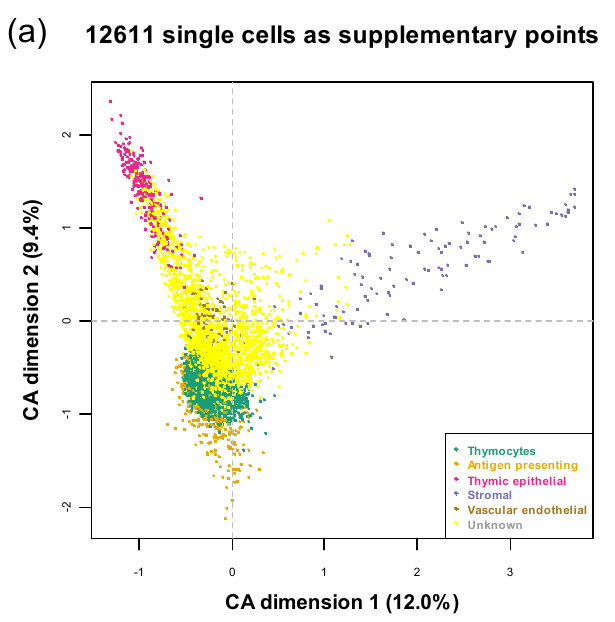}
\vspace{-0.2cm}
\end{figure}
\begin{figure}[h!]
\includegraphics[width=\linewidth]{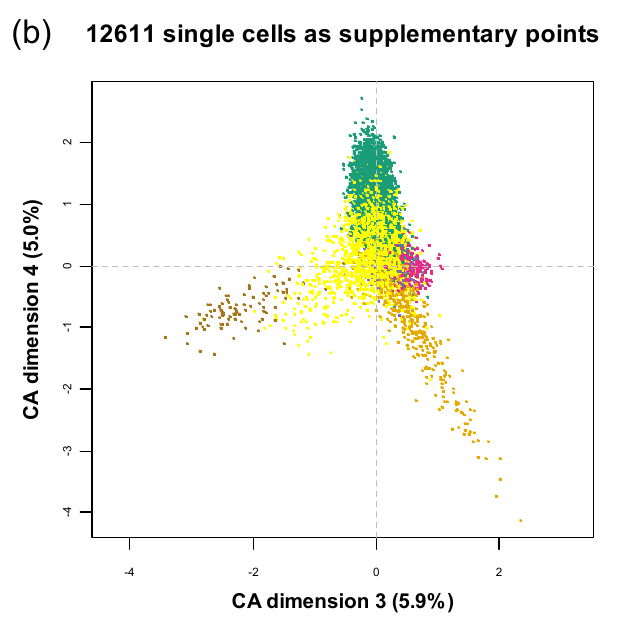}
\vspace{-0.2cm}
\caption{(a) CA gene-principal biplot showing only the original (unpooled) cells as supplementary points. Note that cells without annotation can be plotted as well (in yellow). (b) As in (a), but showing dimensions 3 and 4.}
\label{fig:Ionas4}
\end{figure}


\subsection*{A7. Dilution of incoherence for wide matrices}
\begin{figure}[h!]
\includegraphics[width=0.99\linewidth]{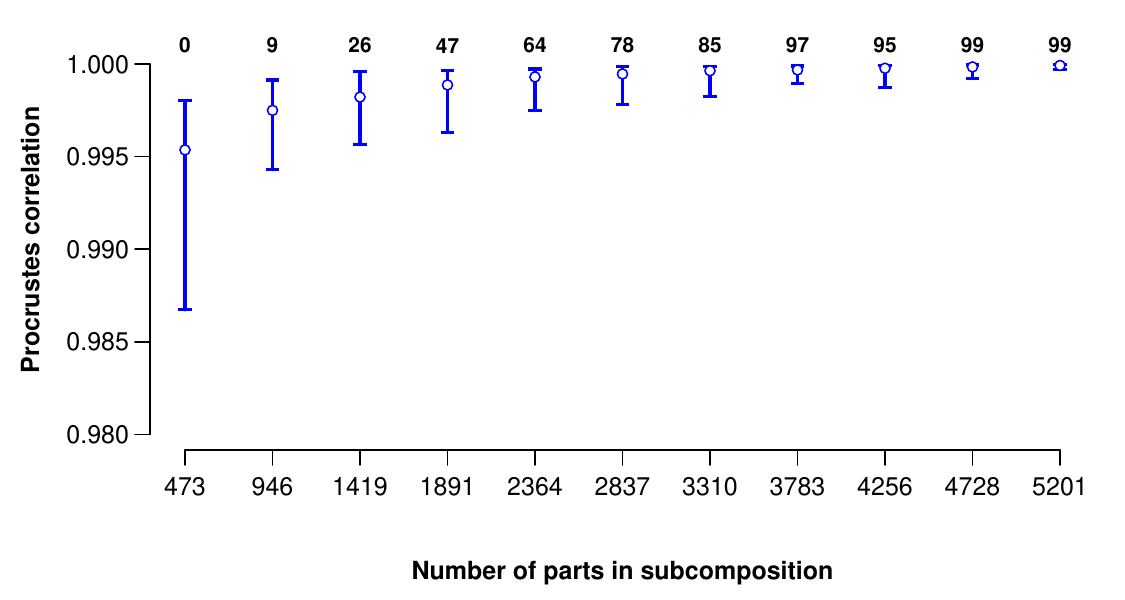}
\vspace{-0.2cm}
    \caption{Procrustes correlations between the chi-square geometry of parts in random subcompositions and their respective chi-square geometry in the original full composition of the genomics data. A 100 random subcompositions are generated of sizes 473 up to 5201, in the same proportions as the subcompositions in Fig.~\ref{Tellus_coherence_CA}. The extents of the 2.5\% and 97.5\% percentiles of these sets of 100 correlations are indicated by vertical lines, as well as their medians, shown as dots. The subcompositions show closer proximity to coherence than those of the Tellus data, which had relatively very few parts. The numbers at the top show the number of times out of the 100 subcompositions that the Procrustes correlations were $>0.999$. Note the much narrower range of the correlation scale on the vertical axis, compared with that of Fig.~\ref{Tellus_coherence_CA}.}
\label{SingleCell_CA_coherence}
\end{figure}

Fig.~\ref{Tellus_coherence_CA} showed, for the 52-part Tellus data the closeness to coherence of random subcompositions using CA, in that case without any power transformation.
The same exercise was applied in Fig.~\ref{SingleCell_CA_coherence} to the present 6147-part data set, with the same percentages of parts in the sequence of subcompositions: for example, the smallest subcomposition contained (4/52) = one-thirteenth of the 6147 parts, i.e., 473 parts. 
Even though they are the same sizes in relative terms, these subcompositions contain many more parts and it turns out that they are much closer to exact coherence when compared with their corresponding parts in the full composition.

\begin{figure}[h!]
\includegraphics[width=0.99\linewidth]{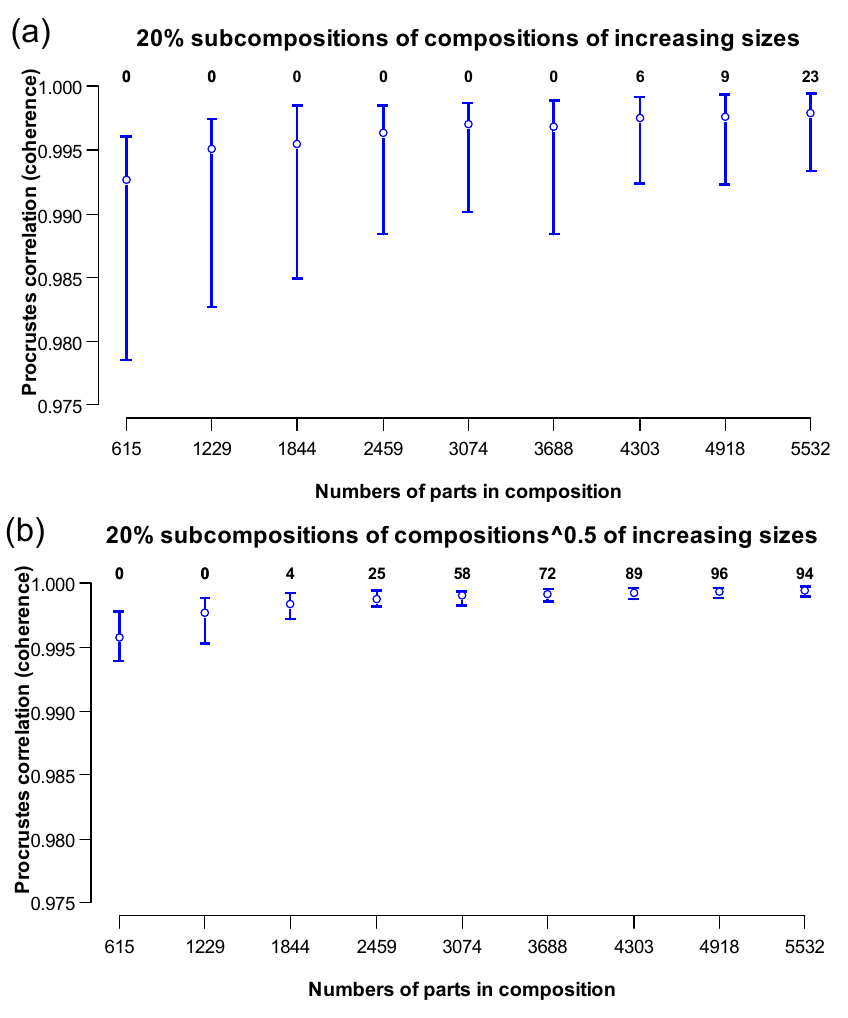}
    \caption{(a) Procrustes correlations between the chi-square geometry of parts in 20\% subcompositions of compositions of increasing sizes, and the corresponding geometry of the same parts in the full composition. A 100 random subcompositions are generated for each composition. The extents of the 2.5\% and 97.5\% percentiles of these sets of 100 correlations are indicated by vertical lines, as well as their medians, shown as dots. The subcompositions show closer proximity to coherence than those of the Tellus data, which had relatively very few parts. The numbers at the top show the number of times out of the 100 subcompositions that the Procrustes correlations were $>0.999$. (b) The same exercise, using the chi-square on Box-Cox power-transformed data (power = 0.5, i.e., square-root), showing much better coherence.}
\label{SingleCell_CA_composition}
\end{figure}

An even more convincing exercise is to consider compositions of increasingly larger sizes, and evaluate the closeness to coherence of subcompositions of a fixed percentage. 
Using the 6147-part genomics data set, Fig.~\ref{SingleCell_CA_composition}, shows the coherence of the chi-square standardization, again measured by the Procrustes correlation in each case, of 20\% subcompositions of random compositions of increasing sizes.

Finally, we also study the dilution effect for the multinomial distribution used to model the single-cell compositions. 
For this, we create a single composition from adding the cell counts of all 7852 cells annotated as thymocytes. 
We then plot the correlation between the two most abundant parts versus the size of increasing subcompositions of the 25648 nonzero genes (Figure S2). 
The multinomial correlation between two parts can be easily computed from the covariance of the multinomial distribution and is given by
\begin{equation}
 \mathrm{corr}(N_i,N_j)=-\sqrt{\frac{p_ip_j}{(1-p_i)(1-p_j)}},
\end{equation}
where $N_i$ denotes the random variable for the counts and $p_i$ their multinomial parameters for the parts in question. 
The size of the composition enters indirectly via the estimated empirical parameters. 
It can be seen that for a subcomposition consisting of more than 2000 genes, the strongest possible correlation due to the unit-sum constraint is around $-0.1$ in our example.

More generally, it can be shown that when considering $D$ randomly picked points from a sphere of radius $D$, when fixing $k$ of them, where $k$ grows no faster than ${\rm o}(D)$, they are independent standard normal random variables for $D$ tending to infinity. 
The spherical constraint is similar to the unit sum, and analogous results can be derived for variables constrained to the simplex.. 
A number of constrained distributions including the multinomial can be shown to be sufficiently close to their independent versions, with precise bounds on their variation distance, see \cite{Diaconis:87}.

\begin{figure}[h!]
\includegraphics[width=0.8\linewidth]{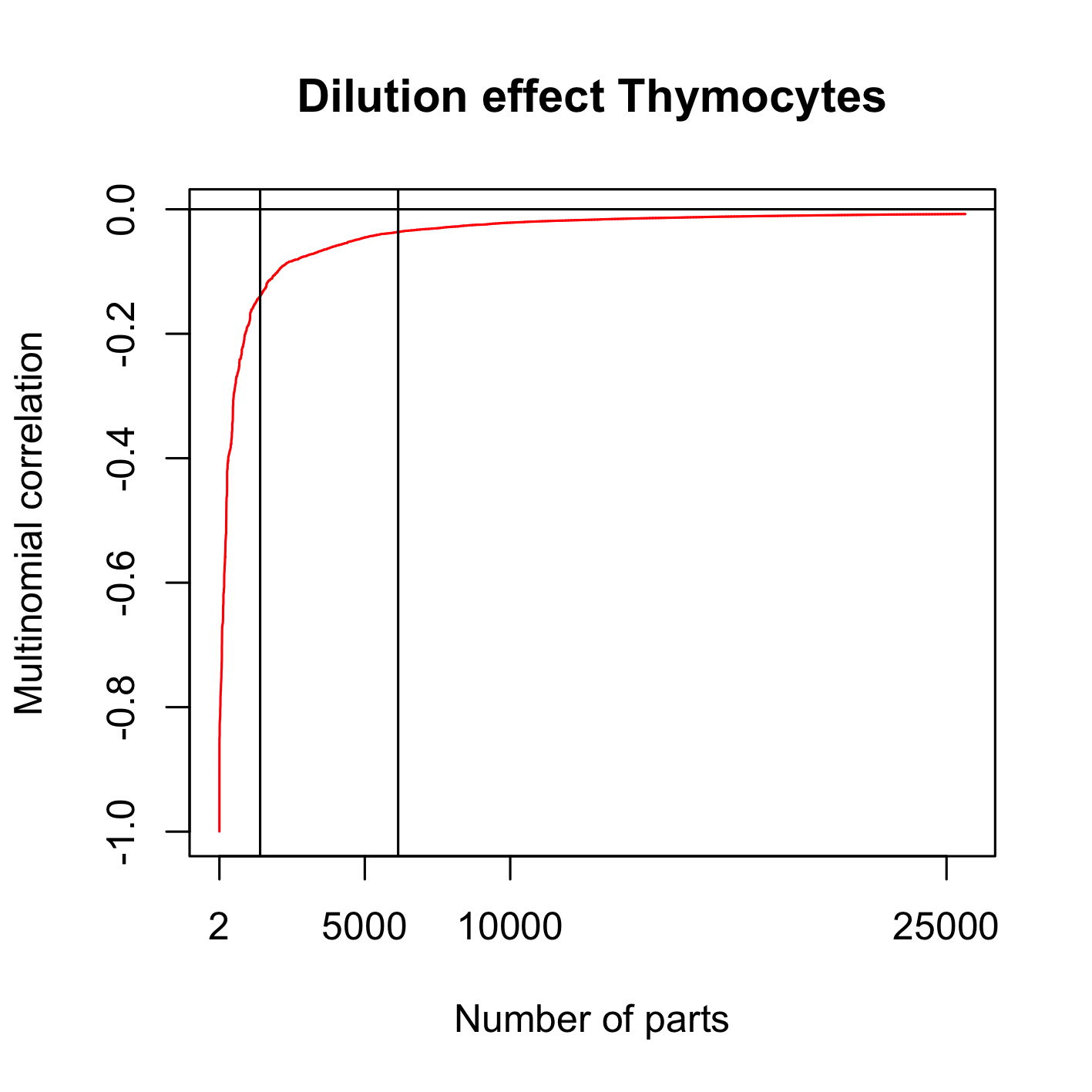}
\vspace{-0.2cm}
\caption{Convergence of multinomial correlation to zero for the example of the two most abundant parts in thymocyte cells. The black vertical lines indicate the sizes of the two subcompositions used in our analysis (1402  and 6147 genes, respectively).}
\label{fig:Dilution}
\end{figure}

\bibliographystyle{imsart-number}
\bibliography{CoDArevision}

\begin{thebibliography}{100}

\bibitem{Aitchison:82}
\begin{barticle}[author]
\bauthor{\bsnm{Aitchison},~\bfnm{J.}\binits{J.}}
(\byear{1982}).
\btitle{The statistical analysis of compositional data (with discussion)}.
\bjournal{J R Stat Soc Ser B}
\bvolume{44}
\bpages{139--77}.
\end{barticle}
\endbibitem

\bibitem{Aitchison:86}
\begin{bbook}[author]
\bauthor{\bsnm{Aitchison},~\bfnm{J.}\binits{J.}}
(\byear{1986}).
\btitle{The Statistical Analysis of Compositional Data}.
\bpublisher{Chapman \& Hall}, \baddress{London}.
\end{bbook}
\endbibitem

\bibitem{Aitchison:97}
\begin{bincollection}[author]
\bauthor{\bsnm{Aitchison},~\bfnm{J.}\binits{J.}}
(\byear{1997}).
\btitle{The one-hour course in compositional data analysis, or compositional
  data analysis is simple}.
In \bbooktitle{Proceedings of IAMG'97}
(\beditor{\bfnm{V.}\binits{V.}~\bsnm{Pawlowsky-Glahn}}, ed.)
\bpages{3--35}.
\bpublisher{International Association for Mathematical Geology}.
\end{bincollection}
\endbibitem

\bibitem{Aitchison:08}
\begin{binproceedings}[author]
\bauthor{\bsnm{Aitchison},~\bfnm{J.}\binits{J.}}
(\byear{2008}).
\btitle{The single principle of compositional data analysis, continuing
  fallacies, confusions and misunderstandings and some suggested remedies}
In \bbooktitle{Proceedings of CodaWork '08, Keynote Address}
\bpages{3--35\ URL: https://core.ac.uk/download/pdf/132548276.pdf}.
\end{binproceedings}
\endbibitem

\bibitem{AitchisonBaconShone:84}
\begin{barticle}[author]
\bauthor{\bsnm{Aitchison},~\bfnm{J.}\binits{J.}} \AND
  \bauthor{\bsnm{Bacon-Shone},~\bfnm{J.}\binits{J.}}
(\byear{1984}).
\btitle{Log constrast models for experiments with mixtures}.
\bjournal{Biometrika}
\bvolume{71}
\bpages{323--330}.
\end{barticle}
\endbibitem

\bibitem{AitchisonGreenacre:02}
\begin{barticle}[author]
\bauthor{\bsnm{Aitchison},~\bfnm{J.}\binits{J.}} \AND
  \bauthor{\bsnm{Greenacre},~\bfnm{M.}\binits{M.}}
(\byear{2002}).
\btitle{Biplots of compositional data}.
\bjournal{J R Stat Soc Ser C (Appl Stat)}
\bvolume{51}
\bpages{375--92}.
\end{barticle}
\endbibitem

\bibitem{AitchisonShen:80}
\begin{barticle}[author]
\bauthor{\bsnm{Aitchison},~\bfnm{J.}\binits{J.}} \AND
  \bauthor{\bsnm{Shen},~\bfnm{S.~M.}\binits{S.~M.}}
(\byear{1980}).
\btitle{Logistic-normal distributions: some properties and uses}.
\bjournal{Biometrika}
\bvolume{67}
\bpages{261--272}.
\end{barticle}
\endbibitem

\bibitem{Amari:16}
\begin{bbook}[author]
\bauthor{\bsnm{Amari},~\bfnm{S.}\binits{S.}}
(\byear{2016}).
\btitle{Information Geometry and Its Applications. Applied Mathematical
  Sciences (vol. 194)}.
\bpublisher{Springer}, \baddress{New York}.
\end{bbook}
\endbibitem

\bibitem{Atkinson:82}
\begin{barticle}[author]
\bauthor{\bsnm{Atkinson},~\bfnm{A.~C.}\binits{A.~C.}}
(\byear{1982}).
\btitle{Discussion of Aitchison, J. (1982), The statistical analysis of
  compositional data (with discussion)}.
\bjournal{J R Stat Soc Ser B}
\bvolume{44}
\bpages{139--177}.
\end{barticle}
\endbibitem

\bibitem{Benzecri:73}
\begin{bbook}[author]
\bauthor{\bsnm{Benzécri},~\bfnm{J.~P.}\binits{J.~P.}}
(\byear{1973}).
\btitle{L'Analyse des Données. Tôme II: L'Analyse des Correspondances}.
\bpublisher{Dunod}, \baddress{Paris}.
\end{bbook}
\endbibitem

\bibitem{Booeshaghi:22}
\begin{barticle}[author]
\bauthor{\bsnm{Booeshaghi},~\bfnm{A~Sina}\binits{A.~S.}},
  \bauthor{\bsnm{Hallgr{\'\i}msd{\'o}ttir},~\bfnm{Ingileif~B}\binits{I.~B.}},
  \bauthor{\bsnm{G{\'a}lvez-Merch{\'a}n},~\bfnm{{\'A}ngel}\binits{{\'A}.}} \AND
  \bauthor{\bsnm{Pachter},~\bfnm{Lior}\binits{L.}}
(\byear{2022}).
\btitle{Depth normalization for single-cell genomics count data}.
\bjournal{bioRxiv}.
\end{barticle}
\endbibitem

\bibitem{Breda:21}
\begin{barticle}[author]
\bauthor{\bsnm{Breda},~\bfnm{J{\'e}r{\'e}mie}\binits{J.}},
  \bauthor{\bsnm{Zavolan},~\bfnm{Mihaela}\binits{M.}} \AND
  \bauthor{\bparticle{van} \bsnm{Nimwegen},~\bfnm{Erik}\binits{E.}}
(\byear{2021}).
\btitle{Bayesian inference of gene expression states from single-cell RNA-seq
  data}.
\bjournal{Nature Biotechnology}
\bvolume{39}
\bpages{1008--1016}.
\end{barticle}
\endbibitem

\bibitem{Buccianti:15}
\begin{barticle}[author]
\bauthor{\bsnm{Buccianti},~\bfnm{A.}\binits{A.}}
(\byear{2015}).
\btitle{The FOREGS repository: Modelling variability in stream water on a
  continental scale revising classical diagrams from CoDA (compositional data
  analysis) perspective}.
\bjournal{J Geochem Expl}
\bvolume{154}
\bpages{94--104}.
\end{barticle}
\endbibitem

\bibitem{Buettner:21}
\begin{barticle}[author]
\bauthor{\bsnm{Buettner},~\bfnm{Maren}\binits{M.}},
  \bauthor{\bsnm{Ostner},~\bfnm{Johannes}\binits{J.}},
  \bauthor{\bsnm{Mueller},~\bfnm{Christian~L}\binits{C.~L.}},
  \bauthor{\bsnm{Theis},~\bfnm{Fabian~J}\binits{F.~J.}} \AND
  \bauthor{\bsnm{Schubert},~\bfnm{Benjamin}\binits{B.}}
(\byear{2021}).
\btitle{scCODA is a Bayesian model for compositional single-cell data
  analysis}.
\bjournal{Nature communications}
\bvolume{12}
\bpages{1--10}.
\end{barticle}
\endbibitem

\bibitem{Butler:08}
\begin{barticle}[author]
\bauthor{\bsnm{Butler},~\bfnm{A.}\binits{A.}} \AND
  \bauthor{\bsnm{Glasbey},~\bfnm{C.}\binits{C.}}
(\byear{2008}).
\btitle{A latent Gaussian model for compositional data with zeros}.
\bjournal{J Roy Stat Soc Ser C (Appl Stat)}
\bvolume{57}
\bpages{505--520}.
\end{barticle}
\endbibitem

\bibitem{Bona:06}
\begin{bbook}[author]
\bauthor{\bsnm{Bóna},~\bfnm{M.}\binits{M.}}
(\byear{2006}).
\btitle{A Walk Through Combinatorics: an Introduction to Enumeration and Graph
  Theory, 2nd Edition}.
\bpublisher{World Scientific Publishing}, \baddress{Singapore}.
\end{bbook}
\endbibitem

\bibitem{Coenders:22}
\begin{barticle}[author]
\bauthor{\bsnm{Coenders},~\bfnm{G.}\binits{G.}} \AND
  \bauthor{\bsnm{Greenacre},~\bfnm{M.}\binits{M.}}
(\byear{2022}).
\btitle{Three approaches to supervised learning for compositional data with
  pairwise logratios}.
\bjournal{J Appl Stat}
\bvolume{49}
\bpages{1--22}.
\bdoi{10.1080/02664763.2022.2108007}
\end{barticle}
\endbibitem

\bibitem{Coenders:20}
\begin{barticle}[author]
\bauthor{\bsnm{Coenders},~\bfnm{G.}\binits{G.}} \AND
  \bauthor{\bsnm{Pawlowsky-Glahn},~\bfnm{V.}\binits{V.}}
(\byear{2020}).
\btitle{On interpretations of tests and effect sizes in regression models with
  a compositional predictor}.
\bjournal{SORT}
\bvolume{44}
\bpages{201--220}.
\bdoi{10.2436/20.8080.02.100}
\end{barticle}
\endbibitem

\bibitem{Combettes:21}
\begin{barticle}[author]
\bauthor{\bsnm{Combettes},~\bfnm{P.~L.}\binits{P.~L.}} \AND
  \bauthor{\bsnm{Müller},~\bfnm{C.~L.}\binits{C.~L.}}
(\byear{2021}).
\btitle{Regression models for compositional data: general log-contrast
  formulations, proximal optimization, and microbiome data applications}.
\bjournal{Statistics in Biosciences}
\bvolume{13}
\bpages{217--242}.
\bdoi{10.1007/s12561-020-09283-2}
\end{barticle}
\endbibitem

\bibitem{Cortes:09}
\begin{barticle}[author]
\bauthor{\bsnm{Cortés},~\bfnm{J.~A.}\binits{J.~A.}}
(\byear{2009}).
\btitle{On the Harker variation diagrams; a comment on ``The statistical
  analysis of compositional data. Where are we and where should we be
  heading?'' by Aitchison and Egozcue (2005)}.
\bjournal{Math Geosc}
\bvolume{41}
\bpages{817--828}.
\bdoi{10.1007/s11004-009-9222-8}
\end{barticle}
\endbibitem

\bibitem{David:77}
\begin{barticle}[author]
\bauthor{\bsnm{David},~\bfnm{M.}\binits{M.}},
  \bauthor{\bsnm{Dagbert},~\bfnm{M.}\binits{M.}} \AND
  \bauthor{\bsnm{Beauchemin},~\bfnm{Y.}\binits{Y.}}
(\byear{1977}).
\btitle{Statistical analysis in geology: correspondence analysis method}.
\bjournal{Quarterly of the Colorado School of Mines}
\bvolume{72}
\bpages{11--57}.
\end{barticle}
\endbibitem

\bibitem{Diaconis:87}
\begin{barticle}[author]
\bauthor{\bsnm{Diaconis},~\bfnm{P.}\binits{P.}} \AND
  \bauthor{\bsnm{Freedman},~\bfnm{D.~A.}\binits{D.~A.}}
(\byear{1987}).
\btitle{A dozen de Finetti-style results in search of a theory}.
\bjournal{Annales de l’Institut Henri Poincaré}
\bvolume{23}
\bpages{397--423}.
\end{barticle}
\endbibitem

\bibitem{Egozcue:05}
\begin{barticle}[author]
\bauthor{\bsnm{Egozcue},~\bfnm{J.~J.}\binits{J.~J.}} \AND
  \bauthor{\bsnm{Pawlowsky-Glahn},~\bfnm{V.}\binits{V.}}
(\byear{2005}).
\btitle{Groups of parts and their balances in compositional data analysis}.
\bjournal{Math Geol}
\bvolume{37}.
\end{barticle}
\endbibitem

\bibitem{Egozcue:19}
\begin{barticle}[author]
\bauthor{\bsnm{Egozcue},~\bfnm{J.~J.}\binits{J.~J.}} \AND
  \bauthor{\bsnm{Pawlowsky-Glahn},~\bfnm{V.}\binits{V.}}
(\byear{2019}).
\btitle{Compositional data: the sample space and its structure}.
\bjournal{TEST}
\bvolume{2019}
\bpages{1-19}.
\end{barticle}
\endbibitem

\bibitem{Erb:21}
\begin{bincollection}[author]
\bauthor{\bsnm{Erb},~\bfnm{I.}\binits{I.}} \AND
  \bauthor{\bsnm{Ay},~\bfnm{N.}\binits{N.}}
(\byear{2021}).
\btitle{The information-geometric perspective of compositional data analysis}.
In \bbooktitle{Advances in Compositional Data Analysis}
(\beditor{\bfnm{P.}\binits{P.}~\bsnm{Filzmoser}},
  \beditor{\bfnm{K.}\binits{K.}~\bsnm{Hron}},
  \beditor{\bfnm{J.~A.}\binits{J.~A.}~\bsnm{Martín-Fernández}} \AND
  \beditor{\bfnm{J.}\binits{J.}~\bsnm{Palarea-Albaladejo}}, eds.)
\bpages{21--43}.
\bpublisher{Springer}, \baddress{New York}.
\end{bincollection}
\endbibitem

\bibitem{Erb:16}
\begin{barticle}[author]
\bauthor{\bsnm{Erb},~\bfnm{I.}\binits{I.}} \AND
  \bauthor{\bsnm{Notredame},~\bfnm{C.}\binits{C.}}
(\byear{2016}).
\btitle{How should we measure proportionality on relative gene expression
  data?}
\bjournal{Theory Biosc}
\bvolume{135}
\bpages{21--36}.
\end{barticle}
\endbibitem

\bibitem{Filzmoser:18}
\begin{bbook}[author]
\bauthor{\bsnm{Filzmoser},~\bfnm{P.}\binits{P.}},
  \bauthor{\bsnm{Hron},~\bfnm{K.}\binits{K.}} \AND
  \bauthor{\bsnm{Templ},~\bfnm{M.}\binits{M.}}
(\byear{2018}).
\btitle{Applied Compositional Data Analysis}.
\bpublisher{Oxford University Press}, \baddress{Oxford}.
\end{bbook}
\endbibitem

\bibitem{Fiserova:11}
\begin{barticle}[author]
\bauthor{\bsnm{Fišerová},~\bfnm{E.}\binits{E.}} \AND
  \bauthor{\bsnm{Hron},~\bfnm{K.}\binits{K.}}
(\byear{2011}).
\btitle{On the interpretation of orthonormal coordinates for compositional
  data}.
\bjournal{Math Geosci}
\bvolume{43}
\bpages{455}.
\end{barticle}
\endbibitem

\bibitem{Gabriel:72}
\begin{barticle}[author]
\bauthor{\bsnm{Gabriel},~\bfnm{K.~R.}\binits{K.~R.}}
(\byear{1972}).
\btitle{Analysis of meteorological data by means of canonical decomposition and
  biplots}.
\bjournal{J Appl Meteor Climat}
\bvolume{11}
\bpages{1071--1077}.
\end{barticle}
\endbibitem

\bibitem{Gordon-Rodriguez:21}
\begin{barticle}[author]
\bauthor{\bsnm{Gordon-Rodriguez},~\bfnm{E.}\binits{E.}},
  \bauthor{\bsnm{Quinn},~\bfnm{T.~P.}\binits{T.~P.}} \AND
  \bauthor{\bsnm{Cunningham},~\bfnm{J.~P.}\binits{J.~P.}}
(\byear{2021}).
\btitle{{Learning sparse log-ratios for high-throughput sequencing data}}.
\bjournal{Bioinformatics}.
\bnote{btab645}.
\bdoi{10.1093/bioinformatics/btab645}
\end{barticle}
\endbibitem

\bibitem{Gower:04}
\begin{bbook}[author]
\bauthor{\bsnm{Gower},~\bfnm{J.}\binits{J.}} \AND
  \bauthor{\bsnm{Dijksterhuis},~\bfnm{G.~B.}\binits{G.~B.}}
(\byear{2004}).
\btitle{Procrustes Problems}.
\bpublisher{Oxford University Press}, \baddress{New York}.
\end{bbook}
\endbibitem

\bibitem{Graeve:20}
\begin{barticle}[author]
\bauthor{\bsnm{Graeve},~\bfnm{M.}\binits{M.}} \AND
  \bauthor{\bsnm{Greenacre},~\bfnm{M.}\binits{M.}}
(\byear{2020}).
\btitle{The selection and analysis of fatty acid ratios: A new approach for the
  univariate and multivariate analysis of fatty acid trophic markers in marine
  organisms}.
\bjournal{Limnol Oceanogr Methods}
\bvolume{18}
\bpages{196--210}.
\end{barticle}
\endbibitem

\bibitem{Gralinska:22}
\begin{barticle}[author]
\bauthor{\bsnm{Gralinska},~\bfnm{Elzbieta}\binits{E.}},
  \bauthor{\bsnm{Kohl},~\bfnm{Clemens}\binits{C.}},
  \bauthor{\bsnm{Sokhandan~Fadakar},~\bfnm{Bita}\binits{B.}} \AND
  \bauthor{\bsnm{Vingron},~\bfnm{Martin}\binits{M.}}
(\byear{2022}).
\btitle{Visualizing cluster-specific genes from single-cell transcriptomics
  data Using association plots}.
\bjournal{Journal of Molecular Biology}
\bvolume{434}
\bpages{167525}.
\end{barticle}
\endbibitem

\bibitem{Greenacre:03}
\begin{barticle}[author]
\bauthor{\bsnm{Greenacre},~\bfnm{M.}\binits{M.}}
(\byear{2003}).
\btitle{Singular value decomposition of matched matrices}.
\bjournal{J Appl Stat}
\bvolume{30}
\bpages{1--13}.
\end{barticle}
\endbibitem

\bibitem{Greenacre:09}
\begin{barticle}[author]
\bauthor{\bsnm{Greenacre},~\bfnm{M.}\binits{M.}}
(\byear{2009}).
\btitle{Power transformations in correspondence analysis}.
\bjournal{Comp Stat Data Anal}
\bvolume{53}
\bpages{3107--16}.
\end{barticle}
\endbibitem

\bibitem{Greenacre:10a}
\begin{barticle}[author]
\bauthor{\bsnm{Greenacre},~\bfnm{M.}\binits{M.}}
(\byear{2010}).
\btitle{Log-ratio analysis is a limiting case of correspondence analysis}.
\bjournal{Math Geosc}
\bvolume{42}
\bpages{129--34}.
\end{barticle}
\endbibitem

\bibitem{Greenacre:11a}
\begin{barticle}[author]
\bauthor{\bsnm{Greenacre},~\bfnm{M.}\binits{M.}}
(\byear{2011}).
\btitle{Measuring subcompositional incoherence}.
\bjournal{Math Geosc}
\bvolume{43}
\bpages{681--93}.
\end{barticle}
\endbibitem

\bibitem{Greenacre:13}
\begin{barticle}[author]
\bauthor{\bsnm{Greenacre},~\bfnm{M.}\binits{M.}}
(\byear{2013}).
\btitle{Contribution biplots}.
\bjournal{J Comput Graph Stat}
\bvolume{22}
\bpages{107--22}.
\end{barticle}
\endbibitem

\bibitem{Greenacre:16b}
\begin{barticle}[author]
\bauthor{\bsnm{Greenacre},~\bfnm{M.}\binits{M.}}
(\byear{2016}).
\btitle{Data reporting and visualization in ecology}.
\bjournal{Polar Biol}
\bvolume{39}
\bpages{2189--2205}.
\end{barticle}
\endbibitem

\bibitem{Greenacre:16a}
\begin{bbook}[author]
\bauthor{\bsnm{Greenacre},~\bfnm{M.}\binits{M.}}
(\byear{2016}).
\btitle{Correspondence Analysis in Practice (3rd edition)}.
\bpublisher{Chapman \& Hall / CRC Press}, \baddress{Boca Raton, Florida}.
\end{bbook}
\endbibitem

\bibitem{Greenacre:17}
\begin{barticle}[author]
\bauthor{\bsnm{Greenacre},~\bfnm{M.}\binits{M.}}
(\byear{2017}).
\btitle{‘Size’ and ‘shape’ in the measurement of multivariate
  proximity}.
\bjournal{Methods in Ecology and Evolution}
\bvolume{8}
\bpages{1415--1424}.
\bdoi{https://doi.org/10.1111/2041-210X.12776}
\end{barticle}
\endbibitem

\bibitem{Greenacre:18}
\begin{bbook}[author]
\bauthor{\bsnm{Greenacre},~\bfnm{M.}\binits{M.}}
(\byear{2018}).
\btitle{Compositional Data Analysis in Practice}.
\bpublisher{Chapman \& Hall / CRC Press}, \baddress{Boca Raton, Florida}.
\end{bbook}
\endbibitem

\bibitem{Greenacre:19}
\begin{barticle}[author]
\bauthor{\bsnm{Greenacre},~\bfnm{M.}\binits{M.}}
(\byear{2019}).
\btitle{Variable selection in compositional data analysis using pairwise
  logratios}.
\bjournal{Math Geosc}
\bvolume{51}
\bpages{649--82}.
\end{barticle}
\endbibitem

\bibitem{Greenacre:19d}
\begin{barticle}[author]
\bauthor{\bsnm{Greenacre},~\bfnm{M.}\binits{M.}}
(\byear{2019}).
\btitle{Discussion of ``Compositional data: the sample space and its
  structure'', by Egozcue and Pawlowsky-Glahn}.
\bjournal{TEST}
\bvolume{2019}
\bpages{20--24}.
\end{barticle}
\endbibitem

\bibitem{Greenacre:20}
\begin{barticle}[author]
\bauthor{\bsnm{Greenacre},~\bfnm{M.}\binits{M.}}
(\byear{2020}).
\btitle{Amalgamations are valid in compositional data analysis, can be used in
  agglomerative clustering, and their logratios have an inverse
  transformation}.
\bjournal{Appl Comput Geosc}
\bvolume{5}
\bpages{100017}.
\end{barticle}
\endbibitem

\bibitem{Greenacre:21}
\begin{barticle}[author]
\bauthor{\bsnm{Greenacre},~\bfnm{M.}\binits{M.}}
(\byear{2021}).
\btitle{Compositional data analysis}.
\bjournal{Annu Rev Stat Appl}
\bvolume{8}
\bpages{271--99}.
\end{barticle}
\endbibitem

\bibitem{Greenacre:21c}
\begin{bincollection}[author]
\bauthor{\bsnm{Greenacre},~\bfnm{M.}\binits{M.}}
(\byear{2022}).
\btitle{Compositional data analysis -- linear algebra, visualization and
  interpretation}.
In \bbooktitle{Innovations in Multivariate Statistical Modelling: Navigating
  Theoretical and Multidisciplinary Domains}
(\beditor{\bfnm{A.}\binits{A.}~\bsnm{Bekker}} \AND
  \beditor{\bfnm{J.}\binits{J.}~\bsnm{Ferreira}}, eds.)
\bpages{https://arxiv.org/abs/2110.12439}.
\bpublisher{Springer}, \baddress{New York}.
\end{bincollection}
\endbibitem

\bibitem{GreenacreGrunskyBaconShone:20}
\begin{barticle}[author]
\bauthor{\bsnm{Greenacre},~\bfnm{M.}\binits{M.}},
  \bauthor{\bsnm{Grunsky},~\bfnm{E.}\binits{E.}} \AND
  \bauthor{\bsnm{Bacon-Shone},~\bfnm{J.}\binits{J.}}
(\byear{2020}).
\btitle{A comparison of amalgamation and isometric logratios in compositional
  data analysis}.
\bjournal{Comput Geosc}
\bvolume{148}
\bpages{104621}.
\end{barticle}
\endbibitem

\bibitem{GreenacreEtAl:23}
\begin{barticle}[author]
\bauthor{\bsnm{Greenacre},~\bfnm{M.}\binits{M.}},
  \bauthor{\bsnm{Grunsky},~\bfnm{E.}\binits{E.}},
  \bauthor{\bsnm{Bacon-Shone},~\bfnm{J.}\binits{J.}},
  \bauthor{\bsnm{Erb},~\bfnm{I.}\binits{I.}} \AND
  \bauthor{\bsnm{Quinn},~\bfnm{T.}\binits{T.}}
\btitle{Supplement to ``Aitchison's compositional data analysis 40 years on: a
  reappraisal"}.
\bjournal{Stat Sci}
\bvolume{0}
\bpages{0}.
\bdoi{10.0000/1234}
\end{barticle}
\endbibitem

\bibitem{GreenacreLewi:09}
\begin{barticle}[author]
\bauthor{\bsnm{Greenacre},~\bfnm{M.}\binits{M.}} \AND
  \bauthor{\bsnm{Lewi},~\bfnm{P.}\binits{P.}}
(\byear{2009}).
\btitle{Distributional equivalence and subcompositional coherence in the
  analysis of compositional data, contingency tables and ratio-scale
  measurements}.
\bjournal{J Classif}
\bvolume{26}
\bpages{29--54}.
\end{barticle}
\endbibitem

\bibitem{GreenacreMartinezBlasco:21}
\begin{barticle}[author]
\bauthor{\bsnm{Greenacre},~\bfnm{M.}\binits{M.}},
  \bauthor{\bsnm{Mártinez-Álvaro},~\bfnm{M.}\binits{M.}} \AND
  \bauthor{\bsnm{Blasco},~\bfnm{A.}\binits{A.}}
(\byear{2021}).
\btitle{Compositional data analysis of microbiome and any-omics datasets: a
  validation of the additive logratio transformation}.
\bjournal{Front Microbiol}
\bvolume{12}
\bpages{2625}.
\bdoi{10.3389/fmicb.2021.727398}
\end{barticle}
\endbibitem

\bibitem{Grunsky:85}
\begin{barticle}[author]
\bauthor{\bsnm{Grunsky},~\bfnm{E.~C.}\binits{E.~C.}}
(\byear{1985}).
\btitle{Recognition of alteration in volcanic rocks using statistical analysis
  of lithogeochemical data}.
\bjournal{J Geochem Explor}
\bvolume{25}
\bpages{157--183}.
\end{barticle}
\endbibitem

\bibitem{Hafemeister:19}
\begin{barticle}[author]
\bauthor{\bsnm{Hafemeister},~\bfnm{Christoph}\binits{C.}} \AND
  \bauthor{\bsnm{Satija},~\bfnm{Rahul}\binits{R.}}
(\byear{2019}).
\btitle{Normalization and variance stabilization of single-cell RNA-seq data
  using regularized negative binomial regression}.
\bjournal{Genome biology}
\bvolume{20}
\bpages{1--15}.
\end{barticle}
\endbibitem

\bibitem{Haug:17}
\begin{barticle}[author]
\bauthor{\bsnm{Haug},~\bfnm{T.}\binits{T.}},
  \bauthor{\bsnm{Falk-Petersen},~\bfnm{S.}\binits{S.}},
  \bauthor{\bsnm{Greenacre},~\bfnm{M.}\binits{M.}} \AND \bauthor{\bparticle{et}
  \bsnm{al.}}
(\byear{2017}).
\btitle{Trophic level and fatty acids in harp seals compared with common minke
  whales in the Barents Sea}.
\bjournal{Marine Biol Res}
\bvolume{13}
\bpages{919H932}.
\bdoi{10.1080/17451000.2017.1313988}
\end{barticle}
\endbibitem

\bibitem{Hausser:09}
\begin{barticle}[author]
\bauthor{\bsnm{Hausser},~\bfnm{J.}\binits{J.}} \AND
  \bauthor{\bsnm{Strimmer},~\bfnm{K.}\binits{K.}}
(\byear{2009}).
\btitle{Entropy inference and the James-Stein estimator, with application to
  nonlinear gene association networks}.
\bjournal{Journal of Machine Learning Research}
\bvolume{10}
\bpages{1469--1484}.
\end{barticle}
\endbibitem

\bibitem{Hausser:21}
\begin{bmanual}[author]
\bauthor{\bsnm{Hausser},~\bfnm{J.}\binits{J.}} \AND
  \bauthor{\bsnm{Strimmer},~\bfnm{K.}\binits{K.}}
(\byear{2021}).
\btitle{entropy: Estimation of Entropy, Mutual Information and Related
  Quantities}
\bnote{R package version 1.3.1}.
\end{bmanual}
\endbibitem

\bibitem{Hellinger:09}
\begin{barticle}[author]
\bauthor{\bsnm{Hellinger},~\bfnm{E.}\binits{E.}}
(\byear{1909}).
\btitle{Neue Begründung der Theorie quadratischer Formen von unendlichvielen
  Veränderlichen.}
\bjournal{Journal für die reine und angewandte Mathematik}
\bvolume{1909}
\bpages{210--271}.
\bdoi{doi:10.1515/crll.1909.136.210}
\end{barticle}
\endbibitem

\bibitem{Hron:21}
\begin{barticle}[author]
\bauthor{\bsnm{Hron},~\bfnm{K.}\binits{K.}},
  \bauthor{\bsnm{Coenders},~\bfnm{G.}\binits{G.}},
  \bauthor{\bsnm{Filzmoser},~\bfnm{P.}\binits{P.}},
  \bauthor{\bsnm{Palarea-Albaladejo},~\bfnm{J.}\binits{J.}},
  \bauthor{\bsnm{Faměra},~\bfnm{M.}\binits{M.}} \AND
  \bauthor{\bsnm{Grygar},~\bfnm{T.~M.}\binits{T.~M.}}
(\byear{2021}).
\btitle{Analysing pairwise logratios revisited}.
\bjournal{Math Geosc}
\bvolume{54}
\bpages{URL: https://www.x-mol.com/paperRedirect/1381133593200320512}.
\end{barticle}
\endbibitem

\bibitem{Hron:17}
\begin{barticle}[author]
\bauthor{\bsnm{Hron},~\bfnm{K.}\binits{K.}},
  \bauthor{\bsnm{Filzmoser},~\bfnm{P.}\binits{P.}} \AND \bauthor{\bparticle{de}
  \bsnm{Caritat},~\bfnm{P.~et~al.}\binits{P.~e.~a.}}
(\byear{2017}).
\btitle{Weighted pivot coordinates for compositional data and their application
  to geochemical mapping}.
\bjournal{Math Geosc}
\bvolume{49}
\bpages{797--814}.
\end{barticle}
\endbibitem

\bibitem{Hsu:22}
\begin{barticle}[author]
\bauthor{\bsnm{Hsu},~\bfnm{Lauren~L}\binits{L.~L.}} \AND
  \bauthor{\bsnm{Culhane},~\bfnm{Aed{\'\i}n~C}\binits{A.~C.}}
(\byear{2022}).
\btitle{Correspondence analysis for dimension reduction, batch integration, and
  visualization of single-cell RNA-seq data}.
\bjournal{bioRxiv}
\bpages{2021--11}.
\end{barticle}
\endbibitem

\bibitem{Hubert:85}
\begin{barticle}[author]
\bauthor{\bsnm{Hubert},~\bfnm{L.}\binits{L.}} \AND
  \bauthor{\bsnm{Arabie},~\bfnm{P.}\binits{P.}}
(\byear{1985}).
\btitle{Comparing partitions}.
\bjournal{Journal of Classification}
\bvolume{2}
\bpages{193--218}.
\end{barticle}
\endbibitem

\bibitem{Jackson:97}
\begin{barticle}[author]
\bauthor{\bsnm{Jackson},~\bfnm{D.~A.}\binits{D.~A.}}
(\byear{1997}).
\btitle{Compositional data in community ecology: the paradigm or peril of
  proportions?}
\bjournal{Ecology}
\bvolume{78}
\bpages{929--940}.
\end{barticle}
\endbibitem

\bibitem{Kraft:17}
\begin{barticle}[author]
\bauthor{\bsnm{Kraft},~\bfnm{A.}\binits{A.}},
  \bauthor{\bsnm{Graeve},~\bfnm{M.}\binits{M.}},
  \bauthor{\bsnm{Janssen},~\bfnm{D.}\binits{D.}} \AND \bauthor{\bparticle{et}
  \bsnm{al.}}
(\byear{2017}).
\btitle{Arctic pelagic amphipods: lipid dynamics and life strategy}.
\bjournal{J Plankton Res}
\bvolume{37}
\bpages{790--807}.
\end{barticle}
\endbibitem

\bibitem{Krzanowski:87}
\begin{barticle}[author]
\bauthor{\bsnm{Krzanowski},~\bfnm{W.}\binits{W.}}
(\byear{1987}).
\btitle{Selection of variables to preserve multivariate data structure, using
  principal components}.
\bjournal{J R Stat Soc Ser C (Appl Stat)}
\bvolume{36}
\bpages{22--33}.
\end{barticle}
\endbibitem

\bibitem{Kynclova:17}
\begin{barticle}[author]
\bauthor{\bsnm{Kynčlova},~\bfnm{P.}\binits{P.}},
  \bauthor{\bsnm{Hron},~\bfnm{K.}\binits{K.}} \AND
  \bauthor{\bsnm{Filzmoser},~\bfnm{P.}\binits{P.}}
(\byear{2017}).
\btitle{Correlation between compositional parts based on symmetric balances}.
\bjournal{Math Geosc}
\bvolume{49}
\bpages{777--796}.
\end{barticle}
\endbibitem

\bibitem{Lewi:76}
\begin{barticle}[author]
\bauthor{\bsnm{Lewi},~\bfnm{P.~J.}\binits{P.~J.}}
(\byear{1976}).
\btitle{Spectral mapping, a technique for classifying biological activity
  profiles of chemical compounds}.
\bjournal{Arz Forsch}
\bvolume{26}
\bpages{1295--300}.
\end{barticle}
\endbibitem

\bibitem{Lewi:86}
\begin{barticle}[author]
\bauthor{\bsnm{Lewi},~\bfnm{P.~J.}\binits{P.~J.}}
(\byear{1986}).
\btitle{Analysis of biological activity profiles by Spectramap}.
\bjournal{Eur J Med Chem}
\bvolume{21}
\bpages{155--62}.
\end{barticle}
\endbibitem

\bibitem{Lewi:05}
\begin{barticle}[author]
\bauthor{\bsnm{Lewi},~\bfnm{P.~J.}\binits{P.~J.}}
(\byear{2005}).
\btitle{Spectral mapping, a personal and historical account of an adventure in
  multivariate data analysis}.
\bjournal{Chem Intell Lab Syst}
\bvolume{77}
\bpages{215--23}.
\end{barticle}
\endbibitem

\bibitem{Lovell:15}
\begin{barticle}[author]
\bauthor{\bsnm{Lovell},~\bfnm{D.}\binits{D.}},
  \bauthor{\bsnm{Pawlowsky-Glahn},~\bfnm{V.}\binits{V.}},
  \bauthor{\bsnm{Egozcue},~\bfnm{J.~J.}\binits{J.~J.}},
  \bauthor{\bsnm{Marguerat},~\bfnm{S.}\binits{S.}} \AND
  \bauthor{\bsnm{B\"ahler},~\bfnm{J.}\binits{J.}}
(\byear{2015}).
\btitle{Proportionality: a valid alternative to correlation for relative data}.
\bjournal{PLoS Comp Biol}
\bvolume{11}
\bpages{e1004075}.
\end{barticle}
\endbibitem

\bibitem{Luecken:19}
\begin{barticle}[author]
\bauthor{\bsnm{Luecken},~\bfnm{Malte~D}\binits{M.~D.}} \AND
  \bauthor{\bsnm{Theis},~\bfnm{Fabian~J}\binits{F.~J.}}
(\byear{2019}).
\btitle{Current best practices in single-cell RNA-seq analysis: a tutorial}.
\bjournal{Molecular systems biology}
\bvolume{15}
\bpages{e8746}.
\end{barticle}
\endbibitem

\bibitem{Martin:18}
\begin{barticle}[author]
\bauthor{\bsnm{Martín-Fernández},~\bfnm{J.~A.}\binits{J.~A.}},
  \bauthor{\bsnm{Pawlowsky-Glahn},~\bfnm{V.}\binits{V.}},
  \bauthor{\bsnm{Egozcue},~\bfnm{J.~J.}\binits{J.~J.}} \AND
  \bauthor{\bsnm{Tolosana-Delgado},~\bfnm{R.}\binits{R.}}
(\byear{2018}).
\btitle{Advances in principal balances for compositional data}.
\bjournal{Math Geosc}
\bvolume{50}
\bpages{273--298}.
\end{barticle}
\endbibitem

\bibitem{Martinez2:21}
\begin{barticle}[author]
\bauthor{\bsnm{Martínez-Álvaro},~\bfnm{M.}\binits{M.}},
  \bauthor{\bsnm{Auffret},~\bfnm{M.~D.}\binits{M.~D.}},
  \bauthor{\bsnm{Duthie},~\bfnm{C.~A.}\binits{C.~A.}},
  \bauthor{\bsnm{Dewhurst},~\bfnm{R.}\binits{R.}},
  \bauthor{\bsnm{Cleveland},~\bfnm{M.}\binits{M.}},
  \bauthor{\bsnm{Watson},~\bfnm{M.}\binits{M.}} \AND
  \bauthor{\bsnm{Roehe},~\bfnm{R.}\binits{R.}}
(\byear{2021}).
\btitle{Bovine host genome acts on specific metabolism, communication and
  genetic processes of rumen microbes host-genomically linked to methane
  emissions}.
\bjournal{\it Submitted for publication}
\bpages{Preprint: https://www.researchsquare.com/article/rs-290150/v1}.
\end{barticle}
\endbibitem

\bibitem{Martinez:21}
\begin{barticle}[author]
\bauthor{\bsnm{Martínez-Álvaro},~\bfnm{M.}\binits{M.}},
  \bauthor{\bsnm{Zubiri-Gaitán},~\bfnm{A.}\binits{A.}},
  \bauthor{\bsnm{Hernández},~\bfnm{P.}\binits{P.}},
  \bauthor{\bsnm{Greenacre},~\bfnm{M.}\binits{M.}},
  \bauthor{\bsnm{Ferrer},~\bfnm{A.}\binits{A.}} \AND
  \bauthor{\bsnm{Blasco},~\bfnm{A.}\binits{A.}}
(\byear{2021}).
\btitle{Comprehensive comparison of the cecum microbiome functional core in
  genetically obese and lean hosts under similar environmental conditions}.
\bjournal{Accepted by {\it Communications Biology}}.
\end{barticle}
\endbibitem

\bibitem{McGregor:20}
\begin{barticle}[author]
\bauthor{\bsnm{McGregor},~\bfnm{Kevin}\binits{K.}},
  \bauthor{\bsnm{Labbé},~\bfnm{Aurélie}\binits{A.}} \AND
  \bauthor{\bsnm{Greenwood},~\bfnm{Celia M.~T.}\binits{C.~M.~T.}}
(\byear{2020}).
\btitle{MDiNE: a model to estimate differential co-occurrence networks in
  microbiome studies}.
\bjournal{Bioinformatics}
\bvolume{36}
\bpages{1840--1847}.
\end{barticle}
\endbibitem

\bibitem{McKinley:17}
\begin{barticle}[author]
\bauthor{\bsnm{McKinley},~\bfnm{J.~M.}\binits{J.~M.}},
  \bauthor{\bsnm{Grunsky},~\bfnm{E.~C.}\binits{E.~C.}} \AND
  \bauthor{\bsnm{Mueller},~\bfnm{J.~A.}\binits{J.~A.}}
(\byear{2017}).
\btitle{Environmental monitoring and peat assessment using a multivariate
  analysis of regional-scale geochemical data}.
\bjournal{Math Geosc}
\bvolume{50}
\bpages{235--246}.
\end{barticle}
\endbibitem

\bibitem{Meier:16}
\begin{barticle}[author]
\bauthor{\bsnm{Meier},~\bfnm{S.}\binits{S.}},
  \bauthor{\bsnm{Falk-Petersen},~\bfnm{S.}\binits{S.}},
  \bauthor{\bsnm{Gade-Sørensen},~\bfnm{L.~A.}\binits{L.~A.}} \AND
  \bauthor{\bparticle{et} \bsnm{al.}}
(\byear{2016}).
\btitle{Fatty acids in common minke whale (Balaenoptera acutorostrata) blubber
  reflect the feeding area and food selection, but also high endogenous
  metabolism}.
\bjournal{Marine Biol}.
\end{barticle}
\endbibitem

\bibitem{Murtagh:84}
\begin{barticle}[author]
\bauthor{\bsnm{Murtagh},~\bfnm{F.}\binits{F.}}
(\byear{1984}).
\btitle{Counting dendrograms: a survey}.
\bjournal{Discrete Appl Math}
\bvolume{7}
\bpages{191--199}.
\end{barticle}
\endbibitem

\bibitem{Palarea:15}
\begin{barticle}[author]
\bauthor{\bsnm{Palarea-Albaladejo},~\bfnm{J}\binits{J.}} \AND
  \bauthor{\bsnm{Martin-Fernandez},~\bfnm{JA}\binits{J.}}
(\byear{2015}).
\btitle{zCompositions -- R package for multivariate imputation of left-censored
  data under a compositional approach}.
\bjournal{Chemometrics and Intelligent Laboratory Systems}
\bvolume{143}
\bpages{85--96}.
\end{barticle}
\endbibitem

\bibitem{PawlowskyBuccianti:11}
\begin{bbook}[author]
\bauthor{\bsnm{Pawlowsky-Glahn},~\bfnm{V.}\binits{V.}} \AND
  \bauthor{\bsnm{Buccianti},~\bfnm{A.}\binits{A.}}
(\byear{2011}).
\btitle{Compositional Data Analysis: Theory and Applications}.
\bpublisher{Wiley}, \baddress{UK}.
\end{bbook}
\endbibitem

\bibitem{PawlowskyEtAl:15}
\begin{bbook}[author]
\bauthor{\bsnm{Pawlowsky-Glahn},~\bfnm{V.}\binits{V.}},
  \bauthor{\bsnm{Egozcue},~\bfnm{J.~J.}\binits{J.~J.}} \AND
  \bauthor{\bsnm{Tolosana-Delgado},~\bfnm{R.}\binits{R.}}
(\byear{2015}).
\btitle{Modeling and Analysis of Compositional Data}.
\bpublisher{Wiley}, \baddress{UK}.
\end{bbook}
\endbibitem

\bibitem{QuinnErb:20}
\begin{barticle}[author]
\bauthor{\bsnm{Quinn},~\bfnm{T.~P.}\binits{T.~P.}} \AND
  \bauthor{\bsnm{Erb},~\bfnm{I.}\binits{I.}}
(\byear{2020}).
\btitle{{Amalgams: data-driven amalgamation for the dimensionality reduction of
  compositional data}}.
\bjournal{NAR Genomics and Bioinformatics}
\bvolume{2}.
\bnote{lqaa076}.
\bdoi{10.1093/nargab/lqaa076}
\end{barticle}
\endbibitem

\bibitem{Quinn:18}
\begin{barticle}[author]
\bauthor{\bsnm{Quinn},~\bfnm{T.~P.}\binits{T.~P.}},
  \bauthor{\bsnm{Erb},~\bfnm{I.}\binits{I.}},
  \bauthor{\bsnm{Richardson},~\bfnm{M.~F.}\binits{M.~F.}} \AND
  \bauthor{\bsnm{Crowley},~\bfnm{T.~M.}\binits{T.~M.}}
(\byear{2018}).
\btitle{Understanding sequencing data as compositions: an outlook and review}.
\bjournal{Bioinformatics}
\bvolume{34}
\bpages{2870--8}.
\bdoi{10.1093/bioinformatics/bty175}
\end{barticle}
\endbibitem

\bibitem{Quinn:17}
\begin{barticle}[author]
\bauthor{\bsnm{Quinn},~\bfnm{T.~P.}\binits{T.~P.}},
  \bauthor{\bsnm{Richardson},~\bfnm{M.~F.}\binits{M.~F.}},
  \bauthor{\bsnm{Lovell},~\bfnm{D.}\binits{D.}} \AND
  \bauthor{\bsnm{Crowley},~\bfnm{T.~M.}\binits{T.~M.}}
(\byear{2017}).
\btitle{propr: an r-package for identifying proportionally abundant features
  using compositional data analysis}.
\bjournal{Sci Rep}
\bvolume{7}
\bpages{16252--16259}.
\end{barticle}
\endbibitem

\bibitem{Rand:71}
\begin{barticle}[author]
\bauthor{\bsnm{Rand},~\bfnm{W.~M.}\binits{W.~M.}}
(\byear{1971}).
\btitle{Objective criteria for the evaluation of clustering methods}.
\bjournal{J Amer Stat Assoc}
\bvolume{66}
\bpages{846--850}.
\end{barticle}
\endbibitem

\bibitem{Ren:17}
\begin{barticle}[author]
\bauthor{\bsnm{Ren},~\bfnm{B.}\binits{B.}},
  \bauthor{\bsnm{Bacallado},~\bfnm{S.}\binits{S.}},
  \bauthor{\bsnm{Favaro},~\bfnm{S.}\binits{S.}},
  \bauthor{\bsnm{Holmes},~\bfnm{S.}\binits{S.}} \AND
  \bauthor{\bsnm{Trippa},~\bfnm{L.}\binits{L.}}
(\byear{2017}).
\btitle{Bayesian nonparametric ordination for the analysis of microbial
  communities}.
\bjournal{J Amer Stat Assoc}
\bvolume{112}
\bpages{1430--1442}.
\bnote{PMID: 29430070}.
\bdoi{10.1080/01621459.2017.1288631}
\end{barticle}
\endbibitem

\bibitem{Rey:21}
\begin{barticle}[author]
\bauthor{\bsnm{Rey},~\bfnm{F.}\binits{F.}},
  \bauthor{\bsnm{Greenacre},~\bfnm{M.}\binits{M.}},
  \bauthor{\bsnm{Silva~Neto},~\bfnm{G~M}\binits{G.~M.}},
  \bauthor{\bsnm{Bueno-Pardo},~\bfnm{J.}\binits{J.}},
  \bauthor{\bsnm{Domingues},~\bfnm{M~R}\binits{M.~R.}} \AND
  \bauthor{\bsnm{Calado},~\bfnm{R.}\binits{R.}}
(\byear{2021}).
\btitle{Fatty acid ratio analysis identifies changes in competent
  meroplanktonic larvae sampled over different supply events}.
\bjournal{Mar Environ Res}
\bvolume{172}
\bpages{accepted for publication}.
\end{barticle}
\endbibitem

\bibitem{ScealyWelsh:11}
\begin{barticle}[author]
\bauthor{\bsnm{Scealy},~\bfnm{J.~L.}\binits{J.~L.}} \AND
  \bauthor{\bsnm{Walsh},~\bfnm{A.~H.}\binits{A.~H.}}
(\byear{2011}).
\btitle{Regression for compositional data by using distributions defined on the
  hypersphere}.
\bjournal{J R Stat Soc Ser B}
\bvolume{73}
\bpages{351--375}.
\end{barticle}
\endbibitem

\bibitem{ScealyWelsh:14}
\begin{barticle}[author]
\bauthor{\bsnm{Scealy},~\bfnm{J.~L.}\binits{J.~L.}} \AND
  \bauthor{\bsnm{Walsh},~\bfnm{A.~H.}\binits{A.~H.}}
(\byear{2014}).
\btitle{Colours and cocktails: Compositional data analysis}.
\bjournal{Aust N Z J Stat}
\bvolume{56(2)}
\bpages{145--169}.
\end{barticle}
\endbibitem

\bibitem{Smithson:22}
\begin{barticle}[author]
\bauthor{\bsnm{Smithson},~\bfnm{M.}\binits{M.}} \AND
  \bauthor{\bsnm{Broomell},~\bfnm{S.~B.}\binits{S.~B.}}
(\byear{2022}).
\btitle{Compositional data analysis tutorial}.
\bjournal{Psych Meth}
\bvolume{27}
\bpages{accepted for publication}.
\end{barticle}
\endbibitem

\bibitem{Smyth:07}
\begin{bmanual}[author]
\bauthor{\bsnm{Smyth},~\bfnm{D.}\binits{D.}}
(\byear{2007}).
\btitle{Methods used in the Tellus Geochemical Mapping of Northern Ireland.}
\bpublisher{British Geological Survey},
\baddress{Open Report, OR/07/022}.
\end{bmanual}
\endbibitem

\bibitem{Stanley:19}
\begin{barticle}[author]
\bauthor{\bsnm{Stanley},~\bfnm{C.~R.}\binits{C.~R.}}
(\byear{2019}).
\btitle{Molar element ratio analysis of lithogeochemical data: a toolbox for
  use in mineral exploration and mining}.
\bjournal{Geochemistry: Exploration, Environment, Analysis}
\bvolume{20}
\bpages{233--256}.
\end{barticle}
\endbibitem

\bibitem{Stephens:82}
\begin{barticle}[author]
\bauthor{\bsnm{Stephens},~\bfnm{M.~A.}\binits{M.~A.}}
(\byear{1982}).
\btitle{Use of the von Mises distribution to analyse continuous proportions}.
\bjournal{Biometrika}
\bvolume{69}
\bpages{197--203}.
\end{barticle}
\endbibitem

\bibitem{Svensson:20}
\begin{barticle}[author]
\bauthor{\bsnm{Svensson},~\bfnm{V.}\binits{V.}}
(\byear{2020}).
\btitle{Droplet scRNA-seq is not zero-inflated}.
\bjournal{Nat Biotechnol}
\bvolume{38}
\bpages{147--150}.
\end{barticle}
\endbibitem

\bibitem{teBeest:21}
\begin{barticle}[author]
\bauthor{\bparticle{te} \bsnm{Beest},~\bfnm{D.~E.}\binits{D.~E.}},
  \bauthor{\bsnm{Nijhuis},~\bfnm{E.~H.}\binits{E.~H.}},
  \bauthor{\bsnm{Möhlmann},~\bfnm{T.~W.~R.}\binits{T.~W.~R.}} \AND
  \bauthor{\bparticle{ter} \bsnm{Braak},~\bfnm{C.~J.~F.}\binits{C.~J.~F.}}
(\byear{2021}).
\btitle{Log-ratio analysis of microbiome data with many zeroes is library size
  dependent}.
\bjournal{Molecular Ecology Resources}
\bvolume{21}
\bpages{1866-1874}.
\bdoi{https://doi.org/10.1111/1755-0998.13391}
\end{barticle}
\endbibitem

\bibitem{R:21}
\begin{bmanual}[author]
\bauthor{\bsnm{{R Core Team}}}
(\byear{2021}).
\btitle{R: A Language and Environment for Statistical Computing}
\bpublisher{R Foundation for Statistical Computing},
\baddress{Vienna, Austria}.
\end{bmanual}
\endbibitem

\bibitem{Townes:19}
\begin{barticle}[author]
\bauthor{\bsnm{Townes},~\bfnm{F.~W.}\binits{F.~W.}},
  \bauthor{\bsnm{Hicks},~\bfnm{S.~C.}\binits{S.~C.}},
  \bauthor{\bsnm{Aryee},~\bfnm{M.~J.}\binits{M.~J.}} \AND
  \bauthor{\bsnm{Irizarry},~\bfnm{R.~A.}\binits{R.~A.}}
(\byear{2019}).
\btitle{Feature selection and dimension reduction for single-cell RNA-Seq based
  on a multinomial model}.
\bjournal{Genome Biol}
\bvolume{20}
\bpages{295}.
\end{barticle}
\endbibitem

\bibitem{BoogaartTolosana:13}
\begin{bbook}[author]
\bauthor{\bparticle{van~den} \bsnm{Boogaart},~\bfnm{K.~G.}\binits{K.~G.}} \AND
  \bauthor{\bsnm{Tolosana-Delgado},~\bfnm{R.}\binits{R.}}
(\byear{2013}).
\btitle{Analyzing Compositional Data with R}.
\bpublisher{Springer-Verlag}, \baddress{Berlin}.
\end{bbook}
\endbibitem

\bibitem{Wollenberg:77}
\begin{barticle}[author]
\bauthor{\bparticle{van~den} \bsnm{Wollenberg},~\bfnm{A.~L.}\binits{A.~L.}}
(\byear{1977}).
\btitle{Redundancy analysis, an alternative for canonical analysis}.
\bjournal{Psychometrika}
\bvolume{42}
\bpages{207--219}.
\end{barticle}
\endbibitem

\bibitem{Wood:21}
\begin{barticle}[author]
\bauthor{\bsnm{Wood},~\bfnm{J.}\binits{J.}} \AND
  \bauthor{\bsnm{Greenacre},~\bfnm{M.}\binits{M.}}
(\byear{2021}).
\btitle{Making the most of expert knowledge to analyse archaeological data: A
  case study on Parthian and Sasanian glazed pottery}.
\bjournal{Archael Anthrop Sci}
\bvolume{13}
\bpages{110}.
\end{barticle}
\endbibitem

\bibitem{Yoo:22}
\begin{barticle}[author]
\bauthor{\bsnm{Yoo},~\bfnm{Jinkyung}\binits{J.}},
  \bauthor{\bsnm{Sun},~\bfnm{Zequn}\binits{Z.}},
  \bauthor{\bsnm{Greenacre},~\bfnm{Michael}\binits{M.}},
  \bauthor{\bsnm{Mad},~\bfnm{Qin}\binits{Q.}},
  \bauthor{\bsnm{Chung},~\bfnm{Dongjun}\binits{D.}} \AND
  \bauthor{\bsnm{Kim},~\bfnm{Young~Min}\binits{Y.~M.}}
(\byear{2022}).
\btitle{A guideline for the statistical analysis of compositional data in
  immunology}.
\bjournal{Communications for Statistical Applications and Methods}
\bvolume{29}
\bpages{453-469}.
\end{barticle}
\endbibitem

\end{thebibliography}

\end{document}